\documentclass{article}

\usepackage{authblk}
\usepackage[square]{natbib}
\usepackage{yhmath}
\usepackage{mathtools, amsmath, amsthm, amssymb, amsbsy, amsfonts, amscd}
\usepackage{euscript}
\usepackage{bm}
\usepackage{color, soul}
\usepackage{empheq}
\usepackage{rotating}
\usepackage{enumerate}
\usepackage{enumitem}
\usepackage{bbold}

\usepackage{ftnxtra}
\usepackage{fnpos}
\usepackage{appendix}

\usepackage{graphicx}
\usepackage{pstool}
\usepackage{psfrag}

\usepackage{comment}

\numberwithin{equation}{section}

\theoremstyle{plain}

\theoremstyle{definition}	
 
 \newtheorem{remark}{Remark}[section]
 \newtheorem{example}{Example}[section]

\usepackage{caption2}

\usepackage{hyperref}
\hypersetup{colorlinks=true, linkcolor=blue}
\hypersetup{colorlinks=true,citecolor=blue}

\DeclareMathAlphabet{\mathpzc}{OT1}{pzc}{m}{it}

\usepackage{mathrsfs}

\definecolor{lighter_purple_mathematica}{rgb}{0.6666666666,0.33333333333,0.666666666666}

\makeatletter
\newsavebox{\@brx}
\newcommand{\llangle}[1][]{\savebox{\@brx}{\(\m@th{#1\langle}\)}%
  \mathopen{\copy\@brx\mkern2mu\kern-0.9\wd\@brx\usebox{\@brx}}}
\newcommand{\rrangle}[1][]{\savebox{\@brx}{\(\m@th{#1\rangle}\)}%
  \mathclose{\copy\@brx\mkern2mu\kern-0.9\wd\@brx\usebox{\@brx}}}%
\let\oldabs\abs
\def\abs{\@ifstar{\oldabs}{\oldabs*}}
\makeatother

\usepackage{accents}
\newcommand{\Fe}{\accentset{e}{\mathbf{F}}}

\newcommand{\Ce}{\accentset{e}{\mathbf{C}}}

\newcommand{\Ie}{\accentset{e}{I}}

\newcommand{\Fr}{\accentset{r}{\mathbf{F}}}
\newcommand{\cFr}{\accentset{r}{F}}
\newcommand{\Jr}{\accentset{r}{J}}
\newcommand{\UFr}{\accentset{r}{\boldsymbol{\mathsf{U}}}}

\newcommand{\phio}{\mathring{\varphi}}
\newcommand{\Jo}{\mathring{J}}
\newcommand{\Co}{\mathring{\mathbf{C}}}
\newcommand{\cCo}{\mathring{\mathrm C}}
\newcommand{\cFo}{\mathring{\mathrm F}}
\newcommand{\rhoo}{\mathring{\rho}}
\newcommand{\Sigmao}{\mathring{\boldsymbol{\Sigma}}}
\newcommand{\cSigmao}{\mathring{\Sigma}}

\newcommand{\uFr}{\accentset{r}{\mathbf{u}}}
\newcommand{\epsr}{\accentset{r}{\boldsymbol\epsilon}}
\newcommand{\vepsr}{\accentset{r}{\boldsymbol\varepsilon}}
\newcommand{\epse}{\accentset{e}{\boldsymbol\epsilon}}
\newcommand{\vepse}{\accentset{e}{\boldsymbol\varepsilon}}
\newcommand{\Wr}{\accentset{r}{W}}

\newcommand{\ccr}{\accentset{r}{\boldsymbol{\mathbb{c}}}}
\newcommand{\CCr}{\accentset{r}{\boldsymbol{\mathbb{C}}}}

\newcommand{\CCG}{\boldsymbol{\mathbb{G}}}
\newcommand{\epsilonr}{\accentset{r}{\boldsymbol{\epsilon}}}

\usepackage[all,cmtip]{xy}
\usepackage{xypic}

\DeclareMathOperator\erf{erf}

\begin{document}

\title{\textbf{Nonlinear Mechanics of Remodeling
}}

\author[1]{Aditya Kumar}
\author[1,2]{Arash Yavari\thanks{Corresponding author, e-mail: arash.yavari@ce.gatech.edu}}
\affil[1]{\small \textit{School of Civil and Environmental Engineering, Georgia Institute of Technology, Atlanta, GA 30332, USA}}
\affil[2]{\small \textit{The George W. Woodruff School of Mechanical Engineering, Georgia Institute of Technology, Atlanta, GA 30332, USA}}

\maketitle

\begin{abstract}
In this paper we present a large-deformation formulation of the mechanics of remodeling. Remodeling is anelasticity with an internal constraint---material evolutions that are mass and volume preserving.  In this special class of material evolutions the explicit time dependence of the energy function is via a remodeling tensor (or a set of remodeling tensors) that is (are) the internal variable(s) of the theory. 
The governing equations of remodeling solids are derived using a two-potential approach and the Lagrange-d'Alembert principle. We consider both isotropic and anisotropic solids and derive their corresponding remodeling equations. We study a particular remodeling of fiber-reinforced solids in which the fiber orientation is time dependent in the reference configuration---$SO(3)$\emph{-remodeling}. We consider an additional \emph{remodeling energy}, which is motivated by the energy spent in living systems to remodel to enhance stiffness in the direction of loading.
 We consider the examples of a solid reinforced with either one or two families of reorienting fibers and derive their remodeling equations. This is a generalization of some of the proposed remodeling equations in the literature.  
We study three examples of material remodeling, namely finite extensions and torsion of solid circular cylinders, which are universal deformation for incompressible isotropic solids and certain anisotropic solids. We consider both displacement and force-control loadings.
Detailed parametric studies are provided for the effects of various material and loading parameters on the fiber remodeling.
 In our numerical examples it is observed that during remodeling there is a competition between the action of internal strain energy function and the remodeling energy (governed by the motivation to provide extra stiffness or strength). 
For a given material, remodeling process dominated by strain energy, e.g., a material under very high loading, works to align fibers in a direction that minimizes strain energy. On the other hand, a remodeling process dominated by the remodeling energy, e.g., a material under small loading, works to align fibers in the direction of maximum principal strain according to a constitutive choice.
We finally linearize the governing equations of the remodeling theory and derive those of linear remodeling mechanics.
\end{abstract}

\begin{description}
\item[Keywords:] Nonlinear elasticity, material aging, remodeling, mechanics of growth, anelasticity, geometric mechanics.
\end{description}

\tableofcontents

\section{Introduction}

The earliest study of remodeling goes back to the nineteenth century and the work of Julius Wolff \citep{Wolff1870,Wolff1873} who suggested that bone optimizes its mechanical properties by remodeling to maximize its resistance to the load---\emph{Wolff's law} (see \citep{Ambrosi2019} for a detailed historical account of the theories of growth and remodeling).
Remodeling, growth, and aging are terms that are often used interchangeably in the literature to describe the evolution of various material properties like mass density, stiffness, strength or the stress-free state. 
The earliest continuum mechanics based model for remodeling---\emph{theory of adaptive elasticity}---is due to Cowin and collaborators \citep{Cowin1976,Hegedus1976,Cowin1978}.
Some researchers have proposed to use the term remodeling to describe the material evolution only when the mass density is not evolving \citep{ambrosi2011}. However, the mechanics of a material with evolving mechanical properties, e.g., stiffness is typically described very differently from the mechanics of a material with evolving stress-free state. The evolution of natural configuration is typically modeled through the introduction of a time-dependent internal variable (`remodeling tensor'), while the evolution of properties like initial stiffness is typically prescribed through empirical relations governing their change based on local stress/strain. In this work, we follow \citet{Epstein2009} to consider remodeling to be the evolution of the stress-free state under conditions of constant mass density, and constant mechanical properties, e.g., stiffness. Such an evolution may be a result of the growth and atrophy or aging processes at smaller length scales, but at continuum length scales we assume to only contribute to an evolution of the stress-free state. 
Thus, we construct a continuum framework to describe the mechanics of remodeling materials. 

No general framework yet exists to describe remodeling. Part of the challenge to develop a mechanical framework is that in biological materials---for which remodeling is most relevant, although natural state evolution without growth can occur in other materials---various energy generation and dissipation processes are occurring at the cellular level that are difficult to model at the continuum level. 
Most of the work in the literature utilizes empirical relations to describe the evolution of remodeling tensor based on a variety of experimental observations.
Although, two concepts often appear in these empirical relations. One is the use of activation and inhibition criteria to establish limits of remodeling. For example, \citet{Rajagopal1992} have previously presented a theory of deformation-induced remodeling that utilizes activation and inhibition criteria as a function of the deformation gradient: $\mathcal{A}(\mathbf{F}(X, t))=0$, and $\mathcal{I}(\mathbf{F}(X, t))=0$. Whether such activation criteria would exist in all cases, and be based on the deformation gradient tensor or stress tensor or strain energy density is unclear. Furthermore, under large deformations, the appropriate measure for stress and strain tensors that governs the two criteria is also unclear. The state of the material before activation and after inhibition is sometimes referred to as ``lazy zone" and the state in between activation and inhibition as ``active zone" \citep{george2018}.  

The second concept that is often used to describe evolution laws for remodeling tensors is the semi-physiological principle of homeostasis in living systems. Homeostasis is the state of steady internal physical and chemical conditions maintained by a living system---a stable equilibrium for the body, e.g., body temperature and $pH$. It is hypothesized to be the central motivation for all organic action. Hence, it has been proposed that a living body remodels to achieve a homeostatic stress, which is a particular preferred value of the stress field regulated through growth and remodeling during regular physiological conditions \citep{Goriely2017}. A linear homeostasis law for a remodeling tensor $\Fr$ may look like $\dot{\Fr}(X, t) =\boldsymbol{\mathcal{C}} \cdot [\boldsymbol{\sigma}(\mathbf{F}, \Fr; X, t) 	- \boldsymbol{\sigma}^*(X)]$, where $\boldsymbol{\mathcal{C}}$ is a fourth-order tensor representing essentially the resistance to remodel, $\boldsymbol{\sigma}$ is the Cauchy stress, and $\boldsymbol{\sigma}^*$ is the homeostatic value of the Cauchy stress.  However, in many different kinds of pathological conditions and during rapid and large changes in mechanical characteristics, homeostasis is clearly violated \citep{Goriely2017}. Hence, directly making use of this principle to derive evolution laws for remodeling is of questionable validity.

We next provide a brief description of a few specific approaches to describe remodeling in various problems (for a comprehensive review of the literature until the min 1990s see \citep{Taber1995}). 
\citet{Rachev1997} studied remodeling of arteries under hypertension. It is known that arteries respond to an increase of their internal pressure by remodeling. \citet{Rachev1997} in his model assumed that arteries remodel under the constraints that strains in the inner and outer layers, and the average circumferential stress remain unchanged due to changes in the internal pressure. His model predicted that both the wall thickness and the inner radius of the artery increase in the remodeling process.
\citet{Driessen2003, Driessen2004} proposed two models for reorientation of collagen fibers in soft tissues.  Both models utilize an empirical first-order rate equation for remodeling that does not depend on any material property.  In the first model which was applied to aortic heart valve, they proposed that fibers reorient to align themselves with the positive principal strain direction. However, they found that this model is not capable of describing the typical helical architecture of collagen fibers in arteries. So in the second model, they proposed that fibers instead align with ``preferred directions" in between principal stretch directions. The preferred directions are independent of the initial fiber orientation and depend only on the magnitude of principal stretch. For large values of stretch, the preferred directions in their model align with the principal directions of stretch. \citet{Hariton2007} proposed a similar model but one where fibers reorient according the principal stress directions.


\citet{Epstein2009,Epstein2015} clearly distinguished between remodeling and other types of material evolution, e.g., aging and pure growth. He defined a `material implant' transformation (we will call it `remodeling tensor') and defined the energy function of a remodeling body using the material implant and an initial energy function. He also studied the effect of the material implant on the symmetry group of a remodeling material. In particular, he showed that under material remodeling the material symmetry group remains essentially unchanged (more precisely, the symmetry group at time $t$ is conjugate to that at time $t=0$ through the remodeling tensor). He defined \emph{morphogenesis} to be an aging process that involves a change in the material symmetry group.

\citet{Melnik2013} considered an incompressible elastic cuboid that is reinforced by two families of mechanically equivalent fibers. The cuboid is under uniform far field normal stresses. Assuming that the fibers reorient themselves along the direction of maximum principal stretch \citep{Menzel2005}, they observed that fibers slowly reorient towards the direction of the larger load. They also showed that the final fiber orientations depend on the applied loads but not on the initial fiber orientations.

\citet{Grillo2016} and \citet{diStefano2019} presented a model for porous biological systems in which the remodeling tensor evolves as a function of the stress. Motivated by the similarity between the anelastic processes of remodeling and plasticity, they assumed that remodeling tensor behaves like a plastic strain in response to the stress. Moreover, in a separate work, \citet{Grillo2018} derived motivation from the evolution of material natural state in various phase-change phenomena to describe the reorientation of tissue fibers in response to external loading with Allen-Cahn type of partial differential equations. Allen-Cahn approach of describing a phase change can be thought of as a balance of linear momentum coupled with a balance of generalized or configurational forces \citep{friedgurtin1994,gurtin1996}. The configurational forces, which act as driving forces for remodeling, can be obtained from a description of energy storage and energy dissipation in the material. \citet{Grillo2018} chose their remodeling internal variable as the mean angle of the fibers and described the free energy change upon remodeling through a remodeling free energy density.
\citet{Topol2019} considered a hollow cylinder made of an incompressible solid with two families of mechanically equivalent fibers in a symmetric helical arrangement. They studied remodeling under a time-dependent inflation. In their formulation they define a fiber survival kernel that models fiber creation and dissolution rates, see also \citep{Topol2014,Topol2017}.

\citet{Chudnovsky1996,Chudnovsky2001} attempted a geometric modeling for material aging. They considered a four-dimensional material manifold whose metric can evolve with time and somehow model the change in the material properties of the body. It is not clear if there is any benefit in using a four dimensional setting as the time parameters in the material and current configurations are assumed to be equal. In order to find the dynamics of the material metric they use a variational approach and assume a Lagrangian density that explicitly depends on the material metric. The Euler-Largange equations corresponding to the variation of the material metric are called \emph{aging equations}.
A metric defines local distances and a material metric corresponds to natural distances in the body. A time-dependent metric has been used in anelasticity since the seminal works of \citet{Eckart1948} and \citet{Kondo1949,kondo1950dislocation}.\footnote{For bulk growth a detailed description can be seen in \citep{Yavari2010}. For accretion (surface growth) see \citep{Sozio2017,Sozio2019}.}
One may wonder if using a time-dependent material metric is the natural object that models aging/remodeling and whether it would be possible to differentiate anelasticity from aging in such a model.

 
This paper is organized as follows. In \S\ref{Sec:Kinematics}, we describe the kinematics and constitutive equations of remodeling. This is done staring from a multiplicative decomposition of the deformation gradient into an elastic and a remodeling deformation gradient, which is volume preserving. Material symmetry is discussed and the constitutive equations are written explicitly for isotropic, transversely isotropic, orthotropic, and monoclinic solids. A dissipation potential is assumed that is convex in the rate of remodeling tensor. The action of the symmetry group on an arbitrary dissipation potential is discussed. 
We also define a remodeling energy that quantifies the tendency of a material to evolve in response to the local state of strain or stress.
Balance laws are derived in \S\ref{Sec:BalanceLaws} using a two-potential approach and the Lagrange-d'Alembert principle. This gives the balance of linear momentum and a remodeling kinetic equation. We explicitly write the remodeling equation for isotropic, transversely isotropic, orthotropic, and monoclinic solids. We also derive the remodeling equation assuming that remodeling involves only reorientation of fibers in an isotropic matrix. We consider both a single family of fibers and two families of fibers that are neither necessarily orthogonal nor are mechanically equivalent. The first and second laws of thermodynamics are briefly discussed and it is shown that convexity of the dissipation potential in the rate of remodeling tensor ensures that the second law of thermodynamics is satisfied.
In \S\ref{Examples} three examples of material remodeling are carefully studied. The first example is finite extension of a solid circular bar reinforced by helical fibers. 
In the second example two families of fibers are considered for the same bar.
The third example is finite torsion of the bar in Example 1. These are all examples of universal deformations. The governing equations of the nonlinear theory are linearized with respect to an initial stress-free configuration in \S\ref{Sec:LinearizedTheory}. Conclusions are given in \S\ref{Sec:Conclusions}.

\section{Material Remodeling} \label{Sec:Kinematics}

\subsection{Kinematics}

\paragraph{Motion, reference and current configurations.}
Let us consider a body that is made of a solid that is undergoing a material evolution. A material evolution can be any time-dependent change in the reference configuration of the body. 
The body is identified with an embedded $3$-submanifold of the Euclidean ambient space $\mathcal{S}$, and is denoted by $\mathcal{B}$. Motion of the body is a one-parameter family of maps $\varphi_t:\mathcal{B}\to\mathcal{C}_t\subset\mathcal{S}$, where $\mathcal{C}_t=\varphi_t(\mathcal{B})$ is the current configuration of the body. 
A material point $X\in\mathcal{B}$ is mapped to $x=x(X,t)=\varphi_t(X)\in\mathcal{C}_t$.

\paragraph{Ambient space metric.}
In a body deformation is understood as the change of local distances between material points. An elastic deformation is locally measured with respect to a local stress-free state.
The body deforms in the Euclidean ambient space which has the flat metric $\mathbf{g}$. 
With respect to a (curvilinear) coordinate system $\{x^a\}$ the metric has the representation $\mathbf{g}=g_{ab}\,dx^a\otimes dx^b$. For example, with respect to the cylindrical coordinates $\{r,\theta,z\}$ this representation reads: $\mathbf{g}=dr\otimes dr+r^2d\theta\otimes d\theta+dz\otimes dz$. If $\{x^a\}$ are the Cartesian coordinates, $\mathbf{g}=\delta_{ab}\,dx^a\otimes dx^b=dx^1\otimes dx^1+dx^2\otimes dx^2+dx^3\otimes dx^3$. The metric tensor on a given tangent space $T_x\mathcal{S}$ is used to calculate the dot product of vectors. More specifically, given two vectors $\mathbf{u}\,,\mathbf{w}\in T_x\mathcal{S}$, their dot product is denoted by $\llangle \mathbf{u},\mathbf{w} \rrangle_{\mathbf{g}}=u^a\,w^b\,g_{ab}$. The inverse of the spatial metric is denoted as $\mathbf{g}^{\sharp}$ with components $g^{ab}$ such that $g^{ac}g_{cb}=\delta^a_b$. A metric induces natural isomorphisms between the tangent space and cotangent space, namely the flat operator that maps a vector to its corresponding co-vector ($1$-form)
\begin{equation} \label{Flat}
\begin{aligned}
  \flat:T_x\mathcal{C} &\longrightarrow T^*_x\mathcal{C} \\
  \mathbf{w}=w^a\frac{\partial}{\partial x^a} &\longmapsto \mathbf{w}^{\flat}=g_{ab}\,w^b\,dx^a\,,
\end{aligned}
\end{equation}
and the sharp operator that maps a co-vector ($1$-form) to its corresponding vector
\begin{equation} \label{Sharp}
\begin{aligned}
  \sharp:T^*_x\mathcal{C} &\longrightarrow T_x\mathcal{C} \\
  \boldsymbol{\omega}=\omega_a\,dx^a &\longmapsto  
  \boldsymbol{\omega}^{\sharp}=g^{ab}\omega_b\,\frac{\partial}{\partial x^a} \,.
\end{aligned}
\end{equation}

\paragraph{Material metric.}
When a body is stress-free in the Euclidean ambient space, the metric $\mathbf{g}$ induces the Euclidean metric $\mathbf{G}$ on $\mathcal{B}$. In this state the natural distances in the body are those that are seen by an observer in the Euclidean space. In the presence of anelastic effects, remodeling, aging, etc., the natural distances in the body may be different from those seen by the Euclidean observer. The natural distances are measured using a material metric $\mathbf{G}$ that is non-flat, in general, and explicitly or implicitly depends on the non-elastic process that the body is undergoing. In a material (curvilinear) coordinate system $\{X^A\}$, the material metric has the representation $\mathbf{G}=G_{AB}\,dX^A\otimes dX^B$. For example, if $\{X^A\}$ are Cartesian coordinates $\mathbf{G}=\delta_{AB}\,dX^A\otimes dX^B=dX^1\otimes dX^1+dX^2\otimes dX^2+dX^3\otimes dX^3$. As another example, in spherical coordinates $\{R,\Theta,\Phi\}$, $\mathbf{G}=dR\otimes dR+R^2\,d\Theta\otimes d\Theta+R^2\sin^2\Theta\,d\Phi\otimes d\Phi$. The flat and sharp operators corresponding to the material metric are defined similarly to \eqref{Flat} and \eqref{Sharp}.
The natural volume element of the Riemannian manifold $(\mathcal{B},\mathbf{G})$ at $X\in\mathcal{B}$ is denoted by $dV(X)$. The corresponding volume element in the current configuration at $x=\varphi(X)\in\mathcal{C}$ is denoted by $dv(x)$.
The Jacobian of deformation relates the deformed and undeformed Riemannian volume elements as $dv(x)=JdV(X)$, where\footnote{The natural volume form of the Riemannian manifold $(\mathcal{B},\mathbf{G})$ is a $3$-form that at $X\in\mathcal{B}$ is denoted by $\boldsymbol{\mu}_{\mathbf{G}}(X)$ and in a coordinate chart $\{X^A\}$ has the representation $\boldsymbol{\mu}_{\mathbf{G}}(X)=\sqrt{\det\mathbf{G}}\,dX^1\wedge dX^2\wedge dX^3$, where $\wedge$ is the wedge product of differential forms. The corresponding volume form in the current configuration at $x=\varphi(X)\in\mathcal{C}$ is denoted by $\boldsymbol{\mu}_{\mathbf{g}}(x)$ and in a coordinate chart $\{x^a\}$ has the representation $\boldsymbol{\mu}_{\mathbf{g}}(x)=\sqrt{\det\mathbf{G}}\,dx^1\wedge dx^2\wedge dx^3$. The Jacobian of deformation relates the deformed and undeformed Riemannian volume forms as  $\varphi^*\boldsymbol{\mu}_{\mathbf{g}}=J\,\boldsymbol{\mu}_{\mathbf{G}}$.}
\begin{equation}
	J=\sqrt{\frac{\det\mathbf{g}}{\det\mathbf{G}}}\det\mathbf{F}\,.
\end{equation}

\paragraph{Covariant derivatives.}
On a general manifold vector fields cannot be intrinsically differentiated (an intrinsic derivative of a tensor field is another tensor field independent of coordinates) unless the manifold is equipped with an extra structure---an affine connection. For a Riemannian manifold there is a unique natural connection---the Levi-Civita connection (natural in the sense that it is metric compatible and has vanishing torsion, i.e., it is symmetric). Let us denote the Levi-Civita connections associated with the metrics $\mathbf{G}$ and $\mathbf{g}$ by $\nabla^{\mathbf{G}}$ and $\nabla^{\mathbf{g}}$, respectively. For example, given vector fields $\mathbf{U}, \mathbf{W}\in T\mathcal{B}$, and $\mathbf{u}, \mathbf{w}\in T\mathcal{S}$, the covariant derivative of $\mathbf{W}$ along $\mathbf{U}$, and the covariant derivative of $\mathbf{w}$ along $\mathbf{u}$ are denoted as $\nabla^{\mathbf{G}}_{\mathbf{U}}\mathbf{W}$ and $\nabla^{\mathbf{g}}_{\mathbf{u}}\mathbf{w}$, respectively. With respect to the the local coordinate charts $\{X^A\}$ and $\{x^a\}$ they have components $W^A{}_{|B}\,U^B$, and $w^a{}_{|b}\,u^b$, respectively, where
\begin{equation}
	W^A{}_{|B}=\frac{\partial W^A}{\partial X^B}+\Gamma^A{}_{BC}\,W^C\,,\qquad
	w^a{}_{|b}=\frac{\partial w^a}{\partial x^b}+\gamma^a{}_{bc}\,w^c
	\,,
\end{equation}
and $\Gamma^A{}_{BC}$ and $\gamma^a{}_{bc}$ are Christoffel symbols of $\nabla^{\mathbf{G}}$ and $\nabla^{\mathbf{g}}$, respectively, and have the following relations with the metrics: $\gamma^a{}_{bc}=\frac{1}{2}g^{ak}\left(g_{kb,c}+g_{kc,b}-g_{bc,k}\right)$, and $\Gamma^A{}_{BC}=\frac{1}{2}G^{AK}\left(G_{KB,C}+G_{KC,B}-G_{BC,K}\right)$.

\paragraph{Velocity and acceleration.}
The material velocity is a vector field $\mathbf{V}:\mathcal{B}\times\mathbb{R}^+\to T\mathcal{C}$, defined as $\mathbf{V}(X,t)=\frac{\partial \varphi(X,t)}{\partial t}\in T_{\varphi_t(X)}\mathcal{C}$, and in components, $V^a(X,t)=\frac{\partial \varphi^a}{\partial t}(X,t)$. We write $\mathbf{V}_t(X)=\mathbf{V}(X,t)$. The spatial velocity is defined as $\mathbf{v}_t(x)=\mathbf{V}_t\circ\varphi_t^{-1}(x)\in T_x\mathcal{C}$, where $x=\varphi_t(X)$. Thus, $\mathbf{v}:\varphi_t(\mathcal{B})\times\mathbb{R}^+\rightarrow T\mathcal{C}$.
The convected velocity is defined as $\pmb{\mathscr{V}}_t=\varphi_t^*\mathbf{v}_t=T\varphi_t^{-1}\circ \mathbf{v}_t\circ\varphi_t=\mathbf{F}^{-1}\cdot\mathbf{V}$.\footnote{For linearization purposes the convected form of the balance of linear momentum is convenient and this is our motivation for reviewing the convected quantities.}
The material acceleration is defined as $\mathbf{A}(X,t)=D^{\mathbf{g}}_{t}\mathbf{V}(X,t)=\nabla^{\mathbf{g}}_{\mathbf{V}(X,t)}\mathbf{V}(X,t)\in T_{\varphi_t(X)}\mathcal{S}$, where $D^{\mathbf{g}}_{t}$ is the covariant derivative along the curve $\varphi_t(X)$ in $\mathcal{C}$. In components, $A^a=\frac{\partial V^a}{\partial t}+\gamma^a{}_{bc}V^bV^c$. 
The spatial acceleration is defined as $\mathbf{a}_t(x)=\mathbf{A}_t\circ\varphi_t^{-1}(x)\in T_x\mathcal{C}$. In components, $a^a=\frac{\partial v^a}{\partial t}+\frac{\partial v^a}{\partial x^b}v^b+\gamma^a{}_{bc}v^bv^c$.
Equivalently, the spatial acceleration can be expressed as the material time derivative of $\mathbf{v}$, i.e., $\mathbf{a}=\dot{\mathbf{v}}=\frac{\partial \mathbf{v}}{\partial t}+\nabla_{\mathbf{v}}^{\mathbf{g}}\mathbf{v}$. 
The convected acceleration is defined as \citep{Simo1988}
\begin{equation}
	\pmb{\mathscr{A}}_t=\varphi_t^*(\mathbf{a}_t)=\frac{\partial \pmb{\mathscr{V}}_t}{\partial t}
	+\nabla^{\varphi_t^*\mathbf{g}}_{\pmb{\mathscr{V}}_t}\pmb{\mathscr{V}}_t
	=\frac{\partial \pmb{\mathscr{V}}_t}{\partial t}
	+\nabla^{\mathbf{C}^{\flat}}_{\pmb{\mathscr{V}}_t}\pmb{\mathscr{V}}_t\,.
\end{equation}

\paragraph{Deformation gradient.}
The so-called deformation gradient, which is the derivative of the deformation mapping is denoted by $\mathbf{F}(X,t)=T\varphi_t(X):T_X\mathcal{B}\to T_x\mathcal{C}_t$, where $T_X\mathcal{B}$ and $T_x\mathcal{C}_t$ are the tangent spaces of $\mathcal{B}$ at $X$ and $\mathcal{C}_t$ at $x$, respectively. 
With respect to local coordinate charts $\{X^A\}$ and $\{x^a\}$ for $\mathcal{B}$ and $\mathcal{C}$, respectively, $\mathbf{F}$ has the following representation 
\begin{equation} 
	\mathbf{F}(X,t)=\frac{\partial\varphi^a(X,t)}{\partial X^A}\,\frac{\partial}{\partial x^a}\otimes dX^A
	\,.
\end{equation}
The adjoint of deformation gradient $\mathbf{F}^{\star}(X,t): T^*_x\mathcal{C}_t\to T^*_X\mathcal{B}$ is defined such that 
\begin{equation} 
	\langle \boldsymbol{\alpha},\mathbf{F}\mathbf{W}\rangle
	=\langle \mathbf{F}^{\star}\boldsymbol{\alpha},\mathbf{W}\rangle\,,\quad 
	\forall \mathbf{W}\in T_X\mathcal{B}\,,~\boldsymbol{\alpha}\in T^*_X\mathcal{C}
	\,,
\end{equation}
where $T^*_X\mathcal{B}$ and $T^*_x\mathcal{C}_t$ are the co-tangent spaces of $\mathcal{B}$ and $X$ and $\mathcal{C}_t$ at $x$, respectively, and $\langle .,. \rangle$ is the natural paring of $1$-forms and vectors, e.g., $\langle \boldsymbol{\omega},\mathbf{w}\rangle=\omega_a\,w^a$. $\mathbf{F}^{\star}$ has the following coordinate representation 
\begin{equation} 
	\mathbf{F}^{\star}(X,t)=\frac{\partial\varphi^a(X,t)}{\partial X^A} \,dX^A 
	\otimes \frac{\partial}{\partial x^a}\,.
\end{equation}
The transpose of the deformation gradient $\mathbf{F}^{\mathsf{T}}(X,t): T_x\mathcal{C}_t\to T_X\mathcal{B}$ is defined as
\begin{equation} 
	\llangle \mathbf{F}\mathbf{U},\mathbf{w} \rrangle_{\mathbf{g}}
	=\llangle \mathbf{U},\mathbf{F}^{\mathsf{T}}\mathbf{w} \rrangle_{\mathbf{G}}\,,\quad 
	\forall \mathbf{U}\in T_X\mathcal{B}\,,~\mathbf{w}\in T_X\mathcal{C}
	\,.
\end{equation}
This implies that in components $\big(F^{\mathsf{T}}\big)^A{}_a=G^{AB}F^b{}_B\,g_{ba}$, or $\mathbf{F}^{\mathsf{T}}=\mathbf{G}^{\sharp}\mathbf{F}^{\star}\mathbf{g}$.

\paragraph{Other measures of strain.}
There are different measures of strain in nonlinear elasticity and anelasticity \citep{MarsdenHughes1983,Ogden1984,Goriely2017,YavariSozio2023}. Consider two vectors in the current (deformed) configuration $\mathbf{u}, \mathbf{w}\in T_x\mathcal{C}$. Their dot product is calculated using the ambient space metric $\mathbf{g}$ as
\begin{equation} 
	\llangle \mathbf{u},\mathbf{w} \rrangle_{\mathbf{g}}
	=\llangle \mathbf{F}\mathbf{U},\mathbf{F}\mathbf{W} \rrangle_{\mathbf{g}}
	=\llangle \mathbf{U},\mathbf{W} \rrangle_{\mathbf{F}^*\mathbf{g}}
	\,,
\end{equation}
where $\mathbf{F}^*\mathbf{g}=\mathbf{F}^{\star}\mathbf{g}\mathbf{F}=\varphi^*\mathbf{g}=\mathbf{C}^{\flat}$ is the pulled-back metric or the right Cauchy-Green strain. To clarify this definition, in components
\begin{equation} 
	\llangle \mathbf{u},\mathbf{w} \rrangle_{\mathbf{g}}
	=u^aw^bg_{ab}=(F^a{}_AF^b{}_B\,g_{ab})U^AW^B
	=C_{AB}U^AW^B
	\,,
\end{equation}
and hence  $C_{AB}=F^a{}_A\,g_{ab}F^b{}_B$, which is the pulled-back metric. Note that
\begin{equation} 
	C^A{}_B=G^{AM}C_{MB}=(G^{AM}F^a{}_M\,g_{ab})F^b{}_B=\big(F^{\mathsf{T}}\big)^A{}_b\,F^b{}_B
	\,,
\end{equation}
and hence $\mathbf{C}=\mathbf{F}^{\mathsf{T}}\mathbf{F}$, which is the familiar definition of the right Cauchy-Green strain.

Next consider two vectors in the reference configuration $\mathbf{U}, \mathbf{W}\in T_X\mathcal{B}$. Their dot product is calculated using the material metric $\mathbf{G}$ as
\begin{equation} 
	\llangle \mathbf{U},\mathbf{W} \rrangle_{\mathbf{G}}
	=\llangle \mathbf{F}^{-1}\mathbf{u},\mathbf{F}^{-1}\mathbf{w} \rrangle_{\mathbf{G}}
	=\llangle \mathbf{u},\mathbf{w} \rrangle_{\mathbf{F}_*\mathbf{G}}
	\,,
\end{equation}
where $\mathbf{F}_*\mathbf{G}=\mathbf{F}^{-\star}\mathbf{G}\mathbf{F}^{-1}$ is the push-forward of the material metric and is denoted as $\mathbf{c}^{\flat}$, which is the spatial analogue of the right Cauchy-Green strain. In components, $c_{ab}=F^{-A}{}_a\,G_{AB}\,F^{-B}{}_b$.

If instead of spatial and material vectors, $1$-forms and their dot products are considered, the left Cauchy-Green strain can be defined as $\mathbf{B}^{\sharp}=\varphi^*\mathbf{g}^{\sharp}$. Its spatial analogue is defined as $\mathbf{b}^{\sharp}=\varphi_*\mathbf{G}^{\sharp}=\mathbf{F}\mathbf{G}^{\sharp}\mathbf{F}^{\star}$. In components, $B^{AB}=F^{-A}{}_a\,F^{-B}{}_b\,g^{ab}$, and $b^{ab}=F^a{}_AF^b{}_B\,G^{AB}$.
The tensor $\mathbf{b}$ is defined as $\mathbf{b}=\mathbf{b}^{\sharp}\mathbf{g}$. Similarly, $\mathbf{c}$ is defined as $\mathbf{c}=\mathbf{g}^{\sharp}\mathbf{c}^{\flat}$. Thus, $\mathbf{c}\mathbf{b}=\mathbf{g}^{\sharp}\mathbf{c}^{\flat}\mathbf{b}^{\sharp}\mathbf{g}=\mathbf{g}^{\sharp}\mathbf{F}^{-\star}\mathbf{G}\mathbf{F}^{-1}  
\mathbf{F}\mathbf{G}^{\sharp}\mathbf{F}^{\star}\mathbf{g}=\mathbf{g}^{\sharp}\mathbf{F}^{-\star}\mathbf{G}\mathbf{G}^{\sharp}\mathbf{F}^{\star}\mathbf{g}=\mathbf{g}^{\sharp}\mathbf{F}^{-\star}\mathbf{F}^{\star}\mathbf{g}=\mathbf{g}^{\sharp}\mathbf{g}=\mathrm{id}_{\mathcal{S}}$.
This means that $\mathbf{b} = \mathbf{c}^{-1}$. Similarly, $\mathbf{B} = \mathbf{C}^{-1}$.

\subsection{Constitutive equations}

For a material undergoing remodeling or aging the energy function is explicitly time dependent. For remodeling/aging solids this set is time dependent, in general.\footnote{As we will see, in remodeling the symmetry of the material is preserved in the sense that the symmetry group is time-dependent according to Noll's rule, i.e., push-forward via the remodeling tensor.} The $X$-dependence of the energy function models inhomogeneity of the body while the explicit dependence on time $t$ models material remodeling/aging. Note that $\mathbf{g}$ is a fixed background metric in the ambient space while $\mathbf{G}=\mathbf{G}(X,t)$ is a time-dependent material metric that is used to calculate the natural local distances in the body and models anelastic effects, e.g., defects, thermal strains, growth, remodeling, etc. Material mass density $\rho_0=\rho_0(X,t)$ can be explicitly time dependent, e.g., in the case of growing or aging materials.

\subsubsection{Material remodeling}

We follow \citet{Epstein2015} and define a time-dependent remodeling tensor $\Fr=\Fr(X,t)$ that at $X\in\mathcal{B}$ is a linear map from the tangent space $T_X\mathcal{B}$ to itself, i.e., $\Fr(X,t):T_X\mathcal{B} \to T_X\mathcal{B}$. It is assumed that the initial body has an energy function $W=W(X,\mathbf{F},\mathring{\mathbf{G}},\mathbf{g})$. The material evolution is called remodeling if \citep{Epstein2015}
\begin{equation} 
	\widetilde{W}(t,X,\mathbf{F},\mathring{\mathbf{G}},\mathbf{g})
	=W(X,\mathbf{F}\Fr^{-1},\mathring{\mathbf{G}},\mathbf{g}) \,.
\end{equation}
This is equivalent to assuming a multiplicative decomposition of the deformation gradient into an elastic and a remodeling part: $\mathbf{F}=\Fe\Fr$, and $W=W(X,\Fe,\mathring{\mathbf{G}},\mathbf{g})$ (see \citep{Sadik2017} and \citep{YavariSozio2023} for a detailed history of this decomposition in anelasticity). 
Notice that $\Fe$ is the push-forward of the total deformation gradient by $\Fr$, i.e., $\Fe=\Fr_*\mathbf{F}$. Thus, $\widetilde{W}(t,X,\mathbf{F},\mathring{\mathbf{G}},\mathbf{g})=W(X,\Fr_*\mathbf{F},\mathring{\mathbf{G}},\mathbf{g})$.

\subsubsection{Material metric}

Suppose the initial body is stress free. Its natural metric is the flat metric $\mathring{\mathbf{G}}$ induced from the Euclidean ambient space. At $X\in\mathcal{B}$ consider two vectors $\mathbf{U}_1$ and $\mathbf{U}_2$ in $T_X\mathcal{B}$. Thier dot product is given as $\llangle \mathbf{U}_1,\mathbf{U}_2 \rrangle_{\mathring{\mathbf{G}}}$. When the body undergoes a remodeling process at time $t$ these vectors are mapped to the vectors $\Fr\mathbf{U}_1$ and $\Fr\mathbf{U}_2$, respectively. The dot product of the new (time-dependent) vectors is calculated as
\begin{equation} 
	\llangle \Fr\mathbf{U}_1,\Fr\mathbf{U}_2 \rrangle_{\mathring{\mathbf{G}}}
	=\llangle \mathbf{U}_1,\mathbf{U}_2 \rrangle_{\Fr^*\mathring{\mathbf{G}}}
	\,.
\end{equation}
This means that $\mathbf{G}=\Fr^*\mathring{\mathbf{G}}=\Fr^{\star}\mathring{\mathbf{G}}\Fr$ (in components, $G_{AB}=\cFr^M{}_A\,\mathring{G}_{MN}\,\cFr^N{}_B$) is the metric that can be used to calculate the natural lengths and angles in the remodeling body. This is the \emph{material metric} of the remodeling body. This metric is identical to the material metric in anelasticity, which is not surprising as remodeling is a special anelastic process.

At $X\in\mathcal{B}$ and at time $t=0$ consider a volume element $dV_0(X)$. If this volume element is allowed to remodel independently of the rest of the body, at time $t$ its volume in the Euclidean ambient space would be $dV_t(X)=\Jr(X,t)\,dV_0(X)$, where
\begin{equation} 
	\Jr(X,t)=\det\Fr(X,t)\sqrt{\frac{\det\mathring{\mathbf{G}}(X)}{\det\mathring{\mathbf{G}}(X)}}=\det\Fr(X,t)
	\,.
\end{equation}
The material tensor $\Fr$ represents a remodeling process if it is volume preserving, i.e., $\Jr(X,t)=\det\Fr(X,t)=1$, for all $X\in\mathcal{B}$ and the entire remodeling time interval. In summary, remodeling is an isochoric anelastic process.

The remodeling tensor $\Fr$ can be understood as a local change of reference configuration and $\Fr_*\mathbf{F}=\mathbf{F}\Fr^{-1}$ is the transformed deformation gradient, or deformation gradient with respect to the new local reference configuration. 
The three local configurations and the linear maps between them are schematically shown in the commutative diagram of Fig.~\ref{Remodeling-Diagram}.
\begin{figure}[h] \centerline{%
\xymatrix@C=+1.50cm{
    (T_X\mathcal{B},\mathbf{G}) \ar[d]^{\Fr} \ar[dr]^{\mathbf{F}} \\
    (T_X\mathcal{B},\mathring{\mathbf{G}})  \ar[r]^{\Fe=\Fr_*\mathbf{F}}  
    &  (T_x\mathcal{C},\mathbf{g})
    }}
\caption{The local remodeling transformation.}\label{Remodeling-Diagram}
\end{figure}

The following summarizes the content of material metric in simple words. Suppose that at time $t$ the remodeling body is partitioned into many small pieces and each piece is allowed to relax independently of the rest of the body. The local relaxation map is $\Fr$. These relaxed pieces cannot be put back together in the Euclidean ambient space and this is due to the incompatibility of $\Fr$. For a local relaxed piece the natural distances and angles are measured using the flat metric of the Euclidean space $\mathring{\mathbf{G}}$. The same lengths and angles can be calculated in the global reference configuration if the pulled-back metric $\Fr^*\mathring{\mathbf{G}}$ is used. This metric has non-vanishing curvature, in general, and hence remodeling may induce residual stresses.

\subsubsection{Material Symmetry}

For the initial elastic body at time $t=0$ the material symmetry group $\mathring{\mathcal{G}}_X$ at $X\in\mathcal{B}$ with respect to the reference configuration $(\mathcal{B},\mathring{\mathbf{G}})$ is defined as 
\begin{equation} \label{Material-Sym0}
	W(X,\mathbf{F}\mathring{\mathbf{K}},\mathring{\mathbf{G}},\mathbf{g})
	=W(X,\mathbf{F},\mathring{\mathbf{G}},\mathbf{g})\,, \qquad \forall\,\,
	\mathring{\mathbf{K}}\in \mathring{\mathcal{G}}_X\leqslant \mathrm{Orth}(\mathring{\mathbf{G}})\,,
\end{equation}
for any deformation gradient $\mathbf{F}$, where $\mathring{\mathbf{K}}:T_X\mathcal{B}\rightarrow T_X\mathcal{B}$ is an invertible linear transformation, and $\mathrm{Orth}(\mathring{\mathbf{G}})=\{\mathbf{Q}: T_X\mathcal{B}\rightarrow T_X\mathcal{B}~|~ \mathbf{Q}^{\star}\mathring{\mathbf{G}}\mathbf{Q}=\mathring{\mathbf{G}} \}$, and $\mathring{\mathcal{G}}_X\leqslant \mathrm{Orth}(\mathring{\mathbf{G}})$ means that $\mathring{\mathcal{G}}_X$ is a subgroup of $\mathrm{Orth}(\mathring{\mathbf{G}})$.

Let us denote the symmetry group of the remodeling body at time $t$ by $\mathcal{G}(X,t)$, and hence
\begin{equation} \label{Material-Sym}
	\widetilde{W}(t,X,\mathbf{F}\mathbf{K},\mathring{\mathbf{G}},\mathbf{g})=
	\widetilde{W}(t,X,\mathbf{F},\mathring{\mathbf{G}},\mathbf{g})\,, \qquad \forall\,\,
	\mathbf{K}\in \mathcal{G}_X\leqslant \mathrm{Orth}(\mathring{\mathbf{G}})\,,
\end{equation}
for any deformation gradient $\mathbf{F}$, where $\mathbf{K}:T_X\mathcal{B}\rightarrow T_X\mathcal{B}$ is an invertible linear transformation. Given $\mathring{\mathbf{K}}\in \mathring{\mathcal{G}}_X$, one can write
\begin{equation} 
\begin{aligned} 
	\widetilde{W}(t,X,\mathbf{F},\mathring{\mathbf{G}},\mathbf{g})
	&=W(X,\mathbf{F}\Fr^{-1},\mathring{\mathbf{G}},\mathbf{g})  \\
	&=W(X,\mathbf{F}\Fr^{-1}\mathring{\mathbf{K}},\mathring{\mathbf{G}},\mathbf{g})  \\
	&=W(X,\mathbf{F}\Fr^{-1}\mathring{\mathbf{K}}\Fr\Fr^{-1},\mathring{\mathbf{G}},\mathbf{g})  \\
	&=\widetilde{W}(t,X,\mathbf{F}(\Fr^{-1}\mathring{\mathbf{K}}\Fr),\mathring{\mathbf{G}},\mathbf{g})  
	\,,
\end{aligned}
\end{equation}
which implies that $\Fr^{-1}\mathring{\mathbf{K}}\Fr\in\mathcal{G}(X,t)$, i.e., $\Fr^{-1}\mathring{\mathcal{G}}_X\Fr\subset\mathcal{G}(X,t)$. Now suppose $\mathbf{K}\in\mathcal{G}(X,t)$, and hence $\widetilde{W}(t,X,\mathbf{F}\mathbf{K},\mathring{\mathbf{G}},\mathbf{g})=\widetilde{W}(t,X,\mathbf{F},\mathring{\mathbf{G}},\mathbf{g})$. Thus
\begin{equation} 
\begin{aligned} 
	W(X,\mathbf{F}\Fr^{-1},\mathring{\mathbf{G}},\mathbf{g})
	&=W(X,\mathbf{F}\mathbf{K}\Fr^{-1},\mathring{\mathbf{G}},\mathbf{g})  \\
	&=W(X,\mathbf{F}\Fr^{-1}\Fr\mathbf{K}\Fr^{-1},\mathring{\mathbf{G}},\mathbf{g}) \\
	&=W(X,\mathbf{F}\Fr^{-1}(\Fr\mathbf{K}\Fr^{-1}),\mathring{\mathbf{G}},\mathbf{g}) 	\,,
\end{aligned}
\end{equation}
which implies that $\Fr\mathbf{K}\Fr^{-1}\in\mathring{\mathcal{G}}_X$, i.e., $\Fr\mathcal{G}(X,t)\Fr^{-1}\subset\mathring{\mathcal{G}}_X$, or equivalently, $\mathcal{G}(X,t)\subset\Fr^{-1}\mathring{\mathcal{G}}_X\Fr$. Therefore
\begin{equation} \label{Noll-Rule}
	\mathcal{G}(X,t) = \Fr^{-1}\mathring{\mathcal{G}}_X\Fr=\Fr^*\mathring{\mathcal{G}}_X\,,
\end{equation}
i.e., the material symmetry group at time $t$ is the pull-back of that at time $t=0$ by the remodeling tensor. This is the so-called Noll's rule \citep{Noll1958,Coleman1959,Coleman1963,Coleman1964} and is identical to what \citet{Epstein2015} obtained.

\subsubsection{Isotropic solids}

For an isotropic solid the energy function is materially covariant, i.e., if $\Xi:\mathcal{B}\to\mathcal{B}$ such that $\Xi(X)=X$, then $W(X,\Xi^*\mathbf{F},\Xi^*\mathring{\mathbf{G}},\mathbf{g})=W(X,\mathbf{F},\mathring{\mathbf{G}},\mathbf{g})$. This is a local property and one can write it as 
\begin{equation} \label{Material-Covariance}
	W(X,\mathbf{A}^*\mathbf{F},\mathbf{A}^*\mathring{\mathbf{G}},\mathbf{g})
	=W(X,\mathbf{F},\mathring{\mathbf{G}},\mathbf{g})\,,
\end{equation}
where $\mathbf{A}:T_X\mathcal{B}\to T_X\mathcal{B}$ is any invertible linear transformation. Thus 
\begin{equation} 
	\widetilde{W}(t,X,\mathbf{F},\mathring{\mathbf{G}},\mathbf{g})
	=W(X,\Fr_*\mathbf{F},\mathring{\mathbf{G}},\mathbf{g}) 
	=W(X,\Fr^*\Fr_*\mathbf{F},\Fr^*\mathring{\mathbf{G}},\mathbf{g})
	=W(X,\mathbf{F},\mathbf{G},\mathbf{g})
	\,,
\end{equation}
where $\mathbf{A}=\Fr$ was chosen and $\mathbf{G}=\Fr^*\mathring{\mathbf{G}}$ is the material metric. In coordinates $G_{AB}=\cFr^M{}_A\,\mathring{G}_{MN}\,\cFr^N{}_B$. 
Objectivity implies that $W=\hat{W}(X,\mathbf{C}^{\flat},\mathbf{G})$, where $\mathbf{C}^{\flat}=\mathbf{F}^*\mathbf{g}=\mathbf{F}^{\star}\mathbf{g}\mathbf{F}$.
Therefore, we have concluded that the energy function of an isotropic remodeling body is identical to its initial energy function if one replaces the flat initial material metric $\mathring{\mathbf{G}}$ by the (evolving) material metric $\mathbf{G}$ \citep{YavariSozio2023}.

For an isotropic solid, $W$ depends only on the principal invariants of $\mathbf{C}^{\flat}$, i.e., $W=\overline{W}(X,\mathring{I}_1,\mathring{I}_2,\mathring{I}_3)$, where
\begin{equation} \label{Remodeled-Energy}
\begin{aligned}
	\mathring{I}_1 &=\operatorname{tr}_{\mathring{\mathbf{G}}}\mathbf{C}=C^A{}_A=C_{AB}\,G^{AB}\,, \\
	\mathring{I}_2 &=\frac{1}{2}\left(I_1^2-\operatorname{tr}_{\mathring{\mathbf{G}}}\mathbf{C}^2\right)
	=\frac{1}{2}\left(I_1^2-C^A{}_B\,C^B{}_A\right)
	=\frac{1}{2}\left(I_1^2-C_{MB}\,C_{NA}\,\mathring{G}^{AM}\mathring{G}^{BN}\right)\,, \\
	\mathring{I}_3 &=\det\mathbf{C}=\frac{\det\mathbf{C}^{\flat}}{\det\mathring{\mathbf{G}}}\,.
\end{aligned}
\end{equation}
From \eqref{Remodeled-Energy}, for an isotropic remodeling body we have $W=\overline{W}(X,I_1,I_2,I_3)$, where
\begin{equation} 
\begin{aligned}
	I_1 &=\operatorname{tr}_{\mathbf{G}}\mathbf{C}^{\flat}=\mathbf{C}^{\flat}\!:\!\mathbf{G}^{\sharp}
	=C_{AB}\,G^{AB}\,, \\
	I_2 &=\frac{1}{2}\left[I_1^2-\operatorname{tr}_{\mathbf{G}}\mathbf{C}^2\right]
	=\frac{1}{2}\left(I_1^2-C_{MB}\,C_{NA}\,G^{AM}G^{BN}\right)\,, \\
	I_3 &=\frac{\det\mathbf{C}^{\flat}}{\det\mathbf{G}}
	=\frac{\det\mathbf{C}^{\flat}}{(\det\Fr)^2\det\mathring{\mathbf{G}}}
	=\frac{\det\mathbf{C}^{\flat}}{\det\mathring{\mathbf{G}}}
	=\mathring{I}_3\,.
\end{aligned}
\end{equation}

For an isotropic solid the Cauchy stress has the following representation \citep{DoyleEricksen1956}\footnote{The standard measures of stress are discussed in Remark \ref{Stress-Measures}.}
\begin{equation}
	\boldsymbol{\sigma} = \frac{2}{\sqrt{I_3}} \left[ \left(I_2\,\overline{W}_2+I_3\,\overline{W}_3\right)
	\mathbf{g}^{\sharp} 
	+\overline{W}_1\,\mathbf{b}^{\sharp}-I_3\,\overline{W}_2\,\mathbf{c}^{\sharp} \right]\,.
\end{equation}
For an incompressible isotropic solid $I_3=1$, and hence
\begin{equation}
	\boldsymbol{\sigma} = -p\,\mathbf{g}^{\sharp} 
	+2\overline{W}_1\,\mathbf{b}^{\sharp}-2\,\overline{W}_2\,\mathbf{c}^{\sharp}\,,
\end{equation}
where $p$ is the Lagrange multiplier associated with the incompressibility constraint $J=\sqrt{I_3}=1$.

\subsubsection{Anisotropic solids}

Material anisotropy can be described by the so-called structural tensors. When structural tensors are added to the list of the arguments of the energy function, it becomes an isotropic function of its arguments \citep{liu1982,boehler1987,zheng1993,zheng1994theory,Lu2000}, or materially covariant in the setting of anelasticity \citep{Lu2012,YavariSozio2023}.
We assume that the initial body has an energy function $W=W(X,\mathbf{F},\mathring{\boldsymbol{\Lambda}},\mathring{\mathbf{G}},\mathbf{g})$, where $\mathring{\boldsymbol{\Lambda}}$ is a collection of structural tensors that describe the anisotropy class of the material.
The time-dependent energy function of the remodeling body is defined as
\begin{equation} 
	\widetilde{W}(t,X,\mathbf{F},\mathring{\mathbf{G}},\mathbf{g})
	=W(X,\mathbf{F}\Fr^{-1},\mathring{\boldsymbol{\Lambda}},\mathring{\mathbf{G}},\mathbf{g}) \,.
\end{equation}
Knowing that $W$ is a materially covariant function we can write
\begin{equation} 
\begin{aligned}
	\widetilde{W}(t,X,\mathbf{F},\mathring{\mathbf{G}},\mathbf{g})
	&=W(X,\Fr_*\mathbf{F},\mathring{\boldsymbol{\Lambda}},\mathring{\mathbf{G}},\mathbf{g}) \\
	&=W(X,\Fr^*\Fr_*\mathbf{F},\Fr^*\mathring{\boldsymbol{\Lambda}}
	,\Fr^*\mathring{\mathbf{G}},\mathbf{g}) \\
	&=W(X,\mathbf{F},\Fr^*\mathring{\boldsymbol{\Lambda}}
	,\Fr^*\mathring{\mathbf{G}},\mathbf{g}) 
	\,.
\end{aligned}
\end{equation}
Therefore, for an anisotropic remodeling body
\begin{equation} \label{Remodeled-Energy-Anisotropic}
	\widetilde{W}(t,X,\mathbf{F},\mathring{\mathbf{G}},\mathbf{g})
	=W(X,\boldsymbol{\Lambda},\mathbf{G},\mathbf{g}) 
	\,,
\end{equation}
where $\boldsymbol{\Lambda}=\Fr^*\mathring{\boldsymbol{\Lambda}}$ and $\mathbf{G}=\Fr^*\mathring{\mathbf{G}}$. In other words, the functional form of the energy function of the remodeling body is identical to that of the initial body. The initial flat material metric and the initial structural tensors are replaced by their pull-backs by the remodeling tensor. This is consistent with what \citet{YavariSozio2023} derived for general anisotropic anelasticity.

\paragraph{Transversely isotropic solids.}
For the initial body the energy function has the form $W=\hat{W}(X,\mathbf{C}^\flat,\mathring{\mathbf{A}},\mathring{\mathbf{G}})$, where $\mathring{\mathbf{A}}=\mathring{\mathbf{N}}\otimes\mathring{\mathbf{N}}$ is a structural tensor \citep{Doyle1956,Spencer1982,Lu2000}. Including the structural tensor, the energy function becomes an isotropic function of its arguments, and can be rewritten as 
\begin{equation} \label{Remodeled-Energy-Anisotropic-TI}
	W=\overline{W}(\mathring{I}_1,\mathring{I}_2,\mathring{I}_3,\mathring{I}_4,\mathring{I}_5)	\,,
\end{equation}
where $\mathring{I}_1$, $\mathring{I}_2$, and $\mathring{I}_3$ are defined in \eqref{Remodeled-Energy} and
\begin{equation} 
	\mathring{I}_4=\mathring{\mathbf{N}}\cdot\mathbf{C}\cdot\mathring{\mathbf{N}}
	=\mathring{N}^A\mathring{N}^B\,C_{AB}\,,
	\qquad \mathring{I}_5=\mathring{\mathbf{N}}\cdot\mathbf{C}^2\cdot\mathring{\mathbf{N}}
	=\mathring{\mathbf{N}}\cdot\mathbf{C}^{\flat}\mathring{\mathbf{G}}^{\sharp}
	\mathbf{C}^{\flat}\cdot\mathring{\mathbf{N}}
	=\mathring{N}^A\mathring{N}^B\,C_{BM}\,C^M{}_A\,.
\end{equation}
At time $t>0$ the remodeling body has the energy function $W=\hat{W}(X,\mathbf{C}^\flat, \mathbf{A},\mathbf{G})$, where $\mathbf{A}=\mathbf{N}\otimes\mathbf{N}=\Fr^*\mathring{\mathbf{A}}=\Fr^{-1}\mathring{\mathbf{N}}\otimes\Fr^{-1}\mathring{\mathbf{N}}$. 
From \eqref{Remodeled-Energy-Anisotropic} and \eqref{Remodeled-Energy-Anisotropic-TI}, the energy function can be written as $W=\overline{W}(I_1,I_2,I_3,I_4,I_5)$, where
\begin{equation} 
\begin{aligned}
	& I_1=\mathrm{tr}\,\mathbf{C}=C^A{}_A\,,&& 
	I_2=\mathrm{det}\,\mathbf{C}~\mathrm{tr}_{\mathbf{G}}\mathbf{C}^{-1}
	=\mathrm{det}(C^A{}_B)(C^{-1})^D{}_D\,,\quad I_3=\mathrm{det}\mathbf{C}=\mathrm{det}(C^A{}_B)\, \\
	& I_4=\mathbf{N}\cdot\mathbf{C}\cdot\mathbf{N}=N^AN^B\,C_{AB}\,,&& 
	I_5=\mathbf{N}\cdot\mathbf{C}^2\cdot\mathbf{N}
	=\mathbf{N}\cdot\mathbf{C}^{\flat}\mathbf{G}^{\sharp}\mathbf{C}^{\flat}\cdot\mathbf{N}
	=N^AN^B\,C_{BM}\,C^M{}_A\,.
\end{aligned}
\end{equation}
Note that $I_4=\llangle \mathbf{N},\mathbf{N}\rrangle_{\mathbf{C}^{\flat}}=\llangle \Fr^{-1}\mathring{\mathbf{N}},\Fr^{-1}\mathring{\mathbf{N}} \rrangle_{\mathbf{C}^{\flat}}$, and $I_5=\llangle \mathbf{N},\mathbf{N}\rrangle_{\mathbf{C}^{\flat}\mathbf{G}^{\sharp}\mathbf{C}^{\flat}}=\llangle \Fr^{-1}\mathring{\mathbf{N}},\Fr^{-1}\mathring{\mathbf{N}} \rrangle_{\mathbf{C}^{\flat}\mathbf{G}^{\sharp}\mathbf{C}^{\flat}}$.

For a transversely isotropic solid, the Cauchy stress has the following representation \citep{Ericksen1954Anisotropic,Golgoon2018a,Golgoon2018b}
\begin{equation}
\begin{aligned}
	\boldsymbol{\sigma} 
	= \frac{2}{\sqrt{I_3}} \left\{ \left(I_2\,\overline{W}_2+I_3\,\overline{W}_3\right)\mathbf{g}^{\sharp} 
	+\overline{W}_1\,\mathbf{b}^{\sharp}
	-I_3\,\overline{W}_2\,\mathbf{c}^{\sharp}
	+\overline{W}_4\,\mathbf{n}\otimes\mathbf{n}
	+\overline{W}_5\left[\mathbf{n}\otimes(\mathbf{b}^{\sharp}\mathbf{g}\mathbf{n})
	+(\mathbf{b}^{\sharp}\mathbf{g}\mathbf{n})\otimes\mathbf{n} \right] \right\}\,.
\end{aligned}
\end{equation}
For an incompressible transversely isotropic solid $I_3=\Ie_3=1$, and hence
\begin{equation}
\begin{aligned}
	\boldsymbol{\sigma} 
	=-p\,\mathbf{g}^{\sharp} 
	+2\overline{W}_1\,\mathbf{b}^{\sharp}-2\overline{W}_2\,\mathbf{c}^{\sharp}
	+2\overline{W}_4\,\mathbf{n}\otimes\mathbf{n}
	+2\overline{W}_5\left[\mathbf{n}\otimes(\mathbf{b}^{\sharp}\mathbf{g}\mathbf{n})
	+(\mathbf{b}^{\sharp}\mathbf{g}\mathbf{n})\otimes\mathbf{n} \right] 
	\,,
\end{aligned}
\end{equation}
where $p$ is the Lagrange multiplier associated with the incompressibility constraint $J=\sqrt{I_3}=1$.

\paragraph{Orthotropic solids.}In an orthotropic solid, at every point there are three mutually orthogonal material preferred directions. In the initial body these are denoted by $\mathring{\mathbf{N}}_1$, $\mathring{\mathbf{N}}_2$, and $\mathring{\mathbf{N}}_3$. A possible choice for structural tensors are  $\mathring{\mathbf{A}}_1=\mathring{\mathbf{N}}_1\otimes\mathring{\mathbf{N}}_1$, $\mathring{\mathbf{A}}_2=\mathring{\mathbf{N}}_2\otimes\mathring{\mathbf{N}}_2$, and $\mathring{\mathbf{A}}_3=\mathring{\mathbf{N}}_3\otimes\mathring{\mathbf{N}}_3$. However, only two of them are independent as $\mathring{\mathbf{A}}_1+\mathring{\mathbf{A}}_2+\mathring{\mathbf{A}}_3=\mathbf{I}$. Without loss of generality we take $\mathring{\mathbf{A}}_1$ and $\mathring{\mathbf{A}}_2$ to be the independent structural tensors.
Including the structural tensors, the energy function becomes an isotropic function of its arguments, and can be rewritten as 
\begin{equation} \label{Remodeled-Energy-Anisotropic-Ort}
	W=\overline{W}(\mathring{I}_1,\mathring{I}_2,\mathring{I}_3,\mathring{I}_4,\mathring{I}_5
	,\mathring{I}_6,\mathring{I}_7)	\,,
\end{equation}
where $\mathring{I}_1$, $\mathring{I}_2$, and $\mathring{I}_3$ are defined in \eqref{Remodeled-Energy} and
\begin{equation} 
\begin{aligned}
	& \mathring{I}_4=\mathring{\mathbf{N}}_1\cdot\mathbf{C}\cdot\mathring{\mathbf{N}}_1
	=\mathring{N}_1^A\mathring{N}_1^B\,C_{AB}\,,
	&& \mathring{I}_5=\mathring{\mathbf{N}}_1\cdot\mathbf{C}^2\cdot\mathring{\mathbf{N}}_1
	=\mathring{\mathbf{N}}_1\cdot\mathbf{C}^{\flat}\mathring{\mathbf{G}}^{\sharp}
	\mathbf{C}^{\flat}\cdot\mathring{\mathbf{N}}_1
	=\mathring{N}^A_1\mathring{N}^B_1\,C_{BM}\,C^M{}_A\,,\\
	& \mathring{I}_6=\mathring{\mathbf{N}}_2\cdot\mathbf{C}\cdot\mathring{\mathbf{N}}_2
	=\mathring{N}^A_2\mathring{N}^B_2\,C_{AB}\,,
	&& \mathring{I}_7=\mathring{\mathbf{N}}_2\cdot\mathbf{C}^2\cdot\mathring{\mathbf{N}}_2
	=\mathring{\mathbf{N}}_2\cdot\mathbf{C}^{\flat}\mathring{\mathbf{G}}^{\sharp}
	\mathbf{C}^{\flat}\cdot\mathring{\mathbf{N}}_2
	=\mathring{N}^A_2\mathring{N}^B_2\,C_{BM}\,C^M{}_A\,.
\end{aligned}
\end{equation}
At time $t>0$ the remodeling body has the energy function $W=\hat{W}(X,\mathbf{C}^\flat, \mathbf{A}_1,\mathbf{A}_2,\mathbf{G})$, where $\mathbf{A}_1=\mathbf{N}_1\otimes\mathbf{N}_1=\Fr^*\mathring{\mathbf{A}}_1=\Fr^{-1}\mathring{\mathbf{N}}_1\otimes\Fr^{-1}\mathring{\mathbf{N}}_1$, and $\mathbf{A}_2=\mathbf{N}_2\otimes\mathbf{N}_2=\Fr^*\mathring{\mathbf{A}}_2=\Fr^{-1}\mathring{\mathbf{N}}_2\otimes\Fr^{-1}\mathring{\mathbf{N}}_2$. 
From \eqref{Remodeled-Energy-Anisotropic} and \eqref{Remodeled-Energy-Anisotropic-Ort}, the energy function can be written as $W=\overline{W}(I_1,I_2,I_3,I_4,I_5,I_6,I_7)$, where
\begin{equation} \label{I1-7-Definitions}
\begin{aligned}
	& I_1=\mathrm{tr}\,\mathbf{C}=C^A{}_A\,,&& 
	I_2=\mathrm{det}\,\mathbf{C}~\mathrm{tr}_{\mathbf{G}}\mathbf{C}^{-1}
	=\mathrm{det}(C^A{}_B)(C^{-1})^D{}_D\,,\quad 
	I_3=\mathrm{det}\mathbf{C}=\mathrm{det}(C^A{}_B)\,, \\
	& I_4=\mathbf{N}\cdot\mathbf{C}\cdot\mathbf{N}_1=N^A_1N^B_1\,C_{AB}\,,&& 
	I_5=\mathbf{N}_1\cdot\mathbf{C}^2\cdot\mathbf{N}_1
	=\mathbf{N}_1\cdot\mathbf{C}^{\flat}\mathbf{G}^{\sharp}\mathbf{C}^{\flat}\cdot\mathbf{N}_1
	=N^A_1N^B_1\,C_{BM}\,C^M{}_A\,, \\
	& I_6=\mathbf{N}_2\cdot\mathbf{C}\cdot\mathbf{N}_2=N^A_2N^B_2\,C_{AB}\,,&& 
	I_7=\mathbf{N}_2\cdot\mathbf{C}^2\cdot\mathbf{N}_2
	=\mathbf{N}_2\cdot\mathbf{C}^{\flat}\mathbf{G}^{\sharp}\mathbf{C}^{\flat}\cdot\mathbf{N}_2
	=N^A_2N^B_2\,C_{BM}\,C^M{}_A
	\,.
\end{aligned}
\end{equation}

For an orthotropic isotropic solid, the Cauchy stress has the following representation \citep{SmithRivlin1958,Spencer1986,Golgoon2018a,Golgoon2018b}
\begin{equation}
\begin{aligned}
	\boldsymbol{\sigma} 
	&= \frac{2}{\sqrt{I_3}} \Big\{ \left(I_2\,\overline{W}_2+I_3\,\overline{W}_3\right)\mathbf{g}^{\sharp} 
	+\overline{W}_1\,\mathbf{b}^{\sharp}
	-I_3\,\overline{W}_2\,\mathbf{c}^{\sharp}
	+\overline{W}_4\,\mathbf{n}_1\otimes\mathbf{n}_1
	+\overline{W}_5\left[\mathbf{n}_1\otimes(\mathbf{b}^{\sharp}\mathbf{g}\mathbf{n}_1)
	+(\mathbf{b}^{\sharp}\mathbf{g}\mathbf{n}_1)\otimes\mathbf{n}_1 \right] \\
	& \qquad +\overline{W}_6\,\mathbf{n}_2\otimes\mathbf{n}_2
	+\overline{W}_7\left[\mathbf{n}_2\otimes(\mathbf{b}^{\sharp}\mathbf{g}\mathbf{n}_2)
	+(\mathbf{b}^{\sharp}\mathbf{g}\mathbf{n}_2)\otimes\mathbf{n}_2 \right] \Big\} \,.
\end{aligned}
\end{equation}
For an incompressible orthotropic solid $I_3=\Ie_3=1$, and hence
\begin{equation}
\begin{aligned}
	\boldsymbol{\sigma} 
	&=-p\,\mathbf{g}^{\sharp} 
	+2\overline{W}_1\,\mathbf{b}^{\sharp}-2\overline{W}_2\,\mathbf{c}^{\sharp}
	+2\overline{W}_4\,\mathbf{n}_1\otimes\mathbf{n}_1
	+2\overline{W}_5\left[\mathbf{n}_1\otimes(\mathbf{b}^{\sharp}\mathbf{g}\mathbf{n}_1)
	+(\mathbf{b}^{\sharp}\mathbf{g}\mathbf{n}_1)\otimes\mathbf{n}_1 \right] \\
	& \qquad +2\overline{W}_6\,\mathbf{n}_2\otimes\mathbf{n}_2
	+2\overline{W}_7\left[\mathbf{n}_2\otimes(\mathbf{b}^{\sharp}\mathbf{g}\mathbf{n}_2)
	+(\mathbf{b}^{\sharp}\mathbf{g}\mathbf{n}_2)\otimes\mathbf{n}_2 \right] 
	\,,
\end{aligned}
\end{equation}
where $p$ is the Lagrange multiplier associated with the incompressibility constraint $J=\sqrt{I_3}=1$.

\paragraph{Monoclinic solids.}A monoclinic solid in its initial state has three material preferred directions $\mathring{\mathbf{N}}_1(X)$, $\mathring{\mathbf{N}}_2(X)$, and $\mathring{\mathbf{N}}_3(X)$ such that $\mathring{\mathbf{N}}_1\cdot\mathring{\mathbf{N}}_2\neq 0$ and $\mathring{\mathbf{N}}_3$ is normal to the plane of $\mathring{\mathbf{N}}_1$ and $\mathring{\mathbf{N}}_2$ \citep{Merodio2020}. 
The energy function of a monoclinic solid depends on nine invariants \citep{Spencer1986}:
\begin{equation}
	W = W(X,\mathring{I}_1,\mathring{I}_2,\mathring{I}_3,\mathring{I}_4,\mathring{I}_5,\mathring{I}_6,
	\mathring{I}_7,\mathring{I}_8,\mathring{I}_9) \,.
\end{equation}
The first seven invariants are identical to those of orthotropic solids. The two extra invariants are defined as
\begin{equation}
	\mathring{I}_8=\mathring{\mathcal{I}}\,\mathring{\mathbf{N}}_1\cdot\mathbf{C}\cdot
	\mathring{\mathbf{N}}_2\,,\qquad
	\mathring{I}_9=\mathring{\mathcal{I}}^2\,,\qquad \mathring{\mathcal{I}}
	=\mathring{\mathbf{N}}_1\cdot\mathring{\mathbf{N}}_2\,.
\end{equation}  
At time $t>0$ the remodeling body has the energy function $W=\overline{W}(I_1,I_2,I_3,I_4,I_5,I_6,I_7,I_8,I_9)$, where the first seven invariants are identical to those given in \eqref{I1-7-Definitions}, and 
\begin{equation}
	I_8=\mathcal{I}\,\mathbf{N}_1\cdot\mathbf{C}\cdot \mathbf{N}_2\,,\qquad
	I_9=\mathcal{I}^2\,,\qquad \mathcal{I}=\mathbf{N}_1\cdot \mathbf{N}_2\,.
\end{equation}  

For a monoclinic solid, the Cauchy stress has the following representation 
\begin{equation}
\begin{aligned}
	\boldsymbol{\sigma} 
	&= \frac{2}{\sqrt{I_3}} \Big\{ \left(I_2\,\overline{W}_2+I_3\,\overline{W}_3\right)\mathbf{g}^{\sharp} 
	+\overline{W}_1\,\mathbf{b}^{\sharp}
	-I_3\,\overline{W}_2\,\mathbf{c}^{\sharp}
	+\overline{W}_4\,\mathbf{n}_1\otimes\mathbf{n}_1
	+\overline{W}_5\left[\mathbf{n}_1\otimes(\mathbf{b}^{\sharp}\mathbf{g}\mathbf{n}_1)
	+(\mathbf{b}^{\sharp}\mathbf{g}\mathbf{n}_1)\otimes\mathbf{n}_1 \right] \\
	& \qquad +\overline{W}_6\,\mathbf{n}_2\otimes\mathbf{n}_2
	+\overline{W}_7\left[\mathbf{n}_2\otimes(\mathbf{b}^{\sharp}\mathbf{g}\mathbf{n}_2)
	+(\mathbf{b}^{\sharp}\mathbf{g}\mathbf{n}_2)\otimes\mathbf{n}_2 \right]
	+\mathcal{I}\,\overline{W}_8\left(\mathbf{n}_1\otimes\mathbf{n}_2+\mathbf{n}_2\otimes\mathbf{n}_1\right) 
	\Big\} \,.
\end{aligned}
\end{equation}
For an incompressible orthotropic solid $I_3=\Ie_3=1$, and hence
\begin{equation}
\begin{aligned}
	\boldsymbol{\sigma} 
	&=-p\,\mathbf{g}^{\sharp} 
	+2\overline{W}_1\,\mathbf{b}^{\sharp}-2\overline{W}_2\,\mathbf{c}^{\sharp}
	+2\overline{W}_4\,\mathbf{n}_1\otimes\mathbf{n}_1
	+2\overline{W}_5\left[\mathbf{n}_1\otimes(\mathbf{b}^{\sharp}\mathbf{g}\mathbf{n}_1)
	+(\mathbf{b}^{\sharp}\mathbf{g}\mathbf{n}_1)\otimes\mathbf{n}_1 \right] \\
	& \qquad +2\overline{W}_6\,\mathbf{n}_2\otimes\mathbf{n}_2
	+2\overline{W}_7\left[\mathbf{n}_2\otimes(\mathbf{b}^{\sharp}\mathbf{g}\mathbf{n}_2)
	+(\mathbf{b}^{\sharp}\mathbf{g}\mathbf{n}_2)\otimes\mathbf{n}_2 \right] 
	+2\mathcal{I}\,\overline{W}_8\left(\mathbf{n}_1\otimes\mathbf{n}_2+\mathbf{n}_2\otimes\mathbf{n}_1\right)
	\,,
\end{aligned}
\end{equation}
where $p$ is the Lagrange multiplier associated with the incompressibility constraint $J=\sqrt{I_3}=1$.

\subsubsection{Dissipation potential}

Remodeling is a dissipative process. This means that in any mechanical formulation of remodeling dissipation due to the evolution of the remodeling tensor must be taken into account. Let us assume the existence of a dissipation potential (or Rayleigh dissipation function) $\phi=\phi(X,\mathbf{F},\Fr,\dot{\Fr},\boldsymbol{G},\boldsymbol{g})$. Objectivity implies that  $\phi=\hat{\phi}(X,\mathbf{C}^{\flat},\Fr,\dot{\Fr},\boldsymbol{G})$.
Let us assume that $\phi$ is a convex function of $\dot{\Fr}$ \citep{ziegler1958attempt,ziegler1987derivation,Germain1983,Goldstein2002,Kumar2016}. The generalized force that corresponds to the evolution of remodeling tensor is related to the dissipation potential as
\begin{equation}
	\boldsymbol{B}_r=- \frac{\partial \phi}{\partial \dot{\Fr}}\,.
\end{equation}
Convexity of $\phi$ in $\dot{\Fr}$ implies that
\begin{equation}
	\phi(X,\mathbf{C}^{\flat},\Fr,\dot{\Fr},\boldsymbol{G})
	+\frac{\partial \phi}{\partial \dot\Fr } (X,\mathbf{C}^{\flat},\Fr,\dot{\Fr},\boldsymbol{G})\!:\!\Delta\dot\Fr 
	\leq \phi(X,\mathbf{C}^{\flat},\Fr,\dot{\Fr}+\Delta\dot{\Fr},\boldsymbol{G})
	\,.
\end{equation}
Let us choose $\Delta\dot\Fr =-\dot\Fr$. Thus
\begin{equation}
	\frac{\partial \phi}{\partial \dot\Fr} (X,\mathbf{C}^{\flat},\Fr,\dot{\Fr},\boldsymbol{G}) \!:\! 
	\dot\Fr  \geq \phi(X,\mathbf{C}^{\flat},\Fr,\dot{\Fr},\boldsymbol{G}) - \phi(\boldsymbol 0)
	\geq 0 \,,
\end{equation}
as $\phi$ attains its minimum for $\dot\Fr=\boldsymbol 0$.
The left hand-side is the entropy production. Therefore, we conclude that the entropy production is non-negative when $\phi$ is convex in $\dot{\Fr}$ (see \S\ref{Sec:2ndLaw} and Eq.~\eqref{Entropy-Production-Remodeling}).

The dissipation potential is invariant under the material symmetry group, i.e.,
\begin{equation} 
	\phi(X,\mathbf{F}\mathbf{K},\mathbf{K}^*\Fr,\dot{\overline{\mathbf{K}^*\Fr}},\mathbf{G},\mathbf{g})
	=\phi(X,\mathbf{F},\Fr,\dot\Fr,\mathbf{G},\mathbf{g})\,,
\quad \forall\,\,\mathbf{K}\in \mathcal{G}_X\leqslant \mathrm{Orth}(\mathbf{G})\,,
\end{equation}
for all deformation gradients $\mathbf{F}$ and remodeling tensors $\Fr\,$, where $\mathrm{Orth}(\mathbf{G})=\{\mathbf{Q}: T_X\mathcal{B}\to T_X\mathcal{B}~|~ \mathbf{Q}^{\star}\mathbf{G}\mathbf{Q}=\mathbf{Q}\mathbf{G}\mathbf{Q}^{\star}=\mathbf{G} \}\,$, $\mathbf{K}^*\Fr=\mathbf{K}^{-1}\Fr \mathbf{K}\,$, and $\dot{\overline{\mathbf{K}^*\Fr}}=\mathbf{K}^*\dot\Fr=\mathbf{K}^{-1}\dot\Fr\mathbf{K}$. If the structural tensors are added to the list of arguments of the dissipation potential, it becomes an isotropic function. Thus, $\phi(X,\mathbf{F},\Fr,\dot\Fr,\mathbf{G},\boldsymbol{\Lambda},\mathbf{g})$ is an isotropic function.

We follow \citet{Kumar2016} and assume the following form for the dissipation potential 
\begin{equation} \label{Dissipation-Potential-Quadratic}
	\phi(X,\mathbf{F},\Fr,\dot\Fr,\mathbf{G},\boldsymbol{\Lambda},\mathbf{g})
	=\frac{1}{2}\,\dot\Fr\cdot\boldsymbol{\mathsf{A}}(X,\mathbf{F},\Fe,\mathbf{G},
	\boldsymbol{\Lambda},
	\mathbf{g})\cdot\dot\Fr
	=\frac{1}{2}\,\dot\cFr^A{}_B\,\dot\cFr^C{}_D\,\mathsf{A}_A{}^B{}_C{}^D\,,
\end{equation}
where $\boldsymbol{\mathsf{A}}(X,\mathbf{F},\Fe,\mathbf{G},\boldsymbol{\Lambda},\mathbf{g})$ is a positive-definite fourth-order tensor.\footnote{Recall that the rate of energy dissipation is written as $\frac{\partial \phi}{\partial \dot\Fr}:\dot\Fr \geq 0$. If the dissipation potential is quadratic, then $\frac{\partial \phi}{\partial \dot\Fr}:\dot\Fr=2\phi$.}
Objectivity implies that $\boldsymbol{\mathsf{A}}(X,\mathbf{F},\Fe,\mathbf{G},\boldsymbol{\Lambda},\mathbf{g})=\hat{\boldsymbol{\mathsf{A}}}(X,\mathbf{C}^{\flat},\Ce^{\flat},\mathbf{G},\boldsymbol{\Lambda})$. Notice that the fourth-order tensor $\boldsymbol{\mathsf{A}}$ has the major symmetries but does not need to have any minor symmetries. 
Isotropy of $\phi(X,\mathbf{F},\Fr,\dot\Fr,\mathbf{G},\boldsymbol{\Lambda},\mathbf{g})$ implies that
\begin{equation}
	\mathbf{K}^{-1}\dot\Fr\mathbf{K}\!:\!\boldsymbol{\mathsf{A}}(\mathbf{F}\mathbf{K},\Fe\mathbf{K}
	,\mathbf{G},\mathbf{K}^*\boldsymbol{\Lambda},\mathbf{g})\!:\!\mathbf{K}^{-1}\dot\Fr\mathbf{K} 
	= \dot\Fr\!:\!\boldsymbol{\mathsf{A}}
	(\mathbf{F},\Fe,\mathbf{G},\boldsymbol{\Lambda},\mathbf{g})\!:\!\dot\Fr\,,\quad\forall\dot\Fr \,.
\end{equation}
As this holds for arbitrary $\dot\Fr $, one concludes that\footnote{In incompetents, $(\mathbf{K}^* \boldsymbol{\mathsf{A}})_{A}{}^B{}_C{}^D= \mathrm{K}^{-I}{}_A \mathrm{K}^{B}{}_J \mathrm{K}^{-K}{}_C \mathrm{K}^{D}{}_L \,{\mathsf{A}}_{I}{}^J{}_K{}^L$.}
\begin{equation}
	\mathbf{K}^* \boldsymbol{\mathsf{A}}(\mathbf{F},\Fe,\mathbf{G},\mathbf{g}) 
	= \boldsymbol{\mathsf{A}}(\mathbf{F},\Fe,\mathbf{G},\mathbf{g})\,,
\end{equation}
i.e., $\boldsymbol{\mathsf{A}}$ is an isotropic tensor. Thus, the most general form for this tensor is \citep{Jog2006}
\begin{equation} \label{Quadratic-phi}
	\mathsf{A}_{A}{}^B{}_C{}^D
	=\eta_1\,\delta_A^B \delta_C^D
	+\eta_2\,\delta _A^D \delta_C^B
	+\eta_3\,G_{AC}G^{BD} 
	\,.
\end{equation}
Or, equivalently 
\begin{equation} 
	\mathsf{A}_{ABCD}
	=\eta_1\,G_{AB}G_{CD}
	+\eta_2\,G_{AD}G_{BC}
	+\eta_3\,G_{AC}G_{BD} 
	\,.
\end{equation}
Thus
\begin{equation}
	\frac{\partial \phi}{\partial \dot\cFr^A{}_B}= \eta_1\,\cFr^M{}_M\,\delta^B_A + \eta_2\,\dot\cFr^B{}_A
	+\eta_3\,G_{AM}\,\dot\cFr^M{}_N\,G^{NB}
	\,.
\end{equation}
Or
\begin{equation} \label{phi-isotropic}
	\frac{\partial \phi}{\partial \dot\Fr}= \eta_1(\operatorname{tr}\dot\Fr)\,\mathbf{I} + \eta_2\,\dot\Fr^{\star}
	+\eta_3\,\mathbf{G}\dot\Fr\mathbf{G}^{\sharp}
	\,,
\end{equation}
where $\eta_i=\eta_i(X,\mathbf{C},\Ce)$, $i=1,2,3$, and $\operatorname{tr}\dot\Fr=\dot\cFr^C{}_C$.
The dissipation potential corresponding to \eqref{Quadratic-phi} is written as
\begin{equation}
\begin{aligned}
	\phi =\frac{1}{2}\eta_1\left(\dot\cFr^A{}_A\right)^2+\frac{1}{2}\eta_2\,\dot\cFr^A{}_B\,\dot\cFr^B{}_A
	+\frac{1}{2}\eta_3\,G_{AC} \dot\cFr^C{}_D G^{DB} \dot\cFr^A{}_B 
	\,.
\end{aligned}
\end{equation}
Let us introduce the new indices $\Gamma=\{AB\}$ such that $\{11,12,13,21,22,23,31,32,33\}\leftrightarrow \{1,2,3,4,5,6,7,8,9\}$. Then the dissipation potential can be rewritten as $\phi=\frac{1}{2}\mathbb{A}^{\Gamma\Lambda}\,\dot{\mathsf{F}}_{\Gamma}\dot{\mathsf{F}}_{\Lambda}$. The tensor $\boldsymbol{\mathsf{A}}$ is positive-definite if and only if the $9\times 9$ matrix $\boldsymbol{\mathbb{A}}$, which has three distinct eigenvalues, is positive-definite. Thus, $\boldsymbol{\mathsf{A}}$ is positive-definite if and only if
\begin{equation} 
	3 \eta_1+\eta_2+\eta_3>0\,,\quad \eta_2+\eta_3>0\,,\quad -\eta_2+\eta_3>0	\,.
\end{equation}

\subsubsection{Remodeling energy}

In addition to the strain energy function and dissipation potential we assume a third energy that quantifies the tendency of the material to remodel in response to strain and stress. We call it the \emph{remodeling energy} and denote it as $\Wr=\Wr(X,\mathbf{C}^{\flat},\Fr,\mathbf{G})$. Here, we discuss it for a specific class of remodeling problems, namely fiber reorientation in solids reinforced with one or two families of fibers.

Let us consider a body that has a distribution of fibers. At $X\in\mathcal{B}$ the fiber has a $\mathbf{G}$-unit tangent that is denoted by $\mathbf{N}(X,t)$. For this class of remodeling solids we assume the following forms for the remodeling energy: $\Wr=\Wr(X,\mathbf{C}^{\flat},\mathbf{N},\mathbf{G})$.
In fiber-reinforced solids fibers tend to reorient themselves in response to applied forces. In the literature it has been postulated that fibers orient themselves along the direction of maximum stretch or normal stress. Let us assume that at $X\in\mathcal{B}$, there is a $\mathbf{G}$-unit material vector $\mathbf{M}(X,t)$ that the fiber tends to rotate towards. We call this the \emph{fiber preferred direction}, which explicitly depends on the state of strain and stress at $X\in\mathcal{B}$. This can be the direction of maximum principal stretch, the direction of maximum tensile stress, etc. Obviously, $\mathbf{M}(X,t)$ depends on $\mathbf{C}^{\flat}$ either directly or indirectly (through the constitutive equations of the material). Thus, one can write the remodeling energy as $\Wr=\Wr(X,\mathbf{M},\mathbf{N},\mathbf{G})$ (with an abuse of notation we are using the same symbol $\Wr$ for this energy). Obviously, this energy is objective as all its arguments are material tensors. 
It should be noted that $\mathbf{N}$ and $-\mathbf{N}$ define the same fiber orientation, i.e., $\mathbf{N}\in\mathbb{RP}^2$---the projective plane. Similarly, $\mathbf{M}$ and $-\mathbf{M}$ define the same fiber preferred direction.
Therefore, the remodeling energy must be invariant under either or both transformations $\mathbf{N}\mapsto-\mathbf{N}$, and $\mathbf{M}\mapsto-\mathbf{M}$. One way to ensure this invariance is to write (again with an abuse of notation) $\Wr=\Wr(X,\mathbf{M}\otimes\mathbf{M},\mathbf{N}\otimes\mathbf{N},\mathbf{G})$.

\begin{example}
As examples of remodeling energy let us consider the following two choices 
\begin{equation} 
	\Wr(X,\mathbf{M},\mathbf{N},\mathbf{G})=\frac{1}{2}\kappa_M\,(\mathbf{M}\cdot\mathbf{N})^2\,,
	\qquad  \Wr(X,\mathbf{M},\mathbf{N},\mathbf{G})=\kappa_M\,|\mathbf{M}\cdot\mathbf{N}|\,,
\end{equation}
where $\kappa_M$ is a scalar that can, in principle, depend on $\mathbf{C}^{\flat}$, i.e., $\kappa_M=\kappa_M(X,\mathbf{C}^{\flat},\mathbf{G})$. In our numerical examples in \S\ref{Examples} we will assume that $\kappa_M$ is a material constant.
\end{example}

Let us next consider a body that is reinforced by two families of fibers that are not necessarily mechanically equivalent. At $X\in\mathcal{B}$ the fibers have the $\mathbf{G}$-unit tangent vectors $\mathbf{N}_1(X,t)$ and $\mathbf{N}_2(X,t)$. Let us denote their corresponding fiber preferred directions by $\mathbf{M}_1(X,t)$ and $\mathbf{M}_2(X,t)$, respectively. The remodeling energy has the following form: $\Wr=\Wr(X,\mathbf{M}_1\otimes\mathbf{M}_1,\mathbf{M}_2\otimes\mathbf{M}_2,\mathbf{N}_1\otimes\mathbf{N}_1,\mathbf{N}_2\otimes\mathbf{N}_2,\mathbf{G})$. 
As examples of remodeling energy let us consider the following two choices 

\begin{equation} 
	\Wr=\frac{1}{2}\kappa_{M1}\,(\mathbf{M}_1\cdot\mathbf{N}_1)^2
	+\frac{1}{2}\kappa_{M2}\,(\mathbf{M}_2\cdot\mathbf{N}_2)^2\,,
	\qquad  \Wr(X,\mathbf{M},\mathbf{N},\mathbf{G})=\kappa_{M1}\,|\mathbf{M}_1\cdot\mathbf{N}_1|
	+\kappa_{M2}\,|\mathbf{M}_2\cdot\mathbf{N}_2|\,,
\end{equation}
where $\kappa_{M1}=\kappa_{M1}(X,\mathbf{C}^{\flat},\mathbf{G})$ and $\kappa_{M2}=\kappa_{M2}(X,\mathbf{C}^{\flat},\mathbf{G})$ are scalars. 
In the numerical examples in \S\ref{Sec:Example2}, we will assume that the two fiber families are mechanically equivalent and $\kappa_{M1}=\kappa_{M2}=\kappa_{M}$ is a constant.

\section{Balance Laws} \label{Sec:BalanceLaws}

In this section we derive the governing equations of remodeling bodies in a variational setting. In addition to the standard governing equations of nonlinear elasticity, a remodeling equation is derived. Its explicit form for different types of remodeling and material anisotropy classes is given in detail.

\subsection{Conservation of mass}

Mass density field in the initial body is denoted as $\rho_0=\rho_0(X)$. At time $t$, mass density at the same material point is denoted by $\rho_0(X,t)$. At $X\in\mathcal{B}$ and at time $t=0$ consider a volume element $dV_0(X)$. Mass of this volume element is $d\mathsf{m}=\rho_0(X)\,dV_0(X)$. 
Under the local change of reference configuration $\Fr$ at $X\in\mathcal{B}$, the volume element is transformed to $dV_t(X)=\Jr(X,t)\,dV_0(X)$, where $\Jr(X,t)=\det\Fr$, and hence, $dV_t(X)=(\det\Fr)\,dV_0(X)=dV_0(X)$.
It is assumed that remodeling is mass conserving, i.e., $d\mathsf{m}=\rho_0(X)\,dV_0(X)=\rho_0(X,t)\,dV_t(X)=\rho_0(X,t)\,dV_0(X)$, and hence, $\rho_0(X,t)=\rho_0(X)$.

\subsection{The Lagrange-d'Alembert principle} \label{Sec:LD-Principle}

The governing equations of a body undergoing finite deformations while remodeling can be derived using the Lagrange-d'Alembert principle. Specifically, one has the two independent variations $(\delta\varphi,\delta\Fr)$.
The Lagrangian density is defined as $\mathcal{L} = \mathcal{T} - W+q\big(\det\Fr-1\big)$, where $\mathcal{T} = \frac{1}{2}\rho_o\Vert\boldsymbol V\Vert^2_{\boldsymbol g}= \frac{1}{2}\rho_o \llangle \mathbf{V},\mathbf{V}\rrangle_{\boldsymbol g}$ is the kinetic energy density, and $q=q(X,t)$ is a Lagrange multiplier field corresponding to the internal constraint $\det\Fr=1$.
According to the Lagrange-d'Alembert variational principle the physical configuration of the remodeling body satisfies the following identity \citep{Lanczos1962,MarsRat2013}:
\begin{equation} \label{LD-Principle}
	\delta \int_{t_1}^{t_2}  \int_{\mathcal B} \mathcal{L} \, dV \, \mathrm dt +
	\int_{t_1}^{t_2} \int_{\mathcal B} 
	\boldsymbol{B}_r\!:\!\delta\Fr  \, dV \, \mathrm dt +
	\int_{t_1}^{t_2} \int_{\mathcal B} \rho_o \llangle \boldsymbol{B} , 
	\delta\varphi \rrangle_{\boldsymbol{g}} \, dV\, \mathrm dt +
	\int_{t_1}^{t_2} \int_{\partial \mathcal B} \llangle \boldsymbol{T} , 
	\delta\varphi \rrangle_{\boldsymbol{g}} \, dA \,\mathrm dt  = 0 \,,
\end{equation}
for any variation fields $\delta\varphi$ and $\delta\Fr$,\footnote{It is assumed that $\delta\varphi(X,t_1)=\delta\varphi(X,t_2)=0$, and $\delta\Fr(X,t_1)=\delta\Fr(X,t_2)=\mathbf{0}$.} where $\boldsymbol{B}$ and $\boldsymbol{T}$ are, respectively, the body force per unit mass and the boundary traction per unit undeformed area. We next find the Euler-Lagrange equations corresponding to $\delta\varphi$ and $\delta\Fr$ separately.


\begin{itemize}[topsep=0pt,noitemsep, leftmargin=10pt]
\item $\delta\varphi$ \textbf{variations:} Note that $\delta\mathcal{L}=\delta\mathcal{T}-\delta W$. It can be shown that
\begin{equation}
	\delta\mathcal{T}= \frac{d}{dt} \left[\rho_o \llangle \mathbf{V},\mathbf{V}\rrangle_{\boldsymbol g}\right]
	-\rho_o \llangle \mathbf{A},\delta\varphi\rrangle_{\boldsymbol g}
	\,,
\end{equation}
where $\mathbf{A}$ is the acceleration vector. Knowing that $\delta\varphi(X,t_1)=\delta\varphi(X,t_2)=0$, the first term on the right-hand side will not contribute to the variational principle. Also, note that \citep{Yavari2019}
\begin{equation}
	\delta W=\frac{\partial W}{\partial \mathbf{F}}:\delta\mathbf{F}=\frac{\partial W}{\partial \mathbf{F}}:\nabla\delta\varphi
	\,,
\end{equation}
where $\nabla\delta\varphi$ is the covariant derivative of $\varphi$.\footnote{$\nabla\delta\varphi$ has coordinates $\delta\varphi^a{}_{|A}=\delta\varphi^a{}_{|b}\,F^b{}_A=F^b{}_A(\delta\varphi^a_{,b}+\gamma^a{}_{bc}\,\delta\varphi^c)=\delta\varphi^a_{,A}+\gamma^a{}_{bc}\,F^b{}_A\,\delta\varphi^c$.} Thus\footnote{$\mathbf{g}^{\sharp}$ is the inverse of the spatial metric with components $g^{ab}$ such that $g^{ac}\,g_{cb}=\delta^a_b$.}
\begin{equation}
\begin{aligned}
	 -\int_{\mathcal B} \delta W \, dV
	 & = -\int_{\mathcal B} \left[\operatorname{Div}\left(\frac{\partial W}{\partial \mathbf{F}}
	 \cdot\delta\varphi\right)
	 +\left(\operatorname{Div}\frac{\partial W}{\partial \mathbf{F}}\right)\cdot\delta\varphi \right] dV \\
	 & =-\int_{\partial\mathcal B} \llangle \mathbf{g}^{\sharp}\frac{\partial W}{\partial \mathbf{F}}
	 \hat{\mathbf{N}},
	 \delta\varphi\rrangle_{\mathbf{g}}\,dA
	+\int_{\mathcal B}  \left(\operatorname{Div}\frac{\partial W}{\partial \mathbf{F}}\right)
	\cdot\delta\varphi\, dV\,,
\end{aligned}
\end{equation}
where $\hat{\mathbf{N}}$ is the $\mathbf{G}$-unit normal vector of $\partial\mathcal{B}$.\footnote{This means that $\llangle \hat{\mathbf{N}},\hat{\mathbf{N}}\rrangle_{\mathbf{G}}=\hat{N}^A\,\hat{N}^B\,G_{AB}=1$.}
Hence, \eqref{LD-Principle} is simplified to read
\begin{equation} 
	\int_{t_1}^{t_2} \int_{\mathcal B}  \llangle \operatorname{Div}\left(\mathbf{g}^{\sharp}
	\frac{\partial W}{\partial \mathbf{F}}\right)
	+\rho_o\mathbf{B}-\rho_o\mathbf{A}  , \delta\varphi \rrangle_{\mathbf{g}} \, dV\, \mathrm dt +
	\int_{t_1}^{t_2} \int_{\partial \mathcal B} \llangle \mathbf{T}
	-\mathbf{g}^{\sharp}\frac{\partial W}{\partial \mathbf{F}}\hat{\mathbf{N}} , 
	\delta\varphi \rrangle_{\boldsymbol{g}} \, dA \,\mathrm dt  = 0 \,.
\end{equation}
On the Dirichlet boundary $\partial_D \mathcal B$, $\delta\varphi=0$, and hence
\begin{equation} 
	\int_{t_1}^{t_2} \int_{\mathcal B}  \llangle \operatorname{Div}\left(\mathbf{g}^{\sharp}
	\frac{\partial W}{\partial \mathbf{F}}\right)
	+\rho_o\mathbf{B}-\rho_o\mathbf{A} , \delta\varphi \rrangle_{\mathbf{g}} \, dV\, \mathrm dt +
	\int_{t_1}^{t_2} \int_{\partial_N \mathcal B} \llangle \mathbf{T}
	-\mathbf{g}^{\sharp}\frac{\partial W}{\partial \mathbf{F}}\hat{\mathbf{N}} , 
	\delta\varphi \rrangle_{\boldsymbol{g}} \, dA \,\mathrm dt 
	 = 0 \,,
\end{equation}
where $\partial_N\mathcal{B}$ is the Neumann boundary.\footnote{It is assumed that the boundary of the body is the disjoint union of the Dirichlet and Neumann boundary, i.e., $\partial\mathcal{B}=\partial_D\mathcal{B}\sqcup\partial_N\mathcal{B}$.} Therefore, the variational principle gives us the balance of linear momentum and the Neumann boundary conditions:
\begin{equation} \label{EL-Compressible}
\begin{dcases}
	\operatorname{Div}\left(\mathbf{g}^{\sharp}\frac{\partial W}{\partial \mathbf{F}}\right)
	+\rho_0\mathbf{B}=\rho_0\mathbf{A}\,, & \text{in~}\mathcal{B}\,,\\
	 \mathbf{g}^{\sharp}\frac{\partial W}{\partial \mathbf{F}}\hat{\mathbf{N}} =\mathbf{T}\,, 
	 & \text{on~} \partial_N \mathcal{B}\,.
\end{dcases}
\end{equation}

\begin{remark}If the remodeling material is incompressible a term $p(J-1)$ is added to the Lagrangian density. In this case, $\delta\mathcal{L}=\delta\mathcal{T}-\delta W+p\,\delta J=\delta\mathcal{T}-\delta W+pJ\,\mathbf{F}^{-1}\!:\!\delta \mathbf{F}$. The Euler-Lagrange equations and natural boundary conditions \eqref{EL-Compressible} are modified to read
\begin{equation} \label{EL-Incompressible}
\begin{dcases}
	\operatorname{Div}\left[-pJ\,\mathbf{F}^{-1}+\mathbf{g}^{\sharp}
	\frac{\partial W}{\partial \mathbf{F}}\right]
	+\rho_0\mathbf{B}=\rho_0\mathbf{A}\,, &\text{in~}\mathcal{B}\,,\\
	\left[-pJ\,\mathbf{F}^{-1}+\mathbf{g}^{\sharp}\frac{\partial W}{\partial \mathbf{F}}\right]\hat{\mathbf{N}} 
	=\mathbf{T}\,, & \text{on~} \partial_N \mathcal{B}\,.
\end{dcases}
\end{equation}
\end{remark}

\begin{remark}\label{Stress-Measures}
As a consequence of the second law of thermodynamics $\mathbf{P}=\mathbf{g}^{\sharp}\frac{\partial W}{\partial \mathbf{F}}$ is the first Piola-Kirchhoff stress with components $P^{aA}=g^{ab}\frac{\partial W}{\partial F^b{}_A}$.\footnote{The second law will be discussed in \S\ref{Sec:2ndLaw}, but it would be more convenient to discuss the balance of linear momentum in terms of different stress measures here.} 
Let us first recall that the Cauchy, the first Piola-Kirchhoff, and the convected stress tensors are related to the energy function as 
\begin{equation} 
	\mathbf{P}= \mathbf{g}^\sharp \frac{\partial W}{\partial \mathbf{F}}\,,\qquad
	\boldsymbol{\sigma}	= \frac{2}{J}\frac{\partial W}{\partial \mathbf{g}}\,,\qquad
	\boldsymbol{\Sigma}= \frac{2}{J}\frac{\partial W}{\partial \mathbf{C}^\flat} \,.
\end{equation}
They are also related as $\mathbf S = \mathbf F^{-1}  \mathbf P = J \boldsymbol \Sigma = J \mathbf F^{-1} \boldsymbol \sigma \mathbf F^{-\star}$.
The balance of linear momentum \eqref{EL-Compressible}$_1$, i.e., $\operatorname{Div}\mathbf{P}+\rho_0\mathbf{B}=\rho_0\mathbf{A}$, in terms of the Cauchy stress reads $
\operatorname{div}_{\mathbf{g}}\boldsymbol{\sigma}+\rho\mathbf{b}=\rho\mathbf{a}$, where $\rho$, $\mathbf{b}=\mathbf{B}\circ\varphi^{-1}$, and $\mathbf{a}$ are spatial mass density, spatial body force, and spatial acceleration.
$\operatorname{div}_{\mathbf{g}}\boldsymbol{\sigma}$ and $\operatorname{Div}\mathbf{P}$ have the components $\sigma^{ab}{}_{|b}$ and $P^{aA}{}_{|A}$, respectively, defined as
\begin{equation}
	\sigma^{ab}{}_{|b}=\sigma^{ab}{}_{,b}+\sigma^{ac}\gamma^b{}_{cb} 
	+\sigma^{cb}\gamma^a{}_{cb} \,,\qquad
	P^{aA}{}_{|A}=P^{aA}{}_{,A}+P^{aB}\Gamma^A{}_{AB} +P^{cA}F^b{}_A\gamma^a{}_{bc}\,.
\end{equation}
One can write the balance of linear momentum entirely with respect to the reference configuration by pulling back the spatial balance of linear momentum to the reference configuration, i.e., $\varphi_t^{*}\left(\operatorname{div}_{\mathbf{g}}\boldsymbol{\sigma}\right)+\varphi_t^{*}\left(\rho\mathbf{b}\right)=\varphi_t^{*}\left(\rho\mathbf{a}\right)$. Thus \citep{Simo1988}
\begin{equation}\label{Linear-Momentum-Convected}
    \operatorname{div}_{\mathbf{C}^{\flat}}\boldsymbol{\Sigma}+\varrho\pmb{\mathscr{B}}_t
    =\varrho\pmb{\mathscr{A}}_t,
\end{equation}
where $\boldsymbol{\Sigma}=\varphi_t^*\boldsymbol{\sigma}$ is the convected stress, $\pmb{\mathscr{B}}_t=\varphi_t^*\mathbf{b}$ is the convected body force, and $\varrho=\rho\circ\varphi_t$.
\end{remark}

\item $\delta\Fr$ \textbf{variations:} 
For these variations, $\delta\mathcal{T}=0$. One can write
\begin{equation} 
	\delta W=\frac{\partial W}{\partial \mathbf{G}}:\delta\mathbf{G}
	+\frac{\partial W}{\partial \boldsymbol{\Lambda}}\!:\!\delta\boldsymbol{\Lambda}
	 \,.
\end{equation}
The collection of structural tenors depends on the type of anisotropy. For example, for transversely isotopic solids $\boldsymbol{\Lambda}=\mathbf{N}\otimes\mathbf{N}$, and hence
\begin{equation} 
	\delta\boldsymbol{\Lambda}=-\Fr^{-1}\delta\Fr \mathbf{N}\otimes\mathbf{N}
	-\mathbf{N}\otimes\Fr^{-1}\delta\Fr \mathbf{N}
	 \,.
\end{equation}
Thus
\begin{equation} 
	\frac{\partial W}{\partial \boldsymbol{\Lambda}}\!:\!\delta\boldsymbol{\Lambda}
	=2\Fr^{-\star}\frac{\partial W}{\partial \boldsymbol{\Lambda}}\mathbf{N}\otimes\mathbf{N}\!:\!\delta\Fr
	 \,.
\end{equation}

Note that $\delta\mathbf{G}=\delta(\Fr^{\star}\mathring{\mathbf{G}}\Fr)=(\delta\Fr)^{\star}\mathring{\mathbf{G}}\Fr+\Fr^{\star}\mathring{\mathbf{G}}\delta\Fr$.
Hence
\begin{equation} 
	\frac{\partial W}{\partial \mathbf{G}}:
	\left[ (\delta\Fr)^{\star}\mathring{\mathbf{G}}\Fr+\Fr^{\star}\mathring{\mathbf{G}}\delta\Fr \right]
	=2\mathring{\mathbf{G}}\Fr^{\star}\frac{\partial W}{\partial \mathbf{G}}:\delta\Fr	 
	=2\Fr^{-\star}\mathbf{G}\frac{\partial W}{\partial \mathbf{G}}:\delta\Fr	 
	\,.
\end{equation}
The variation of $\det\Fr$ is calculated as $\delta(\det\Fr)=(\det\Fr)\Fr^{-\star}\!:\!\delta\Fr=\Fr^{-\star}\!:\!\delta\Fr$.
Thus, \eqref{LD-Principle} is simplified to read
\begin{equation} 
	\int_{t_1}^{t_2}  \int_{\mathcal B} \left[-2\Fr^{-\star}\mathbf{G}\frac{\partial W}{\partial \mathbf{G}}
	-2\Fr^{-\star}\frac{\partial W}{\partial \boldsymbol{\Lambda}}\mathbf{N}\otimes\mathbf{N}
	+q\Fr^{-\star}- \frac{\partial \phi}{\partial \dot{\Fr}} \right]\!:\delta\Fr \, dV \, \mathrm dt  = 0 \,.
\end{equation}
Therefore, the \emph{remodeling equation} for transversely isotropic solids reads 
\begin{equation} 
	\frac{\partial \phi}{\partial \dot{\Fr}}=
	q\Fr^{-\star}-2\Fr^{-\star}\mathbf{G}\frac{\partial W}{\partial \mathbf{G}}
	-2\Fr^{-\star}\frac{\partial W}{\partial \boldsymbol{\Lambda}}\mathbf{N}\otimes\mathbf{N}\,.
\end{equation}
In the case of isotropic solids this is simplified as
\begin{equation} \label{Remodeling-Equation-Metric}
	\frac{\partial \phi}{\partial \dot{\Fr}}=
	q\Fr^{-\star}-2\Fr^{-\star}\mathbf{G}\frac{\partial W}{\partial \mathbf{G}} \,.
\end{equation}
\end{itemize}

Next we rewrite the remodeling equation more explicitly in terms of the integrity bases for isotropic, transversely isotropic, orthotropic, and monoclinic solids.

\subsubsection{Remodeling equation for isotropic solids}

The remodeling equation can be written more explicitly in terms of the principal invariants. One writes
\begin{equation} 
	\frac{\partial W}{\partial \mathbf{G}}
	=\frac{\partial \overline{W}}{\partial I_1}\frac{\partial I_1}{\partial \mathbf{G}}
	+\frac{\partial \overline{W}}{\partial I_2}\frac{\partial I_2}{\partial \mathbf{G}}
	+\frac{\partial \overline{W}}{\partial I_3}\frac{\partial I_3}{\partial \mathbf{G}}
	=W_1\frac{\partial I_1}{\partial \mathbf{G}}
	+W_2\frac{\partial I_2}{\partial \mathbf{G}}
	+W_3\frac{\partial I_3}{\partial \mathbf{G}}
	\,.
\end{equation}
Note that 
\begin{equation} 
	\frac{\partial I_1}{\partial \mathbf{G}}=-\mathbf{G}^{\sharp}\mathbf{C}^{\flat}\mathbf{G}^{\sharp}
	=-\mathbf{C}^{\sharp}
	\,.
\end{equation}
Recall that $I_2=\frac{1}{2}\left(I_1^2-\operatorname{tr}\mathbf{C}^2\right)=\frac{1}{2}\left(I_1^2-\mathrm{C}^A{}_B\,\mathrm{C}^B{}_A\right)$. Thus
\begin{equation} 
	\frac{\partial I_2}{\partial \mathbf{G}}=I_1\frac{\partial I_1}{\partial \mathbf{G}}
	-\frac{1}{2}\frac{\partial \operatorname{tr}\mathbf{C}^2}{\partial \mathbf{G}}
	=-I_1\mathbf{C}^{\sharp}+\mathbf{C}^{2\sharp}
	\,.
\end{equation}
Finally
\begin{equation} 
	\frac{\partial I_3}{\partial \mathbf{G}}=-I_3\mathbf{G}^{\sharp}
	\,.
\end{equation}
Therefore 
\begin{equation} 
	-2\frac{\partial W}{\partial \mathbf{G}}\mathbf{G}\Fr^{-1}
	=2I_3W_3\Fr^{-1}+2(W_1+I_1W_2)\mathbf{C}\Fr^{-1}-2W_2\mathbf{C}^2\Fr^{-1}
	\,.
\end{equation}
Hence, the remodeling equation is simplified to read
\begin{equation} 
	\frac{\partial \phi}{\partial \dot{\Fr}} 
	=\left[(q+2I_3W_3)\mathbf{I}+2(W_1+I_1W_2)\mathbf{C}-2W_2\mathbf{C}^2\right]\Fr^{-1} \,.
\end{equation}

\subsubsection{Remodeling equation for transversely isotropic solids}

For a transversely isotropic solid
\begin{equation} 
	\delta W=\sum_{j=1}^5\frac{\partial \overline{W}}{\partial I_j}\delta I_j=\sum_{j=1}^5W_j\,\delta I_j
	\,,
\end{equation}
where
\begin{equation} 
	\delta I_1 = -\mathbf{C}^{\sharp}\!:\!\delta\mathbf{G}\,,\qquad
	\delta I_2 = \left(-I_1\mathbf{C}^{\sharp}+\mathbf{C}^{2\sharp}\right)\!:\!\delta\mathbf{G} \,,\qquad
	\delta I_3 = -I_3 \mathbf{G}^{\sharp}\!:\!\delta\mathbf{G} 
	\,.
\end{equation}
Note that $\delta \mathbf{G}=\delta\Fr^{\star}\Fr^{-\star}\mathbf{G}+\mathbf{G}\Fr^{-1}\delta\Fr$. Thus
\begin{equation} \label{I1-I3-variation}
	\delta I_1 = -2\Fr^{-\star}\mathbf{C}\!:\!\delta\Fr\,,\qquad
	\delta I_2 = \left[-I_1\Fr^{-\star}\mathbf{C}+\Fr^{-\star}\mathbf{C}^2\right]\!:\!\delta\Fr \,,\qquad
	\delta I_3 = -I_3 \Fr^{-\star}\!:\!\delta\Fr
	\,.
\end{equation}
Also
\begin{equation} \label{I4-I5-variation}
\begin{aligned}
	\delta I_4 &= -2\left[ \Fr^{-\star}\mathbf{C}^{\flat}\mathbf{N}\otimes \mathbf{N} \right]\!:\!\delta\Fr  \, \\
	\delta I_5 &= -2\left[ \Fr^{-\star}\mathbf{C}^{2\flat}\mathbf{N}\otimes \mathbf{N} 
	+\Fr^{-\star}\mathbf{C}^{\flat}\mathbf{N}\otimes \mathbf{C}\mathbf{N}  \right]\!:\!\delta\Fr
	\,.
\end{aligned}
\end{equation}
Thus
\begin{equation} 
\begin{aligned}
	-\delta W &= 2\Big[ I_3W_3 \Fr^{-\star}+(W_1+I_1W_2)\Fr^{-\star}\mathbf{C}-W_2 \Fr^{-\star}\mathbf{C}^2
	+ W_4\, \Fr^{-\star}\mathbf{C}^{\flat}\mathbf{N}\otimes \mathbf{N}   \\
	& \qquad 
	+W_5\left( \Fr^{-\star}\mathbf{C}^{2\flat}\mathbf{N}\otimes \mathbf{N} 
	+\Fr^{-\star}\mathbf{C}^{\flat}\mathbf{N}\otimes \mathbf{C}\mathbf{N}  \right) \Big]\!:\!\delta\Fr
	\,.
\end{aligned}
\end{equation}
Therefore, the remodeling equation for a transversely isotopic solid is written as
\begin{equation} \label{Remodeling-Equation-TI}
\begin{aligned}
	\frac{\partial \phi}{\partial \dot{\Fr}} & = (q+2I_3W_3)\Fr^{-\star}
	+ 2(W_1+I_1W_2)\Fr^{-\star}\mathbf{C}-2W_2 \Fr^{-\star}\mathbf{C}^2
	+ 2W_4\, \Fr^{-\star}\mathbf{C}^{\flat}\mathbf{N}\otimes \mathbf{N} \\
	& \quad +2W_5\left( \Fr^{-\star}\mathbf{C}^{2\flat}\mathbf{N}\otimes \mathbf{N} 
	+\Fr^{-\star}\mathbf{C}^{\flat}\mathbf{N}\otimes \mathbf{C}\mathbf{N}  \right)\,.
\end{aligned}
\end{equation}

\subsubsection{Remodeling equation for orthotropic solids}

For an orthotropic solid
\begin{equation} 
	\delta W=\sum_{j=1}^7\frac{\partial \overline{W}}{\partial I_j}\delta I_j=\sum_{j=1}^7W_j\,\delta I_j
	\,,
\end{equation}
where $\delta I_1$, $\delta I_2$, and $\delta I_3$ are given in \eqref{I1-I3-variation}, and
\begin{equation} \label{I4-I7-variation}
\begin{aligned}
	\delta I_4 &= -2\left[ \Fr^{-\star}\mathbf{C}^{\flat}\mathbf{N}_1\otimes \mathbf{N}_1 \right]\!:\!\delta\Fr\,, && 
	\delta I_5 = -2\left[ \Fr^{-\star}\mathbf{C}^{2\flat}\mathbf{N}_1\otimes \mathbf{N}_1 
	+\Fr^{-\star}\mathbf{C}^{\flat}\mathbf{N}_1\otimes \mathbf{C}\mathbf{N}_1  \right]\!:\!\delta\Fr\,, \\
	\delta I_6 &= -2\left[ \Fr^{-\star}\mathbf{C}^{\flat}\mathbf{N}_2\otimes \mathbf{N}_2 \right]\!:\!\delta\Fr\,, &&
	\delta I_7 = -2\left[ \Fr^{-\star}\mathbf{C}^{2\flat}\mathbf{N}_2\otimes \mathbf{N}_2 
	+\Fr^{-\star}\mathbf{C}^{\flat}\mathbf{N}_2\otimes \mathbf{C}\mathbf{N}_2  \right]\!:\!\delta\Fr 	\,.
\end{aligned}
\end{equation}
Thus
\begin{equation} 
\begin{aligned}
	-\delta W &= 2\Big[ I_3W_3 \Fr^{-\star}+(W_1+I_1W_2)\Fr^{-\star}\mathbf{C}-W_2 \Fr^{-\star}\mathbf{C}^2
	+ W_4\, \Fr^{-\star}\mathbf{C}^{\flat}\mathbf{N}_1\otimes \mathbf{N}_1   \\
	& \qquad 
	+W_5\left( \Fr^{-\star}\mathbf{C}^{2\flat}\mathbf{N}_1\otimes \mathbf{N}_1 
	+\Fr^{-\star}\mathbf{C}^{\flat}\mathbf{N}_1\otimes \mathbf{C}\mathbf{N}_1  \right) 
	+ W_6\, \Fr^{-\star}\mathbf{C}^{\flat}\mathbf{N}_2\otimes \mathbf{N}_2    \\
	& \qquad+W_7\left( \Fr^{-\star}\mathbf{C}^{2\flat}\mathbf{N}_2\otimes \mathbf{N}_2
	+\Fr^{-\star}\mathbf{C}^{\flat}\mathbf{N}_2\otimes \mathbf{C}\mathbf{N}_2 \right)
	\Big]\!:\!\delta\Fr
	\,.
\end{aligned}
\end{equation}
Therefore, the remodeling equation for a transversely isotopic solid is written as
\begin{equation} \label{Remodeling-Equation-Ort}
\begin{aligned}
	\frac{\partial \phi}{\partial \dot{\Fr}} & = (q+2I_3W_3)\Fr^{-\star}
	+ 2(W_1+I_1W_2)\Fr^{-\star}\mathbf{C}-2W_2 \Fr^{-\star}\mathbf{C}^2
	+ 2W_4\, \Fr^{-\star}\mathbf{C}^{\flat}\mathbf{N}_1\otimes \mathbf{N}_1 \\
	& \quad +2W_5\left( \Fr^{-\star}\mathbf{C}^{2\flat}\mathbf{N}_1\otimes \mathbf{N}_1 
	+\Fr^{-\star}\mathbf{C}^{\flat}\mathbf{N}_1\otimes \mathbf{C}\mathbf{N}_1  \right)
	+ 2W_6\, \Fr^{-\star}\mathbf{C}^{\flat}\mathbf{N}_2\otimes \mathbf{N}_2 \\
	& \quad +2W_7\left( \Fr^{-\star}\mathbf{C}^{2\flat}\mathbf{N}_2\otimes \mathbf{N}_2 
	+\Fr^{-\star}\mathbf{C}^{\flat}\mathbf{N}_2\otimes \mathbf{C}\mathbf{N}_2  \right)\,.
\end{aligned}
\end{equation}

\subsubsection{Remodeling equation for monoclinic solids}

For a monoclinic solid
\begin{equation} 
	\delta W=\sum_{j=1}^9\frac{\partial \overline{W}}{\partial I_j}\delta I_j=\sum_{j=1}^9W_j\,\delta I_j
	\,,
\end{equation}
where $\delta I_1$, $\delta I_2$, and $\delta I_3$ are given in \eqref{I1-I3-variation}, $\delta I_4,\hdots,\delta I_7$ are given in \eqref{I4-I7-variation}, and
\begin{equation}
\begin{aligned}
	\delta I_8 &= -\Fr^{-\star} \left[ \frac{I_8}{\mathcal{I}}\left(\mathbf{N}_2^{\flat}\otimes\mathbf{N}_1
	+\mathbf{N}_1^{\flat}\otimes\mathbf{N}_2 \right)
	+\mathcal{I}\,\mathbf{C}^{\flat}\left(\mathbf{N}_2\otimes\mathbf{N}_1
	+\mathbf{N}_1\otimes\mathbf{N}_2 \right) 
	\right]\!:\!\delta\Fr\,, \\
	\delta I_9 &= -2\mathcal{I}\,\Fr^{-\star}\left(\mathbf{N}_2^{\flat}\otimes\mathbf{N}_1
	+\mathbf{N}_1^{\flat}\otimes\mathbf{N}_2 \right)\!:\!\delta\Fr \,.
\end{aligned}
\end{equation}
Therefore, the remodeling equation for a monoclinic solid is written as
\begin{equation} \label{Remodeling-Equation-Mon}
\begin{aligned}
	\frac{\partial \phi}{\partial \dot{\Fr}} & = (q+2I_3W_3)\Fr^{-\star}
	+ 2(W_1+I_1W_2)\Fr^{-\star}\mathbf{C}-2W_2 \Fr^{-\star}\mathbf{C}^2
	+ 2W_4\, \Fr^{-\star}\mathbf{C}^{\flat}\mathbf{N}_1\otimes \mathbf{N}_1 \\
	& \quad +2W_5\left( \Fr^{-\star}\mathbf{C}^{2\flat}\mathbf{N}_1\otimes \mathbf{N}_1 
	+\Fr^{-\star}\mathbf{C}^{\flat}\mathbf{N}_1\otimes \mathbf{C}\mathbf{N}_1  \right)
	+ 2W_6\, \Fr^{-\star}\mathbf{C}^{\flat}\mathbf{N}_2\otimes \mathbf{N}_2 \\
	& \quad +2W_7\left( \Fr^{-\star}\mathbf{C}^{2\flat}\mathbf{N}_2\otimes \mathbf{N}_2 
	+\Fr^{-\star}\mathbf{C}^{\flat}\mathbf{N}_2\otimes \mathbf{C}\mathbf{N}_2  \right)
	+2W_9\,\mathcal{I}\,\Fr^{-\star}\left(\mathbf{N}_2^{\flat}\otimes\mathbf{N}_1
	+\mathbf{N}_1^{\flat}\otimes\mathbf{N}_2 \right)\\
	&\quad +W_8\,\Fr^{-\star} \left[ \frac{I_8}{\mathcal{I}}\left(\mathbf{N}_2^{\flat}\otimes\mathbf{N}_1
	+\mathbf{N}_1^{\flat}\otimes\mathbf{N}_2 \right)
	+\mathcal{I}\,\mathbf{C}^{\flat}\left(\mathbf{N}_2\otimes\mathbf{N}_1
	+\mathbf{N}_1\otimes\mathbf{N}_2 \right) 
	\right]
	\,.
\end{aligned}
\end{equation}

\subsubsection{Remodeling equation for $SO(3)$-remodeling} \label{Sec:SO3}

A special class of remodeling is when at every point the remodeling tensor is a rotation. In this case
$\mathbf{G}=\Fr^{\star}\mathring{\mathbf{G}}\Fr=\mathring{\mathbf{G}}$. 
Thus, $\delta\mathbf{G}=\mathbf{0}$. This, in particular, implies that $W_1$, $W_2$, and $W_3$ do not contribute to the remodeling equation. Notice that $\delta\Fr^{\star}\mathring{\mathbf{G}}\Fr+\Fr^{\star}\mathring{\mathbf{G}}\delta\Fr=\mathbf{0}$, or $(\Fr^{\star}\mathring{\mathbf{G}}\delta\Fr)^{\star}+\Fr^{\star}\mathring{\mathbf{G}}\delta\Fr=\mathbf{0}$. 
Thus, the tensor $\boldsymbol{\Omega}=\Fr^{\star}\mathring{\mathbf{G}}\delta\Fr$ is antisymmetric.

\begin{remark} \label{Stress-Free}
Assuming that the initial body is stress-free, $\mathring{\mathbf{G}}$ is a flat metric (its Riemann curvature vanishes). Therefore, from $\mathbf{G}=\Fr^{\star}\mathring{\mathbf{G}}\Fr=\mathring{\mathbf{G}}$ it is concluded that in $SO(3)$-remodeling the material metric remains flat. This implies that $SO(3)$-remodeling does not induce residual stresses.
\end{remark}

\paragraph{Transversely isotropic solids.}For transversely isotropic solids the elastic energy contributes to the remodeling equation through the invariants $I_4$ and $I_5$.
In order to directly take into account this constraint, \eqref{I4-I5-variation} can be rewritten in terms of $\boldsymbol{\Omega}$.
Note that
\begin{equation} 
\begin{aligned}
	\Fr^{-\star}\mathbf{C}^{\flat}\mathbf{N}\otimes \mathbf{N}\!:\!\delta\Fr
	&=\Fr^{-\star}\mathbf{C}^{\flat}\mathbf{N}\otimes \mathbf{N}\!:\!
	\mathbf{G}^{\sharp}\Fr^{-\star}\Fr^{\star}\mathbf{G}\delta\Fr \\
	& =\Fr^{-\star}\mathbf{G}^{\sharp}\Fr^{-1}\mathbf{C}^{\flat}\mathbf{N}\otimes \mathbf{N}\!:\!
	\boldsymbol{\Omega}\\
	&=\mathbf{G}^{\sharp}\mathbf{C}^{\flat}\mathbf{N}\otimes \mathbf{N}\!:\!\boldsymbol{\Omega} \\
	&=\mathbf{C}\mathbf{N}\otimes \mathbf{N}\!:\!\boldsymbol{\Omega} \\
	& =\frac{1}{2}\left(\mathbf{C}\mathbf{N}\otimes \mathbf{N}
	-\mathbf{N}\otimes \mathbf{C}\mathbf{N}\right)\!:\!\boldsymbol{\Omega}
	\,,
\end{aligned}
\end{equation}
where anti-symmetry of $\boldsymbol{\Omega}$ was used. The two terms that appear in $\delta I_5$ are simplified as follows. The first terms is rewritten as
\begin{equation} 
\begin{aligned}
	\Fr^{-\star}\mathbf{C}^{2\flat}\mathbf{N}\otimes \mathbf{N} \!:\!\delta\Fr
	&=\Fr^{-\star}\mathbf{C}^{2\flat}\mathbf{N}\otimes \mathbf{N} \!:\!
	\mathbf{G}^{\sharp}\Fr^{-\star}\Fr^{\star}\mathbf{G}\delta\Fr \\
	& =\Fr^{-\star}\mathbf{G}^{\sharp}\Fr^{-1}\mathbf{C}^{2\flat}\mathbf{N}\otimes \mathbf{N} \!:\!
	\Fr^{\star}\mathbf{G}\delta\Fr  \\
	& =\mathbf{G}^{\sharp}\mathbf{C}^{2\flat}\mathbf{N}\otimes \mathbf{N} \!:\! \boldsymbol{\Omega} \\
	& =\mathbf{C}^{2}\mathbf{N}\otimes \mathbf{N} \!:\! \boldsymbol{\Omega} \\
	& =\frac{1}{2}\left(\mathbf{C}^{2}\mathbf{N}\otimes \mathbf{N}
	-\mathbf{N} \otimes \mathbf{C}^{2}\mathbf{N} \right)\!:\! \boldsymbol{\Omega}\,.
\end{aligned}
\end{equation}
For the second term
\begin{equation} 
\begin{aligned}
	\Fr^{-\star}\mathbf{C}^{\flat}\mathbf{N}\otimes \mathbf{C}\mathbf{N} \!:\!\delta\Fr
	&=\Fr^{-\star}\mathbf{C}^{\flat}\mathbf{N}\otimes \mathbf{C}\mathbf{N}\!:\!
	\mathbf{G}^{\sharp}\Fr^{-\star}\Fr^{\star}\mathbf{G}\delta\Fr \\
	& =\Fr^{-\star}\mathbf{G}^{\sharp}\Fr^{-1}\mathbf{C}^{\flat}\mathbf{N}\otimes \mathbf{C}\mathbf{N} \!:\!
	\Fr^{\star}\mathbf{G}\delta\Fr  \\
	& =\mathbf{G}^{\sharp}\mathbf{C}^{\flat}\mathbf{N}\otimes \mathbf{C}\mathbf{N} \!:\! \boldsymbol{\Omega} \\
	& =\mathbf{C}\mathbf{N}\otimes \mathbf{C}\mathbf{N} \!:\! \boldsymbol{\Omega} \\
	& =\mathbf{0}\,.
\end{aligned}
\end{equation}
Thus
\begin{equation} 
\begin{aligned}
	\delta I_4 = \left(\mathbf{N}\otimes \mathbf{C}\mathbf{N}
	-\mathbf{C}\mathbf{N}\otimes \mathbf{N}\right)\!:\!\boldsymbol{\Omega}  \,, \qquad
	\delta I_5 =\left(\mathbf{N} \otimes \mathbf{C}^{2}\mathbf{N} 
	-\mathbf{C}^{2}\mathbf{N}\otimes \mathbf{N}\right)\!:\! \boldsymbol{\Omega}
	\,.
\end{aligned}
\end{equation}
The contribution of the dissipation potential to the variational principle is simplified as 
\begin{equation} 
\begin{aligned}
	\frac{\partial \phi}{\partial \dot{\Fr}} \!:\!\delta\Fr 
	&= \frac{\partial \phi}{\partial \dot{\Fr}} \!:\! \mathbf{G}^{\sharp}\Fr^{-\star}\Fr^{\star}\mathbf{G}\delta\Fr \\
	&=\Fr^{-1}\mathbf{G}^{\sharp}\frac{\partial \phi}{\partial \dot{\Fr}} \!:\! \Fr^{\star}\mathbf{G}\delta\Fr \\
	&=\Fr^{-1}\mathbf{G}^{\sharp}\frac{\partial \phi}{\partial \dot{\Fr}} \!:\! \boldsymbol{\Omega} \\
	&=\frac{1}{2}\left[ \Fr^{-1}\mathbf{G}^{\sharp}\frac{\partial \phi}{\partial \dot{\Fr}} 
	- \left(\frac{\partial \phi}{\partial \dot{\Fr}}\right)^{\star}
	\mathbf{G}^{\sharp}\Fr^{-\star} \right]\!:\! \boldsymbol{\Omega}
	\,.
\end{aligned}
\end{equation}
Therefore, the $SO(3)$-remodeling equation is simplified to read
\begin{equation} \label{Remodeling-Equation-SO3}
\begin{aligned}
	\Fr^{-1}\mathbf{G}^{\sharp}\frac{\partial \phi}{\partial \dot{\Fr}} 
	- \left(\frac{\partial \phi}{\partial \dot{\Fr}}\right)^{\!\star}	\mathbf{G}^{\sharp}\Fr^{-\star}  
	+2W_4 \left(\mathbf{N}\otimes \mathbf{C}\mathbf{N}
	-\mathbf{C}\mathbf{N}\otimes \mathbf{N}\right)
	+2W_5\left(\mathbf{N} \otimes \mathbf{C}^{2}\mathbf{N} 
	-\mathbf{C}^{2}\mathbf{N}\otimes \mathbf{N}\right)=\mathbf{0}
	\,.
\end{aligned}
\end{equation}
When \eqref{phi-isotropic} is assumed the remodeling equation is simplified to read
\begin{equation}  \label{Remodeling-Fr-Iso}
\begin{aligned}
	& \eta_1(\operatorname{tr}\dot\Fr)\left( \Fr^{-1}\mathbf{G}^{\sharp}-\mathbf{G}^{\sharp}\Fr^{-\star} \right)
	+\eta_2\left( \Fr^{-1}\mathbf{G}^{\sharp}\dot{\Fr}^{\star}- \dot{\Fr}\mathbf{G}^{\sharp}\Fr^{-\star} \right)
	+\eta_3\left( \Fr^{-1}\dot{\Fr}\mathbf{G}^{\sharp}\dot{\Fr}^{\star}
	- \mathbf{G}^{\sharp}\dot{\Fr}^{\star}\Fr^{-\star} \right)  \\
	& +2W_4 \left(\mathbf{N}\otimes \mathbf{C}\mathbf{N}
	-\mathbf{C}\mathbf{N}\otimes \mathbf{N}\right)
	+2W_5\left(\mathbf{N} \otimes \mathbf{C}^{2}\mathbf{N} 
	-\mathbf{C}^{2}\mathbf{N}\otimes \mathbf{N}\right)=\mathbf{0}
	\,.
\end{aligned}
\end{equation}
The initial condition for the remodeling tensor is $\Fr(X,0)=\mathbf{I}$.

\begin{remark}It should be noted that $\mathbf{N}$ and $-\mathbf{N}$ define the same fiber orientation, i.e., $\mathbf{N}\in\mathbb{RP}^2$---the projective plane. We see that the right-hand side of \eqref{Remodeling-Fr-Iso} is indeed invariant under the transformation $\mathbf{N}\mapsto-\mathbf{N}$.
\end{remark}

\paragraph{Orthotropic solids.}For orthotropic solids the elastic energy contributes to the remodeling equation through the invariants $I_4$, $I_5$, $I_6$, and $I_7$. The kinetic equation reads
\begin{equation} \label{Remodeling-Equation-SO3-Orthotropic}
\begin{aligned}
	 \Fr^{-1}\mathbf{G}^{\sharp}\frac{\partial \phi}{\partial \dot{\Fr}} 
	& - \left(\frac{\partial \phi}{\partial \dot{\Fr}}\right)^{\!\star}	\mathbf{G}^{\sharp}\Fr^{-\star}  
	+2W_4 \left(\mathbf{N}_1\otimes \mathbf{C}\mathbf{N}_1
	-\mathbf{C}\mathbf{N}_1\otimes \mathbf{N}_1\right)
	+2W_5\left(\mathbf{N}_1 \otimes \mathbf{C}^{2}\mathbf{N}_1 
	-\mathbf{C}^{2}\mathbf{N}_1\otimes \mathbf{N}_1 \right) \\
	&  +2W_6 \left(\mathbf{N}_2\otimes \mathbf{C}\mathbf{N}_2
	-\mathbf{C}\mathbf{N}_2\otimes \mathbf{N}_2\right)
	+2W_7\left(\mathbf{N}_2 \otimes \mathbf{C}^{2}\mathbf{N}_2 
	-\mathbf{C}^{2}\mathbf{N}_2\otimes \mathbf{N}_2 \right)
	=\mathbf{0}
	\,.
\end{aligned}
\end{equation}

\paragraph{Monoclinic solids.}It is straightforward to show that $\delta\mathcal{I}=0$, and hence $\delta I_9=0$. Also
\begin{equation} 
\begin{aligned}
	\delta I_8 = \mathcal{I}\left[\mathbf{N}_1\otimes \mathbf{C}\mathbf{N}_2
	+\mathbf{N}_2\otimes \mathbf{C}\mathbf{N}_2
	-\left( \mathbf{C}\mathbf{N}_2\otimes \mathbf{N}_1+ \mathbf{C}\mathbf{N}_1\otimes \mathbf{N}_2 	\right) \right]\!:\!\boldsymbol{\Omega}\,.
\end{aligned}
\end{equation}
The kinetic equation is written as
\begin{equation} \label{Remodeling-Equation-SO3-Monoclinic}
\begin{aligned}
	 \Fr^{-1}\mathbf{G}^{\sharp}\frac{\partial \phi}{\partial \dot{\Fr}} 
	& - \left(\frac{\partial \phi}{\partial \dot{\Fr}}\right)^{\!\star}	\mathbf{G}^{\sharp}\Fr^{-\star}  
	+2W_4 \left(\mathbf{N}_1\otimes \mathbf{C}\mathbf{N}_1
	-\mathbf{C}\mathbf{N}_1\otimes \mathbf{N}_1\right)
	+2W_5\left(\mathbf{N}_1 \otimes \mathbf{C}^{2}\mathbf{N}_1 
	-\mathbf{C}^{2}\mathbf{N}_1\otimes \mathbf{N}_1 \right) \\
	&  +2W_6 \left(\mathbf{N}_2\otimes \mathbf{C}\mathbf{N}_2
	-\mathbf{C}\mathbf{N}_2\otimes \mathbf{N}_2\right)
	+2W_7\left(\mathbf{N}_2 \otimes \mathbf{C}^{2}\mathbf{N}_2 
	-\mathbf{C}^{2}\mathbf{N}_2\otimes \mathbf{N}_2 \right) \\
	& +2\mathcal{I}\, W_8 \left[\mathbf{N}_1\otimes \mathbf{C}\mathbf{N}_2
	+\mathbf{N}_2\otimes \mathbf{C}\mathbf{N}_2
	-\left( \mathbf{C}\mathbf{N}_2\otimes \mathbf{N}_1+ \mathbf{C}\mathbf{N}_1\otimes \mathbf{N}_2 	\right) \right]	=\mathbf{0}	\,.
\end{aligned}
\end{equation}

\subsubsection{Remodeling equation for fiber reorientation: A single family of fibers} \label{Sec:N1-Remodeling}

So far we have written the remodeling equation explicitly in terms of the remodeling tensor $\Fr$. In the literature the following reorientation kinetic equation has been suggested and used \citep{Menzel2005,Melnik2013}
\begin{equation}  \label{Menzel-N}
	\frac{d\mathbf{N}(X,t)}{dt}=\frac{1}{\tau}\Big[\mathbf{N}^{\mathbf{C}}_{\text{max}}(X,t)
	-\left(\mathbf{N}^{\mathbf{C}}_{\text{max}}(X,t)\cdot\mathbf{N}(X,t)\right)\mathbf{N}(X,t) \Big]
	\,,
\end{equation}
where $\tau$ is a relaxation time, and $\mathbf{N}^{\mathbf{C}}_{\text{max}}$ is a unit vector along the maximum stretch at $X\in\mathcal{B}$. Note that $\mathbf{N}=\mathbf{N}^{\mathbf{C}}_{\text{max}}$ is an equilibrium point of the above ODE.
In our formulation $\mathbf{N}(X,t)=\Fr^*(X,t)\,\mathring{\mathbf{N}}(X)=\Fr^{-1}(X,t)\mathring{\mathbf{N}}(X)$, and hence
\begin{equation}  
	\frac{d\mathbf{N}}{dt}=-\Fr^{-1}\dot{\Fr}\Fr^{-1}\mathring{\mathbf{N}}=-\Fr^{-1}\dot{\Fr}\mathbf{N}
	\,.
\end{equation}
Instead of assuming that $\Fr$ is the independent remodeling field, one can use $\mathbf{N}$ directly. In this case, instead of \eqref{Dissipation-Potential-Quadratic} one can assume the following dissipation potential 
\begin{equation} \label{Dissipation-Potential-N}
	\phi(X,\mathbf{F},\mathbf{N},\dot{\mathbf{N}},\mathbf{G},\mathbf{g})
	=\frac{1}{2}\,\dot{\mathbf{N}}\cdot\boldsymbol{\mathsf{B}}(X,\mathbf{F},\Fe,\mathbf{G},
	\mathbf{N}\otimes\mathbf{N},\mathbf{g})
	\cdot\dot{\mathbf{N}}	=\frac{1}{2}\,\dot N^A\,\dot N^B\,\mathsf{B}_{AB}\,,
\end{equation}
where $\boldsymbol{\mathsf{B}}(X,\mathbf{F},\Fe,\mathbf{G},\mathbf{N}\otimes\mathbf{N},\mathbf{g})$ is a positive-definite isotropic second-order tensor.
Objectivity implies that $\boldsymbol{\mathsf{B}}(X,\mathbf{F},\Fe,\mathbf{G},\mathbf{N}\otimes\mathbf{N},\mathbf{g})=\hat{\boldsymbol{\mathsf{B}}}(X,\mathbf{C}^{\flat},\Ce^{\flat},\mathbf{G},\mathbf{N}\otimes\mathbf{N})$.
For the sake of simplicity, we can assume that  $\boldsymbol{\mathsf{B}}=\hat{\boldsymbol{\mathsf{B}}}(X,\mathbf{C}^{\flat},\mathbf{G},\mathbf{N}\otimes\mathbf{N})$. Knowing that $\hat{\boldsymbol{\mathsf{B}}}$ is an isotropic function of its arguments, we conclude that $\hat{\boldsymbol{\mathsf{B}}}=K(I_1,I_2,I_3,I_4,I_5)\,\mathbf{G}$. Thus, $\phi=\frac{1}{2}K(I_1,I_2,I_3,I_4,I_5)\llangle \dot{\mathbf{N}},\dot{\mathbf{N}} \rrangle_{\mathbf{G}}$.

\paragraph{Remodeling energy.}
In addition to the elastic energy let us consider a remodeling energy $\Wr=\Wr(X,\mathbf{C}^{\flat},\mathbf{N},\mathbf{G})$ that quantifies the tendency of the fibers to orient themselves along a particular direction, e.g., the direction of maximum stretch or stress. Thus, the Lagrangian density is defined as $\mathcal{L} = \mathcal{T} - W-\Wr+q_n\big(\mathbf{N}\cdot\mathbf{N}-1\big)$, where $q_n=q_n(X,t)$ is a Lagrange multiplier field corresponding to the internal constraint $\mathbf{N}\cdot\mathbf{N}=1$.
The two independent variations are now $(\delta\varphi,\delta\mathbf{N})$. 

For $\delta\mathbf{N}$ variations, $\delta\mathcal{L} = - \delta W- \delta \Wr+2q_n \mathbf{N}^{\flat}\cdot \delta\mathbf{N}$. Hence, \eqref{LD-Principle} is simplified to read
\begin{equation} 
	\int_{t_1}^{t_2}  \int_{\mathcal B} \left[-W_4\,\frac{\partial I_4}{\partial \mathbf{N}}
	-W_5\,\frac{\partial I_5}{\partial \mathbf{N}}
	-\frac{\partial \Wr}{\partial \mathbf{N}}
	+2q_n \mathbf{N}^{\flat} -\frac{\partial \phi}{\partial \dot{\mathbf{N}}} \right]\!\cdot
	\delta\mathbf{N}\,dV\,\mathrm dt  
	= 0 \,.
\end{equation}
Thus, the remodeling equation reads 
\begin{equation} 
	\frac{\partial \phi}{\partial \dot{\mathbf{N}}}  =
	2q_n \mathbf{N}^{\flat}-\frac{\partial \Wr}{\partial \mathbf{N}}
	-W_4\,\frac{\partial I_4}{\partial \mathbf{N}}
	-W_5\,\frac{\partial I_5}{\partial \mathbf{N}}
	 \,.
\end{equation}
But $\frac{\partial I_4}{\partial \mathbf{N}}=2\mathbf{C}^{\flat}\cdot\mathbf{N}$, $\frac{\partial I_5}{\partial \mathbf{N}}=2\mathbf{C}^{2\flat}\cdot\mathbf{N}$, and hence
\begin{equation} 
	\frac{\partial \phi}{\partial \dot{\mathbf{N}}}  =2q_n \mathbf{N}^{\flat}
	-\frac{\partial \Wr}{\partial \mathbf{N}}
	-2W_4\,\mathbf{C}^{\flat}\cdot\mathbf{N}-2W_5\,\mathbf{C}^{2\flat}\cdot\mathbf{N}	 \,.
\end{equation}
Eliminating $q_n$, the remodeling equation can be rewritten as
\begin{equation} \label{Kinetic-N-General}
	\mathbf{G}^{\sharp}\frac{\partial \phi}{\partial \dot{\mathbf{N}}}
	-\left\langle \frac{\partial \phi}{\partial \dot{\mathbf{N}}}\,,\mathbf{N}\right\rangle \mathbf{N}	
	=\left\langle \frac{\partial \Wr}{\partial \mathbf{N}},\mathbf{N} \right\rangle \mathbf{N}
	-\frac{\partial \Wr}{\partial \mathbf{N}}
	+2W_4(I_4\mathbf{N}-\mathbf{C}\cdot\mathbf{N})
	+2W_5(I_5\mathbf{N}-\mathbf{C}^2\cdot\mathbf{N}) \,.
\end{equation}
For the dissipation potential \eqref{Dissipation-Potential-N}, $\frac{\partial \phi}{\partial \dot{\mathbf{N}}}=\boldsymbol{\mathsf{B}}\cdot\dot{\mathbf{N}}=K\,\dot{\mathbf{N}}^{\flat}$, and one obtains
\begin{equation} 
	K\left[\dot{\mathbf{N}}-\big(\mathbf{N}\!\cdot\dot{\mathbf{N}} \big)\mathbf{N}\right]
	=\left\langle \frac{\partial \Wr}{\partial \mathbf{N}},\mathbf{N} \right\rangle \mathbf{N}
	-\mathbf{G}^{\sharp}\frac{\partial \Wr}{\partial \mathbf{N}}
	+2W_4(I_4\mathbf{N}-\mathbf{C}\cdot\mathbf{N})
	+2W_5(I_5\mathbf{N}-\mathbf{C}^2\cdot\mathbf{N})
	\,.
\end{equation}
As $\mathbf{N}$ is a unit vector, $\mathbf{N}\!\cdot\dot{\mathbf{N}}=0$, and hence the remodeling equation is simplified to read 
\begin{equation} \label{Kinetic-N}
	K\dot{\mathbf{N}}
	=\left\langle \frac{\partial \Wr}{\partial \mathbf{N}},\mathbf{N} \right\rangle \mathbf{N}
	-\mathbf{G}^{\sharp}\frac{\partial \Wr}{\partial \mathbf{N}}
	+2W_4(I_4\mathbf{N}-\mathbf{C}\cdot\mathbf{N})+2W_5(I_5\mathbf{N}
	-\mathbf{C}^2\cdot\mathbf{N})	\,.
\end{equation}

\begin{example}
Let us consider the remodeling energy $\Wr(X,\mathbf{C}^{\flat},\mathbf{N},\mathbf{G})=\kappa_M\,\mathbf{M}\cdot\mathbf{N}$, where $\kappa_M$ is a scalar and $\mathbf{M}$ is some unit vector that explicitly depends on $\mathbf{C}^{\flat}$. For this choice the remodeling equation is simplified to read
\begin{equation} \label{Kinetic-N-M}
	K\dot{\mathbf{N}}
	=\kappa_M \left[(\mathbf{M}\cdot\mathbf{N}) \mathbf{N}-\mathbf{M} \right]
	+2W_4(I_4\mathbf{N}-\mathbf{C}\cdot\mathbf{N})+2W_5(I_5\mathbf{N}
	-\mathbf{C}^2\cdot\mathbf{N})	\,.
\end{equation}
For the choice $\mathbf{M}=\mathbf{N}^{\mathbf{C}}_{\text{max}}$, this is a generalization of the remodeling equation suggested by \citet{Menzel2005}.
It should be noted that the remodeling equation proposed in \citep{Menzel2005} is not invariant under the transformation $\mathbf{N}^{\mathbf{C}}_{\text{max}}\mapsto -\mathbf{N}^{\mathbf{C}}_{\text{max}}$. Similarly, the term $\kappa_M \left[(\mathbf{M}\cdot\mathbf{N}) \mathbf{N}-\mathbf{M} \right]$ is not invariant under the transformation $\mathbf{M}\mapsto -\mathbf{M}$ because $\Wr(X,\mathbf{C}^{\flat},\mathbf{N},\mathbf{G})=\kappa_M\,\mathbf{M}\cdot\mathbf{N}$ is not an acceptable remodeling energy. 
For the remodeling energy $\Wr(X,\mathbf{C}^{\flat},\mathbf{N},\mathbf{G})=\frac{1}{2}\kappa_M\,(\mathbf{M}\cdot\mathbf{N})^2$, the kinetic equation reads
\begin{equation} \label{Kinetic-N-M2}
	K\dot{\mathbf{N}}
	=\kappa_M\,(\mathbf{M}\cdot\mathbf{N}) 
	\left[(\mathbf{M}\cdot\mathbf{N}) \mathbf{N}-\mathbf{M} \right]
	+2W_4(I_4\mathbf{N}-\mathbf{C}\cdot\mathbf{N})+2W_5(I_5\mathbf{N}
	-\mathbf{C}^2\cdot\mathbf{N})	\,.
\end{equation}
Similarly, for the remodeling energy $\Wr(X,\mathbf{C}^{\flat},\mathbf{N},\mathbf{G})=\kappa_M\,|\mathbf{M}\cdot\mathbf{N}|$, the kinetic equation reads
\begin{equation} \label{Kinetic-N-M1}
	K\dot{\mathbf{N}}
	=\kappa_M\,\operatorname{sgn}(\mathbf{M}\cdot\mathbf{N}) 
	\left[(\mathbf{M}\cdot\mathbf{N}) \mathbf{N}-\mathbf{M} \right]
	+2W_4(I_4\mathbf{N}-\mathbf{C}\cdot\mathbf{N})+2W_5(I_5\mathbf{N}
	-\mathbf{C}^2\cdot\mathbf{N})	\,,
\end{equation}
where $\operatorname{sgn}$ is the sign function.
\end{example}

\begin{remark}
In order to understand the remodeling equation better, let us consider the spectral decomposition of $\mathbf{C}^{\flat}$: 
\begin{equation} 
	\mathbf{C}^{\flat}=\lambda_1^2\mathbf{E}_1\otimes\mathbf{E}_1
	+\lambda_2^2\mathbf{E}_2\otimes\mathbf{E}_2+\lambda_1^3\mathbf{E}_3\otimes\mathbf{E}_3
	\,,
\end{equation}
where we assume that $\lambda_1>\lambda_2> \lambda_3$, and hence $\mathbf{N}^{\mathbf{C}}_{\text{max}}=\mathbf{E}_1$. Note that
\begin{equation} 
	I_4=\lambda_1^2\left(\mathbf{N}\cdot\mathbf{E}_1\right)^2
	+\lambda_2^2\left(\mathbf{N}\cdot\mathbf{E}_2\right)^2
	+\lambda_3^2\left(\mathbf{N}\cdot\mathbf{E}_3\right)^2\,,\qquad
	I_5=\lambda_1^4\left(\mathbf{N}\cdot\mathbf{E}_1\right)^2
	+\lambda_2^4\left(\mathbf{N}\cdot\mathbf{E}_2\right)^2
	+\lambda_3^4\left(\mathbf{N}\cdot\mathbf{E}_3\right)^2
	\,.
\end{equation}
Suppose $(\mathbf{M}\cdot\mathbf{N}) \mathbf{N}-\mathbf{M}=\mathbf{0}$, and hence, $\mathbf{M}\cdot\mathbf{N}=\pm 1$. Thus, $\mathbf{N}=\pm \mathbf{M}$. If $\mathbf{N}$ is parallel to any of the principal directions of $\mathbf{C}$, i.e., $\mathbf{N}=\mathbf{E}_i$ for $i=1,2,3$, and $\mathbf{M}=\pm \mathbf{N}$, the right-hand side of \eqref{Kinetic-N-M} vanishes,\footnote{
If $\lambda_1>\lambda_2=\lambda_3$, then
\begin{equation} 
	\mathbf{C}^{\flat}=\lambda_1^2\mathbf{E}_1\otimes\mathbf{E}_1
	+\lambda_2^2\left(\mathbf{I}-\mathbf{E}_1\otimes\mathbf{E}_1\right)	\,.
\end{equation}
In this case, the right-hand side of \eqref{Kinetic-N-General} vanishes for $\mathbf{N}=\mathbf{E}_1$, and any $\mathbf{N}\perp\mathbf{E}_1$.} i.e., the principal directions of $\mathbf{C}$ are equilibrium points for \eqref{Kinetic-N-M}.\footnote{For a general dissipation potential this result holds as long as $\phi$ does not have a linear term in $\dot{\mathbf{N}}$, i.e., if $\frac{\partial \phi}{\partial \dot{\mathbf{N}}}\big|_{\dot{\mathbf{N}}=\mathbf{0}}=\mathbf{0} $.}
\end{remark}

\subsubsection{Remodeling equation for fiber reorientation: Two families of fibers} \label{Sec:N2-Remodeling}

Next let us consider an isotropic solid reinforced by two families of fibers that are not necessarily orthogonal, i.e., effectively a monoclinic solid. The independent fields of the theory are $\varphi$, $\mathbf{N}_1$, and $\mathbf{N}_2$.
We assume the following quadratic dissipation potential 
\begin{equation} \label{dissipation-N2}
	\phi(X,\mathbf{F},\mathbf{N}_1,\mathbf{N}_2,\dot{\mathbf{N}}_1,\dot{\mathbf{N}},\mathbf{G},\mathbf{g})
	=\frac{1}{2}\,\dot{\mathbf{N}}_1\cdot\boldsymbol{\mathsf{B}}_1\cdot\dot{\mathbf{N}}_1
	+\frac{1}{2}\,\dot{\mathbf{N}}_2\cdot\boldsymbol{\mathsf{B}}_2\cdot\dot{\mathbf{N}}_2
	+\dot{\mathbf{N}}_1\cdot\boldsymbol{\mathsf{B}}_3\cdot\dot{\mathbf{N}}_2
	\,,
\end{equation}
where $\boldsymbol{\mathsf{B}}_i$ are symmetric and isotropic functions of their arguments. This implies that
\begin{equation} 
	\phi=\frac{1}{2}K_1 \llangle \dot{\mathbf{N}}_1,\dot{\mathbf{N}}_1 \rrangle_{\mathbf{G}}
	+\frac{1}{2}K_2 \llangle \dot{\mathbf{N}}_2,\dot{\mathbf{N}}_2 \rrangle_{\mathbf{G}}
	+K_3 \llangle \dot{\mathbf{N}}_1,\dot{\mathbf{N}}_2 \rrangle_{\mathbf{G}}
	\,,
\end{equation}
where $K_i=K_i(I_1,\hdots,I_9)$, $i=1,2,3$. The rate of energy dissipation is 
\begin{equation} 
	\frac{\partial \phi}{\partial \dot{\mathbf{N}}_1}\cdot\dot{\mathbf{N}}_1
	+\frac{\partial \phi}{\partial \dot{\mathbf{N}}_2}\cdot\dot{\mathbf{N}}_2\geq 0
	\,.
\end{equation}
As $\dot{\mathbf{N}}_1$ and $\dot{\mathbf{N}}_2$ can vary independently, one concludes that
\begin{equation} 
	\frac{\partial \phi}{\partial \dot{\mathbf{N}}_1}\cdot\dot{\mathbf{N}}_1\geq 0\,,\qquad
	\frac{\partial \phi}{\partial \dot{\mathbf{N}}_2}\cdot\dot{\mathbf{N}}_2\geq 0
	\,.
\end{equation}
For the dissipation potential \eqref{dissipation-N2} this is written as
\begin{equation} 
	\begin{bmatrix}
	\boldsymbol{\mathsf{B}}_1 & \boldsymbol{\mathsf{B}}_3 \\
	\boldsymbol{\mathsf{B}}_3 & \boldsymbol{\mathsf{B}}_2
	\end{bmatrix}
	\begin{bmatrix}
	\dot{\mathbf{N}}_1 \\
	\dot{\mathbf{N}}_2
	\end{bmatrix} \cdot \begin{bmatrix}
	\dot{\mathbf{N}}_1 \\
	\dot{\mathbf{N}}_2
	\end{bmatrix} \geq 0
	\,.
\end{equation}
First, note that $\boldsymbol{\mathsf{B}}_1$ and $\boldsymbol{\mathsf{B}}_2$ are positive-definite, and hence $K_1,K_2>0$. According to Schur's complement condition \citep{De2006}, positive-definiteness of the block matrix is equivalent to positive-definiteness of either $\boldsymbol{\mathsf{B}}_2-\boldsymbol{\mathsf{B}}_3\boldsymbol{\mathsf{B}}_1^{-1}\boldsymbol{\mathsf{B}}_3$ or $\boldsymbol{\mathsf{B}}_1-\boldsymbol{\mathsf{B}}_3\boldsymbol{\mathsf{B}}_2^{-1}\boldsymbol{\mathsf{B}}_3$. This is equivalent to $K_3^2<K_1K_2$.

\paragraph{Remodeling energy.}
The Lagrangian density is written as $\mathcal{L} = \mathcal{T} - W-\Wr+q_{n1}\big(\mathbf{N}_1\cdot\mathbf{N}_1-1\big)+q_{n2}\big(\mathbf{N}_2\cdot\mathbf{N}_2-1\big)$, where $\Wr=\Wr(X,\mathbf{C}^{\flat},\mathbf{N}_1,\mathbf{N}_2,\mathbf{G})$ is the remodeling energy for the two fiber families, and $q_{n1}=q_{n1}(X,t)$ and $q_{n2}=q_{n2}(X,t)$ are the Lagrange multiplier fields corresponding to the internal constraints $\mathbf{N}_1\cdot\mathbf{N}_1=1$ and $\mathbf{N}_2\cdot\mathbf{N}_2=1$.

The three independent variations are now $(\delta\varphi,\delta\mathbf{N}_1,\delta\mathbf{N}_2)$. For $\delta\mathbf{N}_j$ variations, $\delta\mathcal{L} = - \delta W+2q_{nj} \mathbf{N}_j\cdot \delta\mathbf{N}_j$. Hence, \eqref{LD-Principle} is simplified to read
\begin{equation} 
\begin{aligned}
	& \int_{t_1}^{t_2}  \int_{\mathcal B} \left[-W_4\,\frac{\partial I_4}{\partial \mathbf{N}_1}
	-W_5\,\frac{\partial I_5}{\partial \mathbf{N}_1}-W_8\,\frac{\partial I_8}{\partial \mathbf{N}_1}
	-W_9\,\frac{\partial I_9}{\partial \mathbf{N}_1}
	-\frac{\partial \Wr}{\partial \mathbf{N}_1}
	+2q_{n1}\mathbf{N}_1^{\flat}-\frac{\partial \phi}{\partial \dot{\mathbf{N}}_1} \right]
	\!:\delta\mathbf{N}_1\,dV\,\mathrm dt = 0 \,, \\
	& \int_{t_1}^{t_2}  \int_{\mathcal B} \left[-W_6\,\frac{\partial I_6}{\partial \mathbf{N}_2}
	-W_7\,\frac{\partial I_7}{\partial \mathbf{N}_2}-W_8\,\frac{\partial I_8}{\partial \mathbf{N}_2}
	-W_9\,\frac{\partial I_9}{\partial \mathbf{N}_2}
	-\frac{\partial \Wr}{\partial \mathbf{N}_2}
	+2q_{n2} \mathbf{N}_2^{\flat}-\frac{\partial \phi}{\partial \dot{\mathbf{N}}_2} \right]
	\!:\delta\mathbf{N}_2\,dV\,\mathrm dt = 0 \,.
\end{aligned}
\end{equation}
Using the relations
\begin{equation} 
\begin{aligned}
	& \frac{\partial I_4}{\partial \mathbf{N}_1}=2\mathbf{C}^{\flat}\cdot\mathbf{N}_1\,,&&
	\frac{\partial I_5}{\partial \mathbf{N}_1}=2\mathbf{C}^{2\sharp}\cdot\mathbf{N}_1 \,,&& 
	\frac{\partial I_8}{\partial \mathbf{N}_1}=I_8\,\mathcal{I}^{-1}\mathbf{N}_2^{\flat}
	+\mathcal{I}\mathbf{C}^{\flat}\cdot\mathbf{N}_2\,,&&
	\frac{\partial I_9}{\partial \mathbf{N}_1}=2\,\mathcal{I}\,\mathbf{N}_2^{\flat}\,, \\
	& \frac{\partial I_6}{\partial \mathbf{N}_2}=2\mathbf{C}^{\flat}\cdot\mathbf{N}_2\,,&&
	\frac{\partial I_7}{\partial \mathbf{N}_2}=2\mathbf{C}^{2\flat}\cdot\mathbf{N}_2\,, &&
	\frac{\partial I_8}{\partial \mathbf{N}_2}=I_8\,\mathcal{I}^{-1}\,\mathbf{N}_1^{\flat}
	+\mathcal{I}\mathbf{C}^{\flat}\cdot\mathbf{N}_1 \,,&&
	\frac{\partial I_9}{\partial \mathbf{N}_2}=2\,\mathcal{I}\,\mathbf{N}_1^{\flat}\,,
\end{aligned}
\end{equation}
the remodeling equations are written as
\begin{equation} 
\begin{aligned}
	\mathbf{G}^{\sharp}\frac{\partial \phi}{\partial \dot{\mathbf{N}}_1} &= 2q_{n1}\mathbf{N}_1
	-\frac{\partial \Wr}{\partial \mathbf{N}_1}
	-2W_4\,\mathbf{C}\cdot\mathbf{N}_1
	-2W_5\,\mathbf{C}^2\cdot\mathbf{N}_1-W_8\left(I_8\,\mathcal{I}^{-1}\,\mathbf{N}_2
	+\mathcal{I}\mathbf{C}\cdot\mathbf{N}_2\right)
	-2\,\mathcal{I}\,W_9\,\mathbf{N}_2 \,, \\
	\mathbf{G}^{\sharp}\frac{\partial \phi}{\partial \dot{\mathbf{N}}_2} &= 2q_{n2} \mathbf{N}_2 
	-\frac{\partial \Wr}{\partial \mathbf{N}_2}
	-2W_6\,\mathbf{C}\cdot\mathbf{N}_2
	-2W_7\, \mathbf{C}^2\cdot\mathbf{N}_2-W_8\,\left(I_8\,\mathcal{I}^{-1}\,\mathbf{N}_1
	+\mathcal{I}\mathbf{C}\cdot\mathbf{N}_1 \right)
	-2\,\mathcal{I}\,W_9\, \mathbf{N}_1 \,.
\end{aligned}
\end{equation}
Eliminating $q_{n1}$ and $q_{n2}$ from the above equations, one obtains
\begin{equation} 
\begin{aligned}
	\mathbf{G}^{\sharp}\frac{\partial \phi}{\partial \dot{\mathbf{N}}_1}
	-\left(\frac{\partial \phi}{\partial \dot{\mathbf{N}}_1}\cdot \mathbf{N}_1 \right) \mathbf{N}_1
	&= \left\langle \frac{\partial \Wr}{\partial \mathbf{N}_1},\mathbf{N}_1 \right\rangle \mathbf{N}_1
	-\frac{\partial \Wr}{\partial \mathbf{N}_1}
	+2W_4(I_4\mathbf{N}_1-\mathbf{C}\cdot\mathbf{N}_1)
	+2W_5(I_5\mathbf{N}_1-\mathbf{C}^2\cdot\mathbf{N}_1)\\
	& \quad  +W_8\left(2I_8\,\mathbf{N}_1-I_8\,\mathcal{I}^{-1}\,\mathbf{N}_2
	-\mathcal{I}\mathbf{C}\cdot\mathbf{N}_2\right)
	+2\mathcal{I}\,W_9(\mathcal{I}\,\mathbf{N}_1-\mathbf{N}_2) \,, \\
	\mathbf{G}^{\sharp}\frac{\partial \phi}{\partial \dot{\mathbf{N}}_2} 
	-\left(\frac{\partial \phi}{\partial \dot{\mathbf{N}}_2}\cdot \mathbf{N}_2 \right) \mathbf{N}_2
	&= \left\langle \frac{\partial \Wr}{\partial \mathbf{N}_2},\mathbf{N}_2 \right\rangle \mathbf{N}_2
	-\frac{\partial \Wr}{\partial \mathbf{N}_2}
	+2W_6(I_6\mathbf{N}_2-\mathbf{C}\cdot\mathbf{N}_2)
	+2W_7(I_7\mathbf{N}_2-\mathbf{C}^2\cdot\mathbf{N}_2)\\
	& \quad +W_8\left(2I_8\,\mathbf{N}_2-I_8\,\mathcal{I}^{-1}\,\mathbf{N}_1
	-\mathcal{I}\mathbf{C}\cdot\mathbf{N}_1 \right)
	+2\mathcal{I}\,W_9(\mathcal{I}\,\mathbf{N}_2-\mathbf{N}_1) \,.
\end{aligned}
\end{equation}

Let us assume the remodeling energy $\Wr(X,\mathbf{C}^{\flat},\mathbf{N}_1,\mathbf{N}_2,\mathbf{G})=\frac{1}{2}\kappa_{M1}\,(\mathbf{M}\cdot\mathbf{N}_1)^2+\frac{1}{2}\kappa_{M2}\,(\mathbf{M}\cdot\mathbf{N}_2)^2$, where $\kappa_{M1}$ and $\kappa_{M2}$ are scalars, and $\mathbf{M}$ is some unit vector that explicitly depends on $\mathbf{C}^{\flat}$. Let us also assume the quadratic dissipation potential \eqref{dissipation-N2}. The remodeling equations are simplified to read
\begin{equation} \label{Kinetic-N2-2}
\begin{dcases}
	K_1\big(\dot{\mathbf{N}}_1-\mathbf{N}_1\!\cdot\dot{\mathbf{N}}_1\,\mathbf{N}_1 \big)
	+K_3\big(\dot{\mathbf{N}}_2-\mathbf{N}_1\!\cdot\dot{\mathbf{N}}_2\,\mathbf{N}_1 \big)
	= \kappa_{M1}(\mathbf{M}\cdot\mathbf{N}_1)
	 \left[(\mathbf{M}\cdot\mathbf{N}_1)\mathbf{N}_1-\mathbf{M} \right] \\
	  \qquad\qquad\qquad\qquad\qquad\qquad\qquad\qquad 
	  +2W_4(I_4\mathbf{N}_1-\mathbf{C}\cdot\mathbf{N}_1)
	+2W_5(I_5\mathbf{N}_1-\mathbf{C}^2\cdot\mathbf{N}_1)\\
	\qquad\qquad\qquad\qquad\qquad\qquad\qquad\qquad 
	+W_8\left(2I_8\,\mathbf{N}_1-I_8\,\mathcal{I}^{-1}\,\mathbf{N}_2
	-\mathcal{I}\mathbf{C}\cdot\mathbf{N}_2\right)
	+2\mathcal{I}\,W_9(\mathcal{I}\,\mathbf{N}_1-\mathbf{N}_2) \,, \\
	K_2\big(\dot{\mathbf{N}}_2-\mathbf{N}_2\!\cdot\dot{\mathbf{N}}_2\,\mathbf{N}_2 \big)
	+K_3\big(\dot{\mathbf{N}}_1-\mathbf{N}_2\!\cdot\dot{\mathbf{N}}_1\,\mathbf{N}_2 \big)
	= \kappa_{M2}(\mathbf{M}\cdot\mathbf{N}_2)
	 \left[(\mathbf{M}\cdot\mathbf{N}_2)\mathbf{N}_2-\mathbf{M} \right] \\
	 \qquad\qquad\qquad\qquad\qquad\qquad\qquad\qquad 
	 +2W_6(I_6\mathbf{N}_2-\mathbf{C}\cdot\mathbf{N}_2)
	+2W_7(I_7\mathbf{N}_2-\mathbf{C}^2\cdot\mathbf{N}_2)\\
	  \qquad\qquad\qquad\qquad\qquad\qquad\qquad\qquad
	  +W_8\left(2I_8\,\mathbf{N}_2-I_8\,\mathcal{I}^{-1}\,\mathbf{N}_1
	-\mathcal{I}\mathbf{C}\cdot\mathbf{N}_1 \right)
	+2\mathcal{I}\,W_9(\mathcal{I}\,\mathbf{N}_2-\mathbf{N}_1) \,.
\end{dcases}
\end{equation}
Similarly, for the remodeling energy $\Wr(X,\mathbf{C}^{\flat},\mathbf{N}_1,\mathbf{N}_2,\mathbf{G})=\kappa_{M1}\,|\mathbf{M}\cdot\mathbf{N}_1|+\kappa_{M2}\,|\mathbf{M}\cdot\mathbf{N}_2|$, the remodeling equations read
\begin{equation} \label{Kinetic-N2-1}
\begin{dcases}
	K_1\big(\dot{\mathbf{N}}_1-\mathbf{N}_1\!\cdot\dot{\mathbf{N}}_1\,\mathbf{N}_1 \big)
	+K_3\big(\dot{\mathbf{N}}_2-\mathbf{N}_1\!\cdot\dot{\mathbf{N}}_2\,\mathbf{N}_1 \big)
	= \kappa_{M1}\,\operatorname{sgn}(\mathbf{M}\cdot\mathbf{N}_1)
	 \left[(\mathbf{M}\cdot\mathbf{N}_1)\mathbf{N}_1-\mathbf{M} \right] \\
	  \qquad\qquad\qquad\qquad\qquad\qquad\qquad\qquad 
	  +2W_4(I_4\mathbf{N}_1-\mathbf{C}\cdot\mathbf{N}_1)
	+2W_5(I_5\mathbf{N}_1-\mathbf{C}^2\cdot\mathbf{N}_1)\\
	\qquad\qquad\qquad\qquad\qquad\qquad\qquad\qquad 
	+W_8\left(2I_8\,\mathbf{N}_1-I_8\,\mathcal{I}^{-1}\,\mathbf{N}_2
	-\mathcal{I}\mathbf{C}\cdot\mathbf{N}_2\right)
	+2\mathcal{I}\,W_9(\mathcal{I}\,\mathbf{N}_1-\mathbf{N}_2) \,, \\
	K_2\big(\dot{\mathbf{N}}_2-\mathbf{N}_2\!\cdot\dot{\mathbf{N}}_2\,\mathbf{N}_2 \big)
	+K_3\big(\dot{\mathbf{N}}_1-\mathbf{N}_2\!\cdot\dot{\mathbf{N}}_1\,\mathbf{N}_2 \big)
	= \kappa_{M2}\,\operatorname{sgn}(\mathbf{M}\cdot\mathbf{N}_2)
	 \left[(\mathbf{M}\cdot\mathbf{N}_2)\mathbf{N}_2-\mathbf{M} \right] \\
	 \qquad\qquad\qquad\qquad\qquad\qquad\qquad\qquad 
	 +2W_6(I_6\mathbf{N}_2-\mathbf{C}\cdot\mathbf{N}_2)
	+2W_7(I_7\mathbf{N}_2-\mathbf{C}^2\cdot\mathbf{N}_2)\\
	  \qquad\qquad\qquad\qquad\qquad\qquad\qquad\qquad
	  +W_8\left(2I_8\,\mathbf{N}_2-I_8\,\mathcal{I}^{-1}\,\mathbf{N}_1
	-\mathcal{I}\mathbf{C}\cdot\mathbf{N}_1 \right)
	+2\mathcal{I}\,W_9(\mathcal{I}\,\mathbf{N}_2-\mathbf{N}_1) \,.
\end{dcases}
\end{equation}

\subsection{The first law of thermodynamics}

The first law of thermodynamics or the balance of energy reads
\begin{equation}\label{eq:Thermo_First}
	\frac{d}{dt}\int_{\mathcal{U}} \left(W
	+\frac{1}{2}\rho_o \llangle \mathbf{V},\mathbf{V}\rrangle_{\boldsymbol g} \right)dV
	=\int_{\mathcal{U}}\rho_o \left(\llangle\mathbf{B},\mathbf{V}\rrangle_{\mathbf{g}}+R\right)dV
	+\int_{\partial \mathcal{U}}\left(\llangle \mathbf{T},\mathbf{V}\rrangle_{\mathbf{g}}+H\right)dA\,,
\end{equation}
where $\mathcal{U}\subset\mathcal{B}$ is an arbitrary subbody, $W$ is the energy function or the internal energy density, $R=R(X,t)$ is the heat supply per unit mass, $H=-\llangle \mathbf{Q},\hat{\mathbf{N}}\rrangle_{\mathbf{G}}$ is the heat flux, $\mathbf Q=\mathbf Q(X,T,dT,\mathbf C,\mathbf G)$ is the external heat flux per unit area, $\hat{\mathbf{N}}$ is the $\mathbf{G}$-unit normal to the boundary $\partial\mathcal B$, and $T=T(X,t)$ is the absolute temperature field.

The local form of the balance of energy reads
\begin{equation} 
	\dot{W} = \rho_o R+\mathbf{P}\!:\!\nabla^{\mathbf{G}}\mathbf{V}- \operatorname{Div} \mathbf{Q}
	+\llangle \operatorname{Div}\mathbf{P}+ \rho_o(\boldsymbol{\mathsf{B}} 
	- \mathbf{A}),\mathbf{V} \rrangle_{\mathbf{g}}\,.
\end{equation}
At this point we do not know yet that the first Piola-Kirchhoff stress $\mathbf{P}=\frac{\partial W}{\partial \mathbf{F}}$, which is a consequence of the second law of thermodynamics. However, to simplify the calculations we are going to assume this. It is straightforward to see that $\mathbf{P}\!:\!\nabla^{\mathbf{G}}\mathbf{V}=\frac{1}{2}\mathbf{S}\!:\! \dot{\mathbf{C}}^{\flat}$, where $\mathbf{S}=\mathbf{F}^{-1}\mathbf{P}$ is the second Piola-Kirchhoff stress. Thus, the local form of the energy balance reads
\begin{equation} 
	\dot{W} = \rho_o R+\frac{1}{2}\mathbf{S}\!:\! \dot{\mathbf{C}}^{\flat}- \operatorname{Div} \mathbf{Q}
	\,.
\end{equation}

\subsection{The second law of thermodynamics} \label{Sec:2ndLaw}

The second law of thermodynamics can be stated in the form of the material Clausius-Duhem inequality \citep{MarsdenHughes1983} that is written as
\begin{equation} 
	\frac{d}{dt}\int_{\mathcal{U}} \mathcal{N}dV\geq 
	\int_{\mathcal{U}}\rho_o \frac{R}{T}dV+\int_{\partial\mathcal{U}}\frac{H}{T}dA\,,
\end{equation}
where $\mathcal{N}= \hat{\mathcal N}(X,T, \mathbf{C}^\flat,\mathbf{G})$ is the material entropy density (per unit undeformed volume).
The local form of Clausius-Duhem inequality reads
\begin{equation} \label{Local-FirstLaw}
	\dot\eta = T \dot{\mathcal{N}} - \rho_o R+ T \operatorname{Div}\left(\frac{\mathbf Q}{T}\right) 
	 \geq 0\,,
\end{equation}
where $\dot\eta$ is the rate of energy dissipation.

The free energy density is defined as $\Psi = W - T \mathcal N$, and hence, $\Psi = \hat{\Psi}(X,T,\mathbf{C}^{\flat}, \mathbf{G})$. Note that $T \dot{\mathcal{N}}=\dot{W}-\dot{\Psi}-\dot{T}\mathcal{N}$, and thus
\begin{equation} 
	\dot{\eta} = \dot{W} -\dot{\Psi} - \dot{T} \mathcal{N}+ \operatorname{Div}\mathbf{Q}
	- \frac{1}{T} \langle dT, \mathbf{Q} \rangle -\rho_o R \geq 0\,.
\end{equation}
Using \eqref{Local-FirstLaw} in the above inequality one obtains
\begin{equation} \label{eta-inequality}
	\dot{\eta} = \frac{1}{2}\mathbf{S}\!:\! \dot{\mathbf{C}}^{\flat}	 -\dot{\Psi} - \dot{T} \mathcal{N}
	- \frac{1}{T} \langle dT, \mathbf{Q} \rangle  \geq 0\,.
\end{equation}
But
\begin{equation} 
	\dot{\Psi}= \frac{\partial  \hat{\Psi}}{\partial T}\,\dot{T}
	+\frac{\partial  \hat{\Psi}}{\partial \mathbf{C}^{\flat}}\!:\!\dot{\mathbf{C}}^{\flat}
	+\frac{\partial  \hat{\Psi}}{\partial \mathbf{G}}\!:\!\dot{\mathbf{G}}
	= \frac{\partial  \hat{\Psi}}{\partial T}\,\dot{T}
	+\frac{\partial  \hat{\Psi}}{\partial \mathbf{C}^{\flat}}\!:\!\dot{\mathbf{C}}^{\flat}
	+2\Fr^{-\star}\mathbf{G}\frac{\partial \Psi}{\partial \mathbf{G}}\!:\!\dot{\Fr}
	\,.
\end{equation}
As $0=\dot{\overline{\det \Fr}}=\Fr^{-1}\!:\!\dot{\Fr}$, a term $q\Fr^{-1}$ can be added to the last term without changing the equality, where $q$ is a Lagrange multiplier.
Thus, \eqref{eta-inequality} is simplified to read
\begin{equation}
	\dot{\eta} = -\left(\mathcal{N} + \frac{\partial  \hat{\Psi}}{\partial T}\right) \dot{T} 
	+\frac{1}{2} \left(\mathbf{S}-2\frac{\partial  \hat{\Psi}}{\partial \mathbf{C}^{\flat}} \right)\!:\! \dot{\mathbf{C}}^{\flat}
	- \frac{1}{T} \langle dT, \mathbf{Q} \rangle
	+\left(-2\Fr^{-\star}\mathbf{G}\frac{\partial \Psi}{\partial \mathbf{G}}+q\Fr^{-1} \right)\!:\!\dot{\Fr}  \geq 0\,.
\end{equation}
The above inequality must hold for arbitrary $\dot{T} $, and $\dot{\mathbf{C}}^{\flat}$, and hence
\begin{equation} \label{second-law-consequences}
	 \mathcal{N}= -  \frac{\partial \hat{\Psi}}{\partial T}\,,\quad
	 \mathbf{S}=2\frac{\partial  \hat{\Psi}}{\partial \mathbf{C}^{\flat}}\,,\quad
	\dot{\eta} =  - \frac{1}{T} \langle dT, \mathbf{Q} \rangle
	+\left(-2\Fr^{-\star}\mathbf{G}\frac{\partial \Psi}{\partial \mathbf{G}}+q\Fr^{-1} \right)\!:\!\dot{\Fr}  \geq 0\,.
\end{equation}
Note that 
\begin{equation}
	\frac{\partial W}{\partial \mathbf{G}}
	=\frac{\partial W}{\partial \mathbf{G}}\Bigg|_{\mathcal{N},\mathbf{C}^{\flat}}
	=\left[ \frac{\partial \Psi}{\partial \mathbf{G}}+\frac{\partial \Psi}{\partial T}
	\frac{\partial T}{\partial \mathbf{G}}
	 \right]+\frac{\partial T}{\partial \mathbf{G}}\mathcal{N}
	 =\frac{\partial \Psi}{\partial \mathbf{G}}
	\,,
\end{equation}
where use was made of \eqref{second-law-consequences}$_1$. Using the above relation and the remodeling equation \eqref{Remodeling-Equation-Metric} in \eqref{second-law-consequences}$_3$ we obtain\footnote{It is straightforward to show that this inequality has the same form for anisotropic solids.}
\begin{equation} 
	\dot{\eta} =  - \frac{1}{T} \langle dT, \mathbf{Q} \rangle
	+\frac{\partial \phi}{\partial \dot{\Fr}}\!:\!\dot{\Fr}  \geq 0\,.
\end{equation}
If an isothermal process is assumed, i.e., $dT=0$, the entropy production is simplified to read
\begin{equation} \label{Entropy-Production-Remodeling}
	\dot{\eta} =  \frac{\partial \phi}{\partial \dot{\Fr}}\!:\!\dot{\Fr}  \geq 0\,.
\end{equation}

\section{Examples of Material Remodeling}. \label{Examples}

In this section we discuss three concrete examples of material remodeling. Let us consider an incompressible isotropic solid that is reinforced by a family of fibers. At a material point $X\in\mathcal{B}$ in the initial body the unit tangent to the fiber is denoted by $\mathring{\mathbf{N}}(X)$. This material is effectively transversely isotropic and at $X\in\mathcal{B}$ the plane of isotropy is normal to $\mathring{\mathbf{N}}(X)$. The body undergoes a remodeling process during deformation such that the material preferred direction evolves. Let us denote the time-dependent unit tangent to the fiber by $\mathbf{N}(X,t)$, which models reorientation of fibers. A remodeling tensor relates $\mathbf{N}(X,t)$ to $\mathring{\mathbf{N}}(X)=\mathbf{N}(X,0)$, i.e., $\mathbf{N}(X,t)=\Fr^{-1}(X,t)\mathring{\mathbf{N}}(X)$, where $\Fr(X,t)\in SO(3)(T_X\mathcal{B},\mathring{\mathbf{G}})$---the set of rotations. 
This means that $\mathbf{G}=\Fr^*\mathring{\mathbf{G}}=\Fr^{\star}\mathring{\mathbf{G}}\Fr=\mathring{\mathbf{G}}$, i.e., the material metric is flat.

The three examples that are studied in this section are subsets of Family $3$ universal deformations. A universal deformation is one that can be maintained in the absence of body forces for any member of a given class of materials \citep{Ericksen1954,Ericksen1955}. \citet{Ericksen1955} showed that for homogeneous compressible isotropic solids the only universal deformations are homogeneous deformations (and all homogenous deformations are universal). Recently, \citet{Yavari2021a} showed that inhomogeneous compressible isotropic solids do not admit universal deformations. 
For incompressible solids the problem of characterizing universal deformations is much more difficult and interesting \citep{Saccomandi2001,Tadmor2012,Goriely2017}. For homogenous incompressible isotropic solids, \citet{Ericksen1954} found four families of universal deformations (other than volume-preserving homogenous deformations). Later on a fifth family was discovered independently by \citet{SinghPipkin1965} and \citet{KlingbeilShield1966}. This last family is peculiar in the sense that it is inhomogeneous while its principal invariants are constant. Determining all universal deformations with constant principal invariants is still an open problem. Recently, Ericksen's problem was revisited for inhomogeneous and anisotropic solids \citep{Yavari2021,YavariGoriely2021,YavariGoriely2022}.\footnote{Universal displacements are the analogue of universal deformations in linear elasticity \citep{Truesdell1966,Gurtin1972,Yavari2020,YavariGoriely2022,Yavari2022Anelastic-Universality,Yavari2023Fibers}.} 
The three problems that we investigate in this section admit universal deformations for certain universal material preferred directions as was shown in \citep{YavariGoriely2021,YavariGoriely2022}.

\subsection{Example 1: Finite extension of a transversely isotropic circular cylindrical bar}

Consider a solid cylinder with initial radius $R_0$ and length $L$. Assume that for fixed $R\in(0,R_0]$ fibers are along a family of helices. 
Recall that in cylindrical coordinates $(R,\Theta,Z)$ and $(r,\theta,z)$ the initial material metric and the metric of the ambient space have the following representations
\begin{equation} 
	\mathring{\mathbf{G}}=\begin{bmatrix}
	1 & 0 & 0 \\
	0 & R^2 & 0 \\
	0 & 0  & 1
	\end{bmatrix}
	\,,\qquad
	\mathbf{g}=\begin{bmatrix}
	1 & 0 & 0 \\
	0 & r^2 & 0 \\
	0 & 0  & 1
	\end{bmatrix}\,.
\end{equation}
For this body $\mathring{\mathbf{N}}=\mathring{\mathbf{N}}(R,\Theta)$. 
Tangent to a helix in cylindrical coordinates has a vanishing radial coordinate. Also, $\mathring{N}^A\mathring{N}^B\mathring{G}_{AB}=R^2(\mathring{N}^{\Theta})^2+(\mathring{N}^Z)^2=1$. For example, fibers along $Z$ (parallel to the axis of the bar) correspond to $\mathring{N}^{\Theta}=0$ and $\mathring{N}^Z=1$, while for a family of circular fibers $\mathring{N}^{\Theta}=\frac{1}{R}$ and $\mathring{N}^Z=0$.
If $\gamma(R)$ is the angle that $\mathring{\mathbf{N}}(R,\Theta)$ makes with $\mathbf{E}_{\Theta}(\Theta)=\frac{\partial}{\partial \Theta}$, then 
\begin{equation} \label{Helical-Fibers}
	\mathring{\mathbf{N}}(R,\Theta)
	=\frac{\cos\gamma(R)}{R}\,\mathbf{E}_{\Theta}(\Theta)+\sin\gamma(R)\,\mathbf{E}_Z\,,
\end{equation}
where $\mathbf{E}_Z=\frac{\partial}{\partial Z}$.
Assume that in a remodeling process this family of helices is transformed to another family of helices. At a given point with coordinates $(R,\Theta,Z)$ this corresponds to rotating $\mathring{\mathbf{N}}$ along the $\mathbf{E}_R=\frac{\partial}{\partial R}$ axis. Thus, we have the following representation for $\Fr$:\footnote{Note that $\Fr$ is written such that its physical components are dimensionless, i.e.,
\begin{equation} 
	\widehat{\Fr}(R,t)=\begin{bmatrix}
	1 & 0 & 0 \\
	0 & \cos\alpha(R,t) & -\sin\alpha(R,t) \\
	0 & \sin\alpha(R,t)  & \cos\alpha(R,t)
	\end{bmatrix}
	\,.
\end{equation}
Recall that the physical and curvilinear components are related as $\widehat{\Fr}^A{}_B=\sqrt{G_{AA}}\sqrt{G^{BB}}\,\Fr^A{}_B$ (no summation) \citep{Truesdell1953}.
}
\begin{equation} \label{Cylinder-Remodeling-Tensor}
	\Fr(R,t)=\begin{bmatrix}
	1 & 0 & 0 \\
	0 & \cos\alpha(R,t) & -\frac{1}{R}\sin\alpha(R,t) \\
	0 & R\,\sin\alpha(R,t)  & \cos\alpha(R,t)
	\end{bmatrix}
	\,,
\end{equation}
where $\alpha(R,t)$ is the angle of rotation.
Thus
\begin{equation} \label{Fiber-Initial}
	\mathbf{N}(R,t)=\begin{bmatrix}
	0 \\
	\frac{\cos(\gamma(R)-\alpha(R,t))}{R} \\
	\sin(\gamma(R)-\alpha(R,t)) 
	\end{bmatrix}
	\,.
\end{equation}
We will write the remodeling equation directly in terms of $\dot{\mathbf{N}}(R,t)$, and not $\dot{\Fr}(R,t)$.
The initial condition is $\alpha(R,0)=0$.

Let us consider radial deformations and assume the following kinematics ansatz 
\begin{equation} \label{Deformation}
   r=r(R,t)\,,\quad \theta=\Theta\,,\quad z=\lambda(t) Z\,,
\end{equation}
where $\lambda(t)$ is the axial stretch.\footnote{It should be noted that \eqref{Deformation} is a subset of Family $3$ universal deformations \citep{Ericksen1954}, and the fiber distribution \eqref{Fiber-Initial} are universal material preferred directions \citep{YavariGoriely2021}. This means that the deformations \eqref{Deformation} can be maintained in the absence of body forces for any incompressible isotropic solid cylinder reinforced by fibers with distribution given in \eqref{Fiber-Initial}.} 
In a force-control loading $\lambda(t)$ is an unknown function to be determined, while in a displacement-control loading $\lambda(t)$ is given. Let us assume that loading is slow enough so that the inertial effects can be neglected.
The deformation gradient reads
\begin{equation} \label{F-Example1}
   \mathbf{F}=\mathbf{F}(R,t)=\begin{bmatrix}
  r_{,R}(R,t) & 0  & 0  \\
  0 & 1  & 0  \\
  0 & 0  & \lambda(t)
\end{bmatrix}\,.
\end{equation}
Incompressibility implies that $r(R,t)=\frac{R}{\sqrt{\lambda(t)}}$. 
The principal invariants read
\begin{equation}
\begin{aligned}
	& I_1=\lambda^2(t)+2\lambda^{-1}(t)\,,\\ 
	& I_2=2\lambda(t)+\lambda^{-2}(t)\,,\\
	& I_4=\lambda^2(t)\sin^2(\alpha(R,t)-\gamma(R))+\lambda^{-1}(t)
	\cos^2(\alpha(R,t)-\gamma(R))\,,\\
	& I_5=\lambda^4(t)\sin^2(\alpha(R,t)-\gamma(R))+\lambda^{-2}(t)
	\cos^2(\alpha(R,t)-\gamma(R))
	\,.
\end{aligned}
\end{equation}

\paragraph{Stress and equilibrium equations.}
The non-zero components of the Cauchy stress are:
\begin{equation}
\begin{aligned}
	\sigma^{rr}(R,t) &=-p(R,t)+2\lambda^{-1}(t)\,W_1-2\lambda(t)\,W_2\,,\\ 
	\sigma^{\theta\theta}(R,t) & = -\frac{p(R,t)\lambda(t)}{R^2}+\frac{2W_1}{R^2}
	-\frac{2\lambda^2(t)\,W_2}{R^2}
	+\frac{2\left[\lambda(t)\,W_4+2W_5\right]}{R^2\lambda(t)}\cos^2(\alpha(R,t)-\gamma(R)) \,,\\
	\sigma^{zz}(R,t) & =-p(R,t)+2\lambda^2(t)\,W_1-2\lambda^{-2}(t)\,W_2
	+2\lambda^2(t)\left[W_4+2\lambda^2(t)\,W_5\right]\sin^2(\alpha(R,t)-\gamma(R)) \,,\\
	\sigma^{\theta z}(R,t) & 
	= -\frac{\lambda(t)\,W_4+\left[1+\lambda^3(t)\right]W_5}{R} \sin\left[2(\alpha(R,t)-\gamma(R))\right] 
	\,.
\end{aligned}
\end{equation}
The only nontrivial equilibrium equation is $\sigma^{rr}{}_{,r}+\frac{1}{r}\sigma^{rr}-r\sigma^{\theta\theta}=0$. In terms of the referential coordinates this reads 
\begin{equation}
	\frac{\partial}{\partial R}\sigma^{rr}(R,t) 
	=\frac{2\left[\lambda(t)\,W_4+2W_5\right]}{R\lambda^2(t)}\cos^2(\alpha(R,t)-\gamma(R))\,.
\end{equation}
We assume the boundary condition $\sigma^{rr}(R_0,t)=0$. Thus
\begin{equation} \label{srr}
	\sigma^{rr}(R,t) = -\frac{2}{\lambda^2(t)}\int_{R}^{R_0}\frac{\lambda(t)\,W_4+2W_5}{\xi}
	\cos^2(\alpha(\xi,t)-\gamma(\xi))\,d\xi\,.
\end{equation}
This, in particular, implies that
\begin{equation}
	-p(R,t) = -\frac{2}{\lambda^2(t)}\int_{R}^{R_0}\frac{\lambda(t)\,W_4+2W_5}{\xi}
	\cos^2(\alpha(\xi,t)-\gamma(\xi))\,d\xi
	-2\lambda^{-1}(t)\,W_1+2\lambda(t)\,W_2
	\,.
\end{equation}
Now the physical components of the other three stresses are simplified to read\footnote{Note that $\hat{\sigma}^{rr}=\sigma^{rr}$, $\hat{\sigma}^{\theta\theta}=r^2\sigma^{\theta\theta}$, $\hat{\sigma}^{\theta }=r\sigma^{\theta\theta}$, and $\hat{\sigma}^{zz}=\sigma^{zz}$.} 
\begin{equation}\label{scomp}
\begin{aligned}
	\hat{\sigma}^{\theta\theta}(R,t) & =
	\frac{2\left[\lambda(t)\,W_4+2W_5\right]}{\lambda^2(t)}\cos^2(\alpha(R,t)-\gamma(R))
	-\frac{2}{\lambda^2(t)}\int_{R}^{R_0}\frac{\lambda(t)\,W_4+2W_5}{\xi}
	\cos^2(\alpha(\xi,t)-\gamma(\xi))\,d\xi \,,\\
	\hat{\sigma}^{zz}(R,t) & =2\left[\lambda^2(t)-\lambda^{-1}(t)\right]W_1
	+2\left[\lambda(t)-\lambda^{-2}(t)\right]W_2
	+2\lambda^2(t)\left[W_4+2\lambda^2(t)\,W_5\right]\sin^2(\alpha(R,t)-\gamma(R)) \\
	& \quad -\frac{2}{\lambda^2(t)}\int_{R}^{R_0}\frac{\lambda(t)\,W_4+2W_5}{\xi}
	\cos^2(\alpha(\xi,t)-\gamma(\xi))\,d\xi \,,\\
	\hat{\sigma}^{\theta z}(R,t) & 
	=-\frac{\lambda(t)\,W_4+\left[1+\lambda^3(t)\right]W_5}{\lambda^{\frac{1}{2}}(t)}
	\sin\left[2(\alpha(R,t)-\gamma(R))\right] 
	\,.
\end{aligned}
\end{equation}

\paragraph{The axial force.}
For displacement-control loading $\lambda(t)$ is a given function and the only unknown of the problem is $\alpha(R,t)$, which is governed by the remodeling equation. For force-control loadings, the unknowns of the problem are $\lambda(t)$ and $\alpha(R,t)$. In this case at the two ends of the bar ($Z=0,L$), the axial force required to maintain the deformation is  
\begin{equation} \label{Axial-Force-Example1}
	F(t)=2\pi \int_{0}^{R_0}P^{zZ}(R,t)R\,dR\,,
\end{equation}
where $P^{zZ}(R,t)=\lambda^{-1}(t)\,\sigma^{zz}(R,t)$ is the $zZ$-component of the first Piola-Kirchhoff stress.
This is simplified to read
\begin{equation} \label{Example1-AxialForce}
\begin{aligned}
	& -2\lambda^{-3}(t) \int_{0}^{R_0} R \int_{R}^{R_0}\frac{\lambda(t)\,W_4+2W_5}{\xi}
	\cos^2(\alpha(\xi,t)-\gamma(\xi))\,d\xi\,dR \\
	&\qquad +2\left[\lambda(t)-\lambda^{-2}(t)\right] \int_{0}^{R_0} W_1\,R\,dR
	+2\left[1-\lambda^{-3}(t)\right] \int_{0}^{R_0}  W_2\,R\,dR \\
	&\qquad +2\lambda(t) \int_{0}^{R_0} R\left[W_4+2\lambda^2(t)\,W_5\right]\sin^2(\alpha(R,t)-\gamma(R))\,dR 
	= \frac{F(t)}{2\pi}	\,.
\end{aligned}
\end{equation}
It is assumed that $F(0)=0$, and hence $\lambda(0)=1$.
We will consider both displacement-control and force-control cases.

\paragraph{The remodeling equation.}
The remodeling equation \eqref{Kinetic-N} is used. For this problem it is written as
\begin{equation} 
\begin{aligned} 
	& \sin(\alpha (R,t)-\gamma(R)) \Big\{\left(\lambda^3(t)-1\right) \left[\lambda(t) W_4
	+(1+\lambda^3(t)) W_5\right]
	 \sin(2(\gamma(R)-\alpha(R,t)))-K\lambda^2(t) \,\dot{\alpha}(R,t)\Big\}=0\,, \\
	 & \cos(\alpha (R,t)-\gamma(R)) \Big\{\left(\lambda^3(t)-1\right) \left[\lambda(t) W_4
	 +(1+\lambda^3(t)) W_5\right]
	 \sin(2(\gamma(R)-\alpha(R,t)))-K\lambda^2(t) \,\dot{\alpha}(R,t)\Big\}=0\,.
\end{aligned}
\end{equation}
Knowing that the sine and cosine cannot vanish simultaneously, the remodeling equation reads
\begin{equation} 
\begin{aligned} 
	K\,\dot{\alpha}(R,t)=\left(\lambda^{-2}(t)-\lambda(t)\right) \left[\lambda(t) W_4
	+(1+\lambda^3(t)) W_5\right] \sin(2(\alpha(R,t)-\gamma(R)))\,.
\end{aligned}
\end{equation}
Choosing $\mathbf{M}=\mathbf{N}^{\mathbf{C}}_{\text{max}}= \mathbf{E}_Z$,
the remodeling equation \eqref{Kinetic-N-M2} is simplified to read:
\begin{equation} \label{Ex1-Kin2}
\begin{aligned}
	K\, \dot{\alpha}
	= \left\{-\frac{1}{2}\kappa_M
	+(\lambda^{-2}-\lambda)\left[\lambda W_4+\left(\lambda^3+1\right) W_5\right] \right\}
	\sin2(\alpha-\gamma) \,.
\end{aligned}
\end{equation}
Similarly, the remodeling equation \eqref{Kinetic-N-M1} is simplified to read:
\begin{equation}  \label{Ex1-Kin1}
\begin{aligned}
	K\, \dot{\alpha}
	= \kappa_M\,\operatorname{sgn}\left[\sin(\gamma -\alpha)\right]\, \cos(\alpha-\gamma) 
	+(\lambda^{-2}-\lambda)\left[\lambda W_4+\left(\lambda^3+1\right) W_5\right] 
	\sin2(\alpha-\gamma) \,.
\end{aligned}
\end{equation}


For our numerical examples we consider an incompressible Mooney-Rivlin reinforced model ($I_4$ and $I_5$ reinforcements) for which \citep{TriantafyllidisAbeyaratne1983,MerodioOgden2003, MerodioOgden2005}
\begin{equation} \label{MR-Material}
	W(I_1,I_2,I_4,I_5)=C_1(I_1-3)+C_2(I_2-3)+\frac{\mu_1}{2}(I_4-1)^2+\frac{\mu_2}{2}(I_5-1)^2\,,
\end{equation}	
where $C_1$, $C_2$, $\mu_1$, and $\mu_2$ are positive constants.
Thus, $W_1=C_1$, $W_2=C_2$, $W_4=\mu_1(I_4-1)$, and $W_5=\mu_1(I_5-1)$. For this material the remodeling equation \eqref{Ex1-Kin2} is simplified as
\begin{equation} \label{KineticEQ-Eexample1}
\begin{aligned}
	\dot{\alpha} &= \frac{\kappa_M}{K}\,\cos(\gamma-\alpha)
	-\frac{1}{2 \lambda^8}
	\Bigg\{\tau_1^{-1}\,\lambda^4 \left[\left(\lambda^6-1\right)^2 \cos2(\gamma-\alpha)
	-\lambda^{12}+2 \lambda^8-2 \lambda^2+1\right]\\
	&\qquad +\tau_2^{-1} \left[\left(\lambda^{12}-1\right)^2\cos2(\gamma-\alpha)-\lambda^{24}
	+2 \lambda^{16}-2 \lambda^4+1\right]  \Bigg\} \sin2(\gamma-\alpha)	\,,
\end{aligned}
\end{equation}
where $\tau_1=K/\mu_1$, and $\tau_2=K/\mu_2$ are relaxation times of this material.
Similarly, the kinetic equation \eqref{Ex1-Kin1} is simplified to read
\begin{equation} 
\begin{aligned}
	\dot{\alpha}
	&= \frac{\kappa_M}{K}  \operatorname{sgn}[\sin(\gamma-\alpha)] \cos (\gamma-\alpha) \\
	& \quad +\left(\lambda^{-1}-\lambda^{-4}\right) 
	\Big[\tau_1^{-1}\,\lambda^2 
	 \left(\lambda^3 \sin^2(\gamma-\alpha)+\cos^2(\gamma-\alpha)-\lambda \right) \\
	& \qquad\qquad\qquad\qquad 
	+\tau_2^{-1} \left(\lambda^3+1\right) \left(\lambda^6 \sin^2(\gamma-\alpha)
	+\cos^2(\gamma-\alpha)-\lambda^2\right)\Big]\sin2(\gamma-\alpha) \,.
\end{aligned}
\end{equation}


\paragraph{Displacement-control loading.}
Let us first consider a displacement-control loading. It is assumed that $\lambda(t)=1+(\lambda_0-1)\,\erf\big(\frac{t}{t_0}\big)$, where $\erf$ is the error function and $t_0$ is some characteristic time. Thus, $\lambda(0)=1$, and for $t>t_0$, $\lambda(t)\approx \lambda_0$. 
In summary, the following initial-value problem needs to be solved:\footnote{Our numerical results show that the two kinetic equations will give very similar results and we choose to work with \eqref{KineticEQ-Eexample1}.}
\begin{equation}
\begin{dcases}
	\dot{\alpha} = \frac{\kappa_M}{K}\,\cos(\gamma-\alpha)
	-\frac{1}{2 \lambda^8}
	\Bigg\{\tau_1^{-1}\,\lambda^4 \left[\left(\lambda^6-1\right)^2 \cos2(\gamma-\alpha)
	-\lambda^{12}+2 \lambda^8-2 \lambda^2+1\right]\\
	\qquad +\tau_2^{-1} \left[\left(\lambda^{12}-1\right)^2\cos2(\gamma-\alpha)-\lambda^{24}
	+2 \lambda^{16}-2 \lambda^4+1\right]  \Bigg\} \sin2(\gamma-\alpha)	\,,\\
	 \alpha(R,0)=0\,.
\end{dcases}
\end{equation}

\paragraph{Force-control loading.}
Next, it is assumed that the axial force $F(t)$ is given while both $\lambda(t)$ and $\alpha(R,t)$ are unknowns to be determined.  
For the reinforced Mooney-Rivlin material \eqref{Example1-AxialForce} is simplified as
\begin{equation}
\begin{aligned}
	& \left[\lambda(t)-\lambda^{-2}(t)\right]C_1	+\left[1-\lambda^{-3}(t)\right]C_2
	+\frac{2\lambda(t)}{R_0^2} 
	\int_{0}^{R_0}\ R \left[W_4+2\lambda^2(t)\,W_5\right]\sin^2(\left[\alpha(R,t)-\gamma(R)\right]\,dR \\
	& \quad -\frac{2}{\lambda^3(t)\,R_0^2} \int_{0}^{R_0} R \int_{R}^{R_0}\frac{\lambda(t)\,W_4+2W_5}{\xi}
	\cos^2\left[\alpha(\xi,t)-\gamma(\xi)\right]\,d\xi\,dR=\frac{F(t)}{2\pi R_0^2}\,.
\end{aligned}
\end{equation}
In summary, the following initial-value problem needs to be solved:
\begin{equation}
\begin{dcases}
	\left[\lambda(t)-\lambda^{-2}(t)\right]C_1	+\left[1-\lambda^{-3}(t)\right]C_2
	+\frac{2\lambda(t)}{R_0^2} 
	\int_{0}^{R_0}\ \left[W_4+2\lambda^2(t)\,W_5\right]\sin^2(\left[\alpha(R,t)-\gamma(R)\right]\,dR \\
	\quad -\frac{2}{\lambda^3(t)\,R_0^2} \int_{0}^{R_0} \int_{R}^{R_0}\frac{\lambda(t)\,W_4+2W_5}{\xi}
	\cos^2\left[\alpha(\xi,t)-\gamma(\xi)\right]\,d\xi\,dR=\frac{F(t)}{2\pi R_0^2}\,, \\
	\dot{\alpha} = \frac{\kappa_M}{K}\,\cos(\gamma-\alpha)
	-\frac{1}{2 \lambda^8}
	\Bigg\{\tau_1^{-1}\,\lambda^4 \left[\left(\lambda^6-1\right)^2 \cos2(\gamma-\alpha)
	-\lambda^{12}+2 \lambda^8-2 \lambda^2+1\right]\\
	\qquad +\tau_2^{-1} \left[\left(\lambda^{12}-1\right)^2\cos2(\gamma-\alpha)-\lambda^{24}
	+2 \lambda^{16}-2 \lambda^4+1\right]  \Bigg\} \sin2(\gamma-\alpha)	\,,\\
	 \lambda(0)=1\,,~ \alpha(R,0)=0\,.
\end{dcases}
\end{equation}


\begin{figure}[h!]
	\begin{center}
		\includegraphics[width=0.9\textwidth]{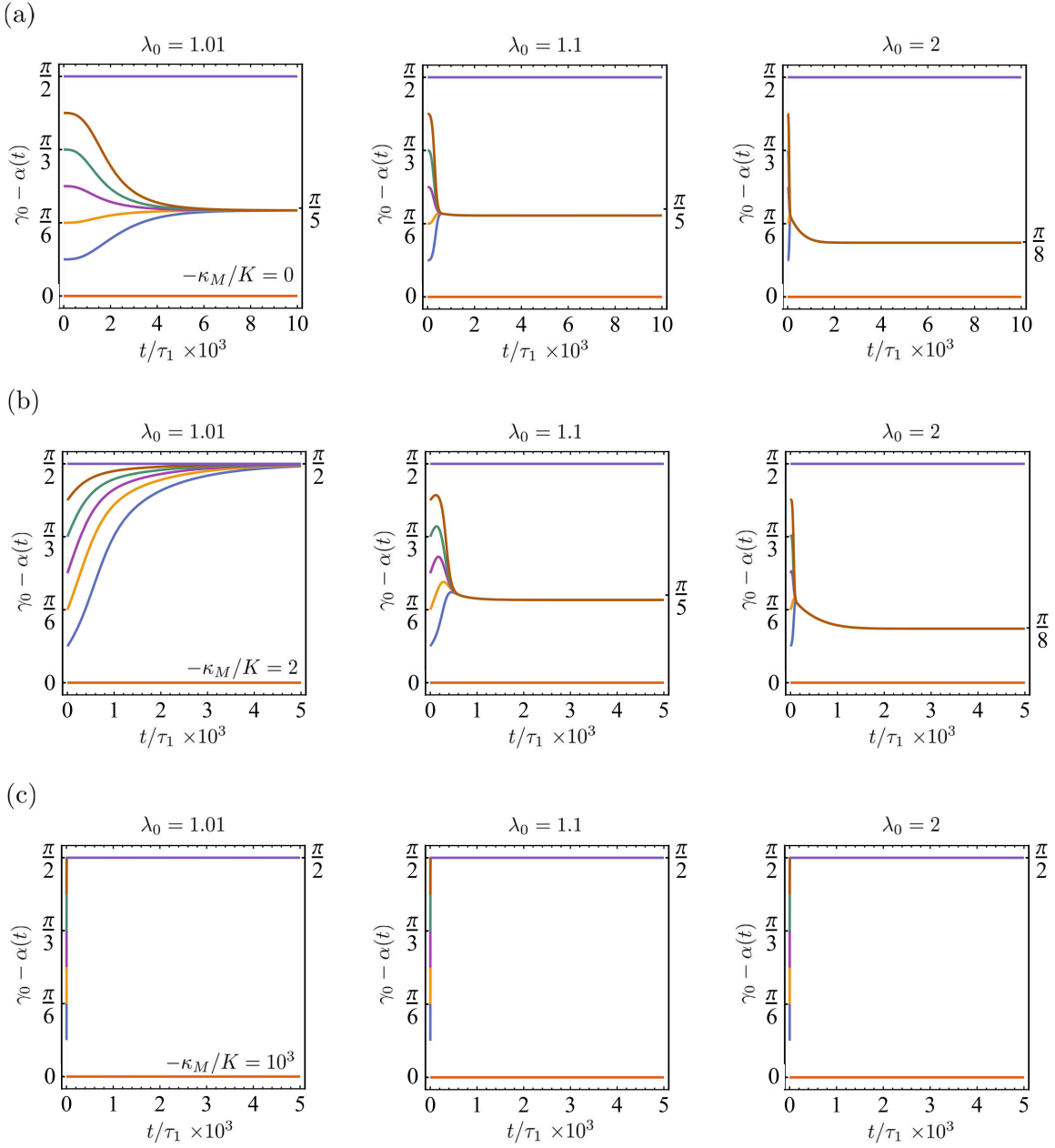}
	\end{center}
	\vspace{-0.2cm} \caption{Finite extension of a transversely isotropic cylindrical bar under displacement-control loading. The remodeled fiber orientation $\gamma_0-\alpha(t)$ is plotted as a function of $t/\tau_1$, where $\tau_1$ is the material's relaxation time. Seven initial fiber orientations, $\gamma_0-\alpha(0)=\gamma_0$ (corresponding to different colors) are investigated in equal increments of $\pi/12$ ranging from $0$ to $\pi/2$. (a) corresponds to $-\kappa_M/K= 0$, (b) to $-\kappa_M/K= 2$, and (c) to $-\kappa_M/K= 10^{3}$. For each case, the displacement function $\lambda(t)=1+(\lambda_0-1)\,\erf\big(\frac{t}{t_0}\big)$ is applied  with three different values of the maximum stretch $\lambda_{0}=1.01, 1.1,$ and $2$.}
	\label{Fig4-1-1}
\end{figure}

\begin{figure}[h!]
	\begin{center}
		\includegraphics[width=2.5in]{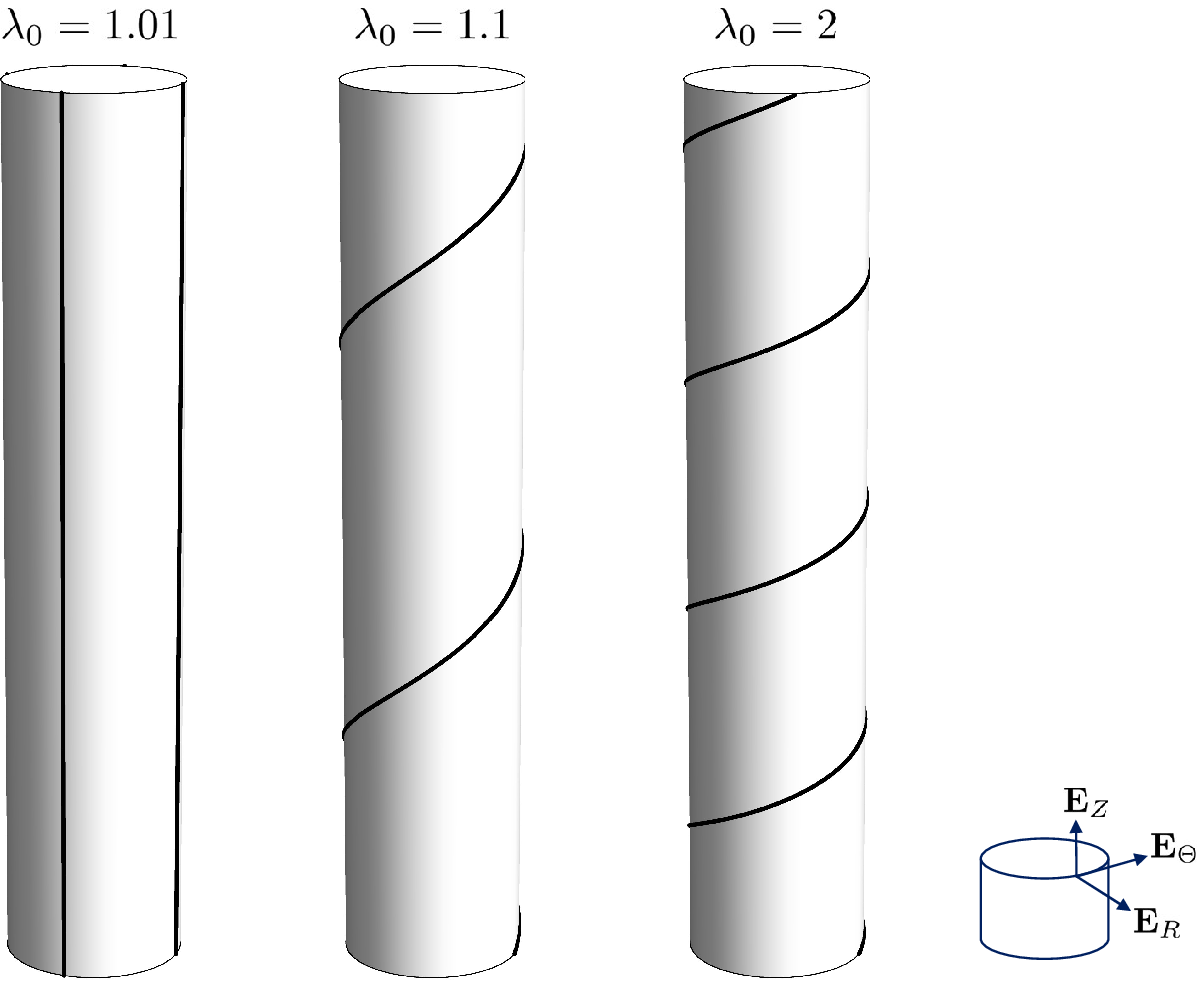}
	\end{center}
	\vspace{-0.2cm} \caption{Remodeled fiber orientation for a transversely isotropic bar under finite extension with $-\kappa_M/K= 2$ shown in the reference configuration. For $\lambda_0=1.01$, fibers align along the direction of loading. For $\lambda_0=1.1$ and $2$, they align at an angle of $\pi/5$ and $\pi/8$, respectively.}
	\label{Fig4-1-2}
\end{figure}

\paragraph{Numerical results.}
We first consider a displacement-control loading. The material constants in the Mooney-Rivlin model are taken to be $C_1=0.01$, $C_2=0$, $\mu_1=1$, and $\mu_2=0$. The relaxation times for the material are chosen to be $\tau_{1}=0.001$, and $\tau_{2}=0.0$. For all $R\in(0,R_0]$, fibers are assumed to have the same helix angle, that is, $\gamma(R)=\gamma_0$. Furthermore, it is assumed that the preferred orientation for fibers is in the direction of maximum principal strain, namely, $\mathbf{M}=\mathbf{N}^{\mathbf{C}}_{\text{max}}= \mathbf{E}_Z$.
Then, the parameters $\gamma_0$, $\kappa_M$, and  $\lambda_0$  are varied to investigate their effects on the fiber remodeling. Fig.~\ref{Fig4-1-1} shows the results for the applied loading $\lambda(t)=1+(\lambda_0-1)\,\erf\big(\frac{t}{t_0}\big)$ with $t_0=1$.
The orientation of remodeled fibers is plotted in terms of the helix angle $\gamma_0-\alpha(t)$ for a wide range of values for $\gamma_0$, $\kappa_M$, and  $\lambda_0$. Seven values of initial fiber orientation are chosen from $0$ to $\pi/2$ in equal increments of $\pi/12$. Recall that $\alpha(0)=0$. Three values of the parameter $\kappa_M= 0, -2 K$, and $-10^{3} K$ are studied in parts (a), (b), and (c) of the figure respectively. Furthermore, for each value of $\kappa_M$, three values of maximum applied stretch $\lambda_{0}=1.01, 1.1$, and $2$ are studied. 

\begin{figure}[h!]
	\begin{center}
	\vskip 0.3in
		\includegraphics[width=0.9\textwidth]{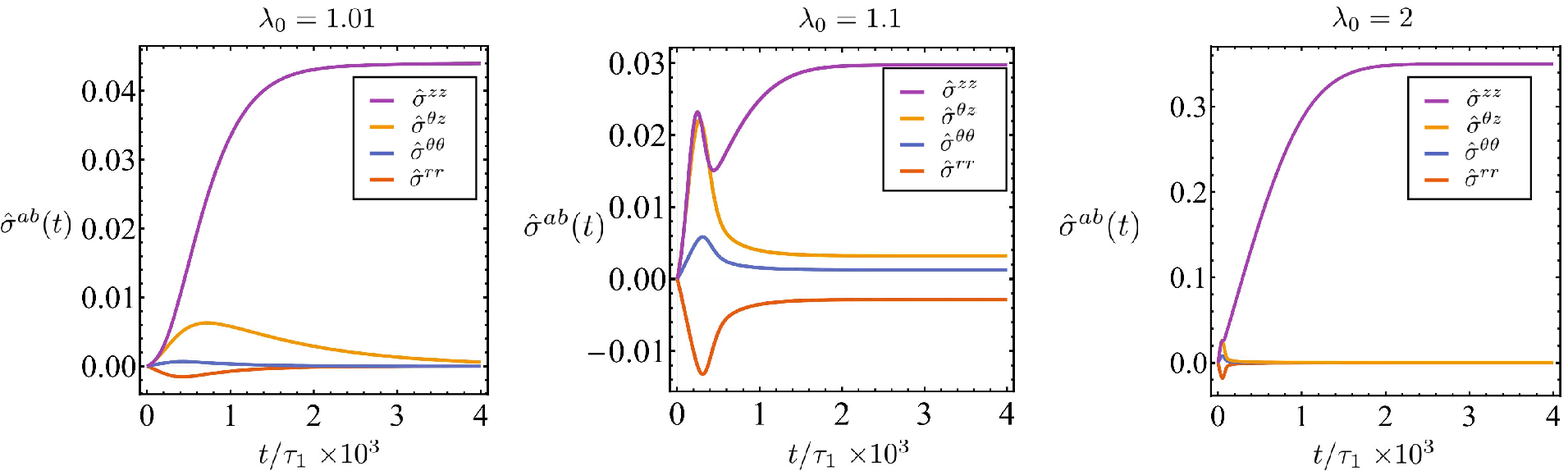}
	\end{center}
	\vspace{-0.2cm} \caption{Evolution of stress components (\ref{srr},\ref{scomp}) with time at $R/R_0=0.5$ in a transversely isotropic bar under finite extension with $-\kappa_M/K= 2$ and three values of maximum stretch $\lambda_0=1.1$ and $2$.}
	\label{Fig4-1-3}
\end{figure}

The first observation which can be made from Fig.~\ref{Fig4-1-1} is that the final remodeled fiber orientation is independent of the initial fiber orientation angle $\gamma_0$ in the range $0 < \gamma_{0} < \pi/2$. Note though, the remodeling process is not monotonic in time as is clearly observed for $-\kappa_M/K= 2$ and $\lambda_{0}=1.1$. The orientation angles $\gamma_0=0, \pi/2$ are found to be equilibrium helix angles as expected and fibers oriented in those directions do not remodel. 
Now, when $\kappa_M= 0$, the fibers should remodel to minimize the energy function $W$, hence move further away from the angle $\pi/2$. This is observed in Fig.~\ref{Fig4-1-1}(a) where the final fiber orientation decreases from $\pi/5$ to $\pi/8$ as the maximum stretch is increased from $1$ to $2$. On the other hand, when $-\kappa_M \gg K$, the fibers should remodel along $\mathbf{E}_Z$. This is seen in Fig.~\ref{Fig4-1-1}(c) where the orientation angle evolves to $\pi/2$ for all values of $\lambda_0$. This case corresponds to the classical remodeling equation studied by \citet{Menzel2005} and others. When $-\kappa_M \sim K$, there should exist a competition between strain energy and remodeling energy. Fig. \ref{Fig4-1-1}(b) shows that for small values of the maximum stretch, fibers orient themselves along $\mathbf{E}_Z$, while for larger values of stretch, they orient themselves along a direction according to the strain energy minimization. A visual representation of the final orientation of fibers as a function of $\lambda_0$ is shown in Fig.~\ref{Fig4-1-2}. More insight into this case is also provided by the evolution of stress components (\ref{srr},\ref{scomp}) shown in Figure \ref{Fig4-1-3} for $R/R_{0}=0.5$. For small and large values of maximum stretch, all stress components except $\hat{\sigma}^{zz}$ are seen to evolve to zero presumably due to remodeling energy and strain energy dominating in respective cases. However, for intermediate values of stretch, the stress components can evolve to a non-zero value indicating a strong competition between the two energies.

\begin{figure}[h!]
	\begin{center}
	\vskip 0.0in	
		\includegraphics[width=0.6\textwidth]{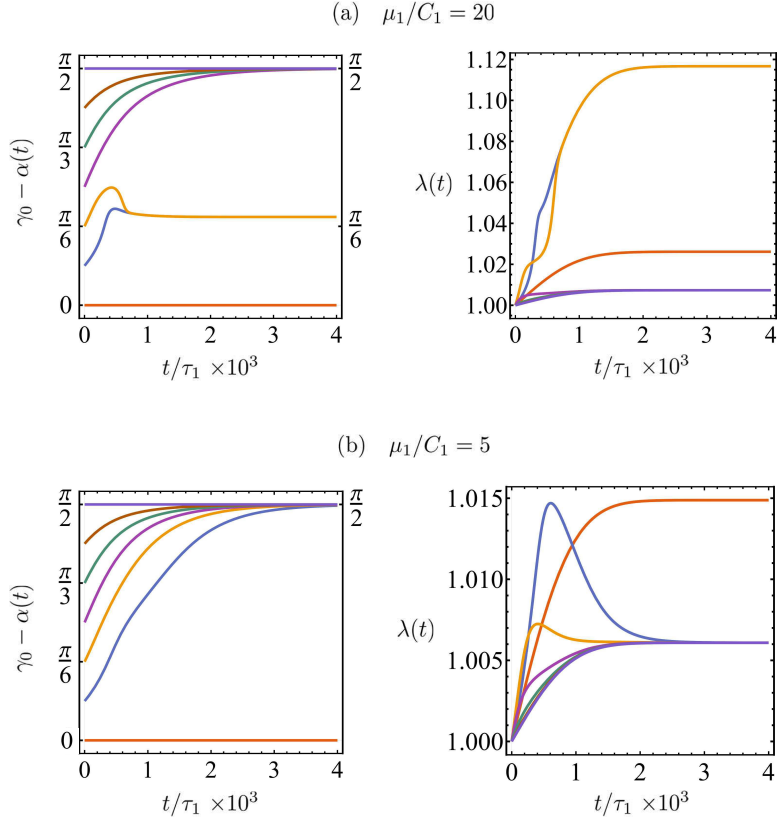}
	\end{center}
	\vspace{-0.2cm} \caption{Finite extension of a transversely isotropic cylindrical bar under force-control loading for two fiber-to-matrix modulus ratios: (a) $C_{1}/\mu_{1}=5$ and (b) $C_{1}/\mu_{1}=5$. The remodeled fiber orientation $\gamma_0-\alpha(t)$ and stretch $\lambda(t)$ are plotted  with normalized time $t/\tau_{1}$ for $-\kappa_M/K= 2$ and seven initial fiber orientations, $\gamma_0$, in equal increments of $\pi/12$ from $0$ to $\pi/2$.}
	\label{Fig4-1-4}
\end{figure}

We next consider a force-control loading. Applying an axial force $F(t)=1+(F_0-1)\,\erf\big(\frac{t}{t_0}\big)$ with $F_0=0.1$ and $t_0=1$, we examine the evolution of $\gamma_0-\alpha(t)$ and $\lambda(t)$ for $-\kappa_M/K= 2$ and two values of fiber-to-matrix modulus ratios, namely, $C_{1}/\mu_{1}=20, 5$. The results are shown in Fig.~\ref{Fig4-1-4}. We observe that for $C_{1}/\mu_{1}=20$, final fiber orientation is not independent of initial fiber orientation. While for some initial orientations, the fibers align along $\mathbf{E}_Z$ according to remodeling energy, for others, they align according to the strain energy. For $C_{1}/\mu_{1}=20$, the effect of remodeling energy is stronger, and fibers for all initial orientations (except $0$) align along $\mathbf{E}_Z$. 

\begin{figure}[h!]
	\vskip 0.3in
	\begin{center}
		\includegraphics[width=0.95\textwidth]{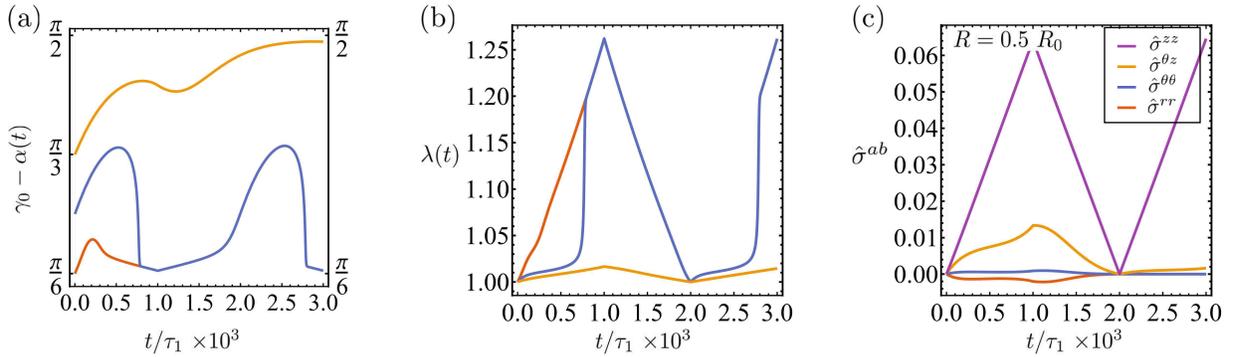}
	\end{center}
	\vspace{-0.2cm} \caption{Evolution of (a) remodeled fiber  orientation $\gamma_0-\alpha(t)$, (b) stretch $\lambda(t)$, and (c) stress components $\hat{\sigma}^{ab} (0.5 R_0, t)$, during one and a half cycles of loading-unloading as defined by \eqref{F-loadingunloading}.}
	\label{Fig4-1-5}
\end{figure}

Lastly, to investigate how the fiber-remodeling affects the maximum stretch during cycles of loading-unloading, we consider the following axial force loading with one cycle of loading-unloading followed by a second loading:
\begin{equation}
	F(t)=\left\{ \begin{array}{ll}
		F_0 t\,, & {\rm if} \quad  t \leq 1\,, \\
		F_0 - F_0 (t-1)\,, & {\rm if} \quad  t>1\,, \quad \text{and} \quad t \leq 2\,, \\
		F_0 (t-2)\,, & {\rm if} \quad  t>2\,, \quad \text{and} \quad   t \leq 3\,, 
	\end{array} 
	\right.\label{F-loadingunloading}
\end{equation}
with $F_0=0.2$. Fig.~\ref{Fig4-1-5} shows the results for remodeled fiber orientation, stretch, and stress components for $\gamma_0=\pi/6, \pi/4$, and $\pi/3$. We make two key observations. First, at $t/\tau_1=2 \times 10^3$, when $F(t)=0$ after one cycle of loading and unloading, $\lambda(t=2 \times 10^3)=1$, and all the stress components are zero. Thus, as expected (see Remark \ref{Stress-Free}), there are no residual stresses observed. Second, the remodeled fiber orientation for $\gamma_{0}=\pi/3$ at the end of second loading phase, $t/\tau_1=3 \times 10^3$, is different from that at the end of first loading phase, $t/\tau_1=3 \times 10^3$, while for the other two values of $\gamma_{0}$, it remains the same. This shows that the remodeling process can be loading history-dependent.   

\subsection{Example 2: Finite extension of a monoclinic circular cylindrical bar} \label{Sec:Example2}

Let us consider the circular cylindrical bar of the previous example, however, with two families of helical fibers. 
For this monoclinic solid cylinder we have two unit vector fields $\mathring{\mathbf{N}}_1=\mathring{\mathbf{N}}_1(R,\Theta)$, and $\mathring{\mathbf{N}}_2=\mathring{\mathbf{N}}_2(R,\Theta)$. 
Suppose $\gamma_1(R)$ and $\gamma_2(R)$ are the angles that $\mathring{\mathbf{N}}_1(R,\Theta)$ and $\mathring{\mathbf{N}}_2(R,\Theta)$ make with $\mathbf{E}_{\Theta}$, i.e., 
\begin{equation}
	\mathring{\mathbf{N}}_1(R,\Theta)=\frac{\cos\gamma_1(R)}{R}\,\mathbf{E}_{\Theta}(\Theta)
	+\sin\gamma_1(R)\,\mathbf{E}_Z\,,\qquad
	\mathring{\mathbf{N}}_2(R,\Theta)=\frac{\cos\gamma_2(R)}{R}\,\mathbf{E}_{\Theta}(\Theta)
	+\sin\gamma_2(R)\,\mathbf{E}_Z\,.
\end{equation}
During a remodeling process these vectors are transformed to the following two vectors
\begin{equation} \label{Fibers-Initial}
	\mathbf{N}_1(R,t)=\begin{bmatrix}
	0 \\
	\frac{\cos(\gamma_1(R)-\alpha_1(R,t))}{R} \\
	\sin(\gamma_1(R)-\alpha_1(R,t)) 
	\end{bmatrix}\,,\qquad
	\mathbf{N}_2(R,t)=\begin{bmatrix}
	0 \\
	\frac{\cos(\gamma_2(R)-\alpha_2(R,t))}{R} \\
	\sin(\gamma_2(R)-\alpha_2(R,t)) 
	\end{bmatrix}
	\,,
\end{equation}
where $\alpha_1(R,t)$ and $\alpha_2(R,t)$ are the angles of rotation to be determined.
We assume the kinematics ansatz \eqref{Deformation},\footnote{Notice that \eqref{Deformation} is a subset of Family $3$ universal deformations \citep{Ericksen1954}, and the fiber distributions \eqref{Fibers-Initial} are universal material preferred directions \citep{YavariGoriely2021}. This implies that the deformations \eqref{Deformation} can be maintained in the absence of body forces for any incompressible isotropic solid cylinder reinforced by the two families of fibers with distributions given in \eqref{Fibers-Initial}.}  and hence the deformation gradient is given in \eqref{F-Example1}.
From incompressibility $r(R,t)=\frac{R}{\sqrt{\lambda(t)}}$, and the nine monoclinic invariants read
\begin{equation}
\begin{aligned}
	& I_1=\lambda^2+2\lambda^{-1}\,,\\ 
	& I_2=2\lambda+\lambda^{-2}\,,\\
	& I_4=\lambda^2\sin^2(\alpha_1-\gamma_1)+\lambda^{-1}
	\cos^2(\alpha_1-\gamma_1)\,,\\
	& I_5=\lambda^4 \sin^2(\alpha_1-\gamma_1)+\lambda^{-2}
	\cos^2(\alpha_1-\gamma_1)\,,\\
	& I_6=\lambda^2 \sin^2(\alpha_2-\gamma_2)+\lambda^{-1}
	\cos^2(\alpha_2-\gamma_2)\,,\\
	& I_7=\lambda^4 \sin^2(\alpha_2-\gamma_2)+\lambda^{-2}
	\cos^2(\alpha_2 -\gamma_2)\,,\\
	& I_8 = \frac{1}{\lambda} \cos(\alpha_2-\alpha_1+\gamma_1-\gamma_2) 
	\left[\lambda^3 \sin(\alpha_1-\gamma_1) \sin(\alpha_2-\gamma_2)
	+\cos(\alpha_1-\gamma_1) \cos(\alpha_2-\gamma_2)\right] \,,\\
	& I_9 = \cos^2(\gamma_1-\gamma_2+\alpha_2-\alpha_1)
	\,.
\end{aligned}
\end{equation}

\paragraph{Stress and equilibrium equations.}
The non-zero components of the Cauchy stress are:
\begin{equation}
\begin{aligned}
	\sigma^{rr}(R,t) &=-p+2\lambda^{-1}\,W_1-2\lambda\,W_2\,,
\end{aligned}
\end{equation}
\begin{equation}
\begin{aligned}
	\sigma^{\theta\theta}(R,t) & = -p\frac{\lambda}{R^2}+\frac{2W_1}{R^2}
	-\frac{2\lambda^2\,W_2}{R^2}
	+\frac{2\left(\lambda\,W_4+2W_5\right)}{R^2\lambda(t)}\cos^2(\alpha_1-\gamma_1)
	+\frac{2\left(\lambda\,W_6+2W_7\right)}{R^2\lambda(t)}\cos^2(\alpha_2-\gamma_2) \\
	& \quad +\frac{4}{R^2}
	\cos (\alpha_1-\gamma_1) \cos(\alpha_2-\gamma_2) 
	\cos(\alpha_2-\alpha_1+\gamma_1-\gamma_2)\,W_8
	 \,,
\end{aligned}
\end{equation}
\begin{equation}
\begin{aligned}
	\sigma^{zz}(R,t) & =-p+2\lambda^2\,W_1-2\lambda^{-2}(t)\,W_2
	+2\lambda^2\left[W_4+2\lambda^2\,W_5\right]\sin^2(\alpha_1-\gamma_1) \\
	& \quad +2\lambda^2\left[W_6+2\lambda^2\,W_7\right]\sin^2(\alpha_2-\gamma_2) \\
	& \quad +4 \lambda^2\,\sin(\alpha_1-\gamma_1) \sin(\alpha_2-\gamma_2)  
	\cos(\alpha_2-\alpha_1+\gamma_1-\gamma_2)\,W_8  \,,
\end{aligned}
\end{equation}
and
\begin{equation}
\begin{aligned}
	\sigma^{\theta z}(R,t) & 
	= -\frac{\lambda\,W_4+\left(1+\lambda^3\right)W_5}{R} \sin\left[2(\alpha_1-\gamma_1)\right]
	-\frac{\lambda\,W_6+\left(1+\lambda^3\right)W_7}{R} \sin\left[2(\alpha_2-\gamma_2)\right] \\
	&\quad -\frac{\lambda\,W_8}{R} 
	\Big\{\sin\left[2(\alpha_1-\gamma_1)\right]+\sin\left[2(\alpha_2-\gamma_2)\right]\Big\}
	\,.
\end{aligned}
\end{equation}
The radial equilibrium equation is written as (the other two equilibrium equations imply that $p=p(R,t)$)
\begin{equation}
\begin{aligned}
	\frac{\partial}{\partial R}\sigma^{rr}(R,t) 
	&=\frac{2\left(\lambda\,W_4+2W_5\right)}{R\lambda^2}\cos^2(\alpha_1-\gamma_1)
	+\frac{2\left(\lambda\,W_6+2W_7\right)}{R\lambda^2}\cos^2(\alpha_2-\gamma_2)\\
	& \quad +\frac{4\,W_8}{R\lambda}
	\cos(\alpha_1-\gamma_1) \cos(\alpha_2-\gamma_2) 
	\cos(\alpha_2-\alpha_1+\gamma_1-\gamma_2)
	\,.
\end{aligned}
\end{equation}
Using the boundary condition $\sigma^{rr}(R_0,t)=0$, one obtains
\begin{equation}
\begin{aligned}
	\sigma^{rr}(R,t) &= -\frac{2}{\lambda^2}\int_{R}^{R_0}\frac{\lambda\,W_4+2W_5}{\xi}
	\cos^2(\alpha_1-\gamma_1)\,d\xi 
	-\frac{2}{\lambda^2}\int_{R}^{R_0}\frac{\lambda\,W_6+2W_7}{\xi}
	\cos^2(\alpha_2-\gamma_2)\,d\xi \\
	&\quad -\frac{4}{\lambda} \int_{R}^{R_0} \frac{W_8}{\xi}
	\cos(\alpha_1-\gamma_1) \cos(\alpha_2-\gamma_2) 
	\cos(\alpha_2-\alpha_1+\gamma_1-\gamma_2)\,d\xi 
	\,.
\end{aligned}
\end{equation}
This, in particular, implies that
\begin{equation}
\begin{aligned}
	-p & = -\frac{2}{\lambda^2}\int_{R}^{R_0}\frac{\lambda\,W_4+2W_5}{\xi}
	\cos^2(\alpha_1-\gamma_1)\,d\xi
	-\frac{2}{\lambda^2}\int_{R}^{R_0}\frac{\lambda\,W_6+2W_7}{\xi}
	\cos^2(\alpha_2-\gamma_2)\,d\xi \\
	& \quad -\frac{4}{\lambda} \int_{R}^{R_0} \frac{W_8}{\xi}
	\cos(\alpha_1-\gamma_1) \cos(\alpha_2-\gamma_2) 
	\cos(\alpha_2-\alpha_1+\gamma_1-\gamma_2)\,d\xi 
	-2\lambda^{-1}\,W_1+2\lambda\,W_2
	\,.
\end{aligned}
\end{equation}
Thus, the non-zero physical components of stress read (recall that $\hat{\sigma}^{rr}=\sigma^{rr}$)
\begin{equation}
\begin{aligned}
	\hat{\sigma}^{\theta\theta}(R,t) & =
	\frac{2\left(\lambda\,W_4+2W_5\right)}{\lambda^2}\cos^2(\alpha_1-\gamma_1)
	+\frac{2\left(\lambda\,W_6+2W_7\right)}{\lambda^2}\cos^2(\alpha_2-\gamma_2) \\
	& \quad +\frac{4}{\lambda}
	\cos (\alpha_1-\gamma_1) \cos(\alpha_2-\gamma_2) 
	\cos(\alpha_2-\alpha_1+\gamma_1-\gamma_2)\,W_8 \\
	& \quad -\frac{2}{\lambda^2}\int_{R}^{R_0}\frac{\lambda\,W_4+2W_5}{\xi}
	\cos^2(\alpha_1-\gamma_1)\,d\xi  -\frac{2}{\lambda^2}\int_{R}^{R_0}\frac{\lambda\,W_6+2W_7}{\xi}
	\cos^2(\alpha_2-\gamma_2)\,d\xi \\
	&\quad -\frac{4}{\lambda} \int_{R}^{R_0} \frac{W_8}{\xi}
	\cos(\alpha_1-\gamma_1) \cos(\alpha_2-\gamma_2) 
	\cos(\alpha_2-\alpha_1+\gamma_1-\gamma_2)\,d\xi \,,
\end{aligned}
\end{equation}
\begin{equation}
\begin{aligned}
	\hat{\sigma}^{zz}(R,t) & =2\left[\lambda^2-\lambda^{-1}\right]W_1
	+2\left[\lambda-\lambda^{-2}\right]W_2
	+2\lambda^2\left[W_4+2\lambda^2\,W_5\right]\sin^2(\alpha_1-\gamma_1) \\
	& +2\lambda^2\left[W_6+2\lambda^2\,W_7\right]\sin^2(\alpha_2-\gamma_2) \\
	& \quad -\frac{2}{\lambda^2}\int_{R}^{R_0}\frac{\lambda\,W_4+2W_5}{\xi}
	\cos^2(\alpha_1-\gamma_1)\,d\xi-\frac{2}{\lambda^2}\int_{R}^{R_0}\frac{\lambda\,W_6+2W_7}{\xi}
	\cos^2(\alpha_2-\gamma_2)\,d\xi \\
	&\quad -\frac{4}{\lambda} \int_{R}^{R_0} \frac{W_8}{\xi}
	\cos(\alpha_1-\gamma_1) \cos(\alpha_2-\gamma_2) 
	\cos(\alpha_2-\alpha_1+\gamma_1-\gamma_2)\,d\xi \,,
\end{aligned}
\end{equation}
and
\begin{equation}
\begin{aligned}
	\hat{\sigma}^{\theta z}(R,t) & 
	=-\frac{\lambda\,W_4+\left[1+\lambda^3\right]W_5}{\lambda^{\frac{1}{2}}}
	\sin\left[2(\alpha_1-\gamma_1)\right] 
	-\frac{\lambda\,W_6+\left[1+\lambda^3\right]W_7}{\lambda^{\frac{1}{2}}}
	\sin\left[2(\alpha_2-\gamma_2)\right] \\
	& \quad -\lambda^{\frac{1}{2}}\, 
	\Big\{\sin\left[2(\alpha_1-\gamma_1)\right]+\sin\left[2(\alpha_2-\gamma_2)\right]\Big\}\,W_8
	\,.
\end{aligned}
\end{equation}

\paragraph{The axial force.}
For displacement-control loading $\lambda(t)$ is given while $\alpha_1(R,t)$ and $\alpha_2(R,t)$ are unknowns that are governed by the remodeling equations. For force-control loadings, the unknowns of the problem are $\lambda(t)$, $\alpha_1(R,t)$, and $\alpha_2(R,t)$. In this case at the two ends of the bar ($Z=0,L$), the axial force required to maintain the deformation is given in \eqref{Axial-Force-Example1}.
Thus
\begin{equation} \label{Example2-AxialForce}
\begin{aligned}
	& 2\left(\lambda-\lambda^{-2}\right) \int_{0}^{R_0} W_1\,R\,dR
	+2\left(1-\lambda^{-3}\right) \int_{0}^{R_0}  W_2\,R\,dR \ \\
	& \quad 
	+2\lambda \int_{0}^{R_0} R\left(W_4+2\lambda^2\,W_5\right) \sin^2(\alpha_1-\gamma_1)\,dR 
	+2\lambda \int_{0}^{R_0} R\left(W_6+2\lambda^2\,W_7\right) \sin^2(\alpha_2-\gamma_2)\,dR  \\
        & \quad -\frac{2}{\lambda^3} \int_{0}^{R_0} R \int_{R}^{R_0}\frac{\lambda\,W_4+2W_5}{\xi}
	\cos^2(\alpha_1-\gamma_1)\,d\xi\,dR
	-\frac{2}{\lambda^3} \int_{0}^{R_0} R \int_{R}^{R_0}\frac{\lambda\,W_6+2W_7}{\xi}
	\cos^2(\alpha_2-\gamma_2)\,d\xi\,dR \\	
	& \quad -\frac{4}{\lambda^2} \int_{0}^{R_0} R \int_{R}^{R_0} \frac{W_8}{\xi}
	\cos(\alpha_1-\gamma_1) \cos(\alpha_2-\gamma_2) 
	\cos(\alpha_2-\alpha_1+\gamma_1-\gamma_2)\,d\xi \,dR   = \frac{F(t)}{2\pi}	\,.
\end{aligned}
\end{equation}

\paragraph{The remodeling equation.}
The remodeling equations \eqref{Kinetic-N2-2} are simplified to read
\begin{equation} \label{KineticEq1-Example2}
\begin{aligned}
	K_1\, \dot{\alpha}_1
	+K_3 \cos(-\alpha_1+\alpha_2+\gamma_1-\gamma_2)\, \dot{\alpha}_2
	& =	 -\frac{1}{2}\kappa_1 \sin2(\alpha_1-\gamma_1)) \\
	& \quad
	+\frac{1-\lambda^3}{2 \lambda^2} \left[\lambda (2W_4+W_8)+2\left(\lambda^3+1\right) W_5\right] 
	\sin2(\alpha_1-\gamma_1) \\
	& \quad -\frac{1}{2\lambda} \left[(\lambda^3+1) W_8+2 \lambda W_9 \right]
	\sin2(\alpha_2-\alpha_1+\gamma_1-\gamma_2)\,,\\
	 K_2 \,\dot{\alpha}_2
	+K_3 \cos(-\alpha_1+\alpha_2+\gamma_1-\gamma_2) \,\dot{\alpha}_1
	&=  -\frac{1}{2}\kappa_2  \sin2(\alpha_2-\gamma_2) \\
	& \quad
	+\frac{1-\lambda^3}{2 \lambda^2}
	\left(\lambda  (2 W_6+W_8)+2 \left(\lambda ^3+1\right)W_7\right) \sin2(\alpha_2-\gamma_2)\\
	&\quad	+\frac{1}{2 \lambda}  \left(\lambda^3 W_8+W_8+2 \lambda  W_9\right) 
	\sin(2(\alpha_2-\alpha_1+\gamma_1-\gamma_2))\,.
\end{aligned}
\end{equation}
Similarly, the remodeling equations \eqref{Kinetic-N2-1} read\footnote{Our numerical results show that the two remodeling equations \eqref{KineticEq1-Example2} and \eqref{KineticEq2-Example2} give very similar results. We use \eqref{KineticEq1-Example2} in our numerical examples.}
\begin{equation} \label{KineticEq2-Example2}
\begin{aligned}
	K_1\, \dot{\alpha}_1
	+K_3 \cos(-\alpha_1+\alpha_2+\gamma_1-\gamma_2)\, \dot{\alpha}_2
	& = \frac{1}{2} \kappa_1\operatorname{sgn}(\sin(\gamma_1-\alpha_1))\,\sin2(\gamma_1-\alpha_1) \\
	& \quad
	+\frac{1-\lambda^3}{2 \lambda^2} \left[\lambda (2W_4+W_8)+2\left(\lambda^3+1\right) W_5\right] 
	\sin2(\alpha_1-\gamma_1) \\
	& \quad -\frac{1}{2\lambda} \left[(\lambda^3+1) W_8+2 \lambda W_9 \right]
	\sin2(\alpha_2-\alpha_1+\gamma_1-\gamma_2)\,,\\
	 K_2 \,\dot{\alpha}_2
	+K_3 \cos(-\alpha_1+\alpha_2+\gamma_1-\gamma_2) \,\dot{\alpha}_1
	&=  \frac{1}{2} \kappa_2\operatorname{sgn}(\sin(\gamma_2-\alpha_2))\,\sin2(\gamma_2-\alpha_2) \\
	& \quad
	+\frac{1-\lambda^3}{2 \lambda^2}
	\left(\lambda  (2 W_6+W_8)+2 \left(\lambda ^3+1\right)W_7\right) \sin2(\alpha_2-\gamma_2)\\
	&\quad	+\frac{1}{2 \lambda}  \left(\lambda^3 W_8+W_8+2 \lambda  W_9\right) 
	\sin(2(\alpha_2-\alpha_1+\gamma_1-\gamma_2))\,.
\end{aligned}
\end{equation}


For the numerical examples we consider the following generalized incompressible Mooney-Rivlin reinforced model for which 
\begin{equation} \label{MR-Material2}
	W=C_1(I_1-3)+C_2(I_2-3)
	+\frac{\mu_1}{2}\left[(I_4-1)^2+(I_6-1)^2\right]
	+\frac{\mu_2}{2}\left[(I_5-1)^2+(I_7-1)^2\right]
	+\frac{\mu_3}{2}(I_8-I_9)^2
	\,,
\end{equation}	
where $C_1$, $C_2$, $\mu_1$, $\mu_2$, and $\mu_3$ are positive constants.
Thus, $W_1=C_1$, $W_2=C_2$, $W_4=\mu_1(I_4-1)$, $W_5=\mu_2(I_5-1)$, $W_6=\mu_1(I_6-1)$, $W_7=\mu_2(I_7-1)$, $W_8=\mu_3(I_8-I_9)$, and $W_9=\mu_3\left(I_9-I_8\right)$.

\paragraph{Numerical results.}
Similar to the last example, we consider a displacement-control loading: $\lambda(t)=1+(\lambda_0-1)\,\erf\big(\frac{t}{t_0}\big)$ with $t_0=1$.
The material constants in the constitutive model are taken to be $C_1=0.05$, $C_2=0$, $\mu_1=1$, and $\mu_2=0$. The parameters $K_1$, $K_2$, $K_3$ are fixed at $K_{1}=K_{2}=K=0.001$ and $K_3=0.0001$. We again define a relaxation time $\tau_1=K_{1}/\mu_1$. We assume that $\gamma_1 (R)=\gamma_0$ and $\gamma_2 (R)=-\gamma_0-\pi/12$. 
Again, it is assumed that the preferred orientation for fibers is the direction of maximum principal strain, namely, $\mathbf{M}=\mathbf{N}^{\mathbf{C}}_{\text{max}}= \mathbf{E}_Z$. The parameters $\kappa_{M1}$ and $\kappa_{M2}$ are taken to be equal: $\kappa_{M1}=\kappa_{M2}=\kappa_{M}$.
The parameters $\gamma_0$, $\kappa_{M}$, and  $\lambda_0$  are varied in order to investigate their impact on the fiber reorientation. 
Similar behavior is observed as the previous example. The final remodeled fiber orientation is independent of the initial fiber orientation angle $\gamma_0$. Larger value for $-\kappa_{M}$ or smaller value for $\lambda_0$ results in an remodeling-energy-dominant remodeling, whereas smaller value of $-\kappa_{M}$ or larger value of $\lambda_0$ results in strain-energy-dominant remodeling.

\begin{figure}[h!]
	\begin{center}
		\includegraphics[width=0.8\textwidth]{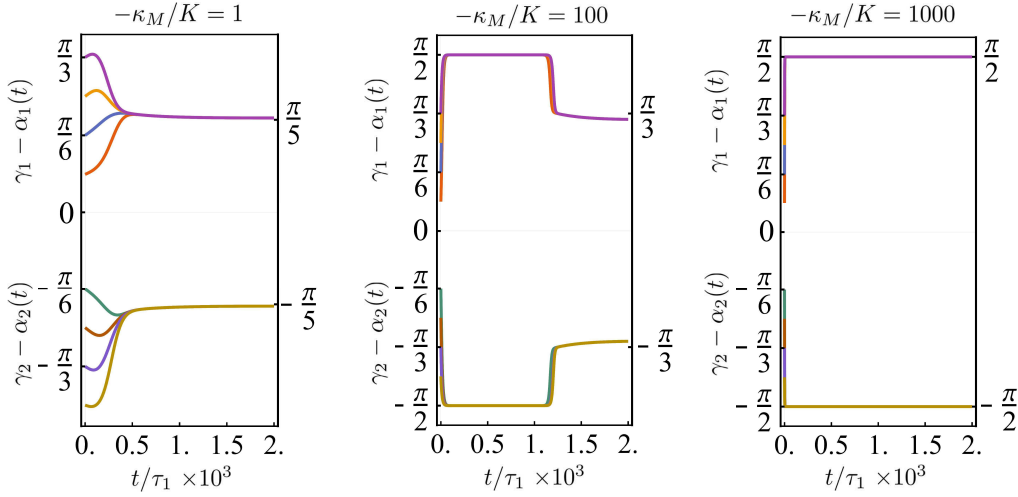}
	\end{center}
	\vspace{-0.2cm} \caption{Finite extension of a monoclinic cylindrical bar under displacement-control loading. The remodeled fiber orientation of the two fiber families: $\gamma_1-\alpha1(t)$ and $\gamma_2-\alpha2(t)$ are plotted as a function of $t/\tau_1$, where $\tau_1$ is the material's relaxation time. Three values of the ratio $-\kappa_{M}/K= 1, 100$, and $1000$ are chosen. The initial fiber orientation of the two families are $\gamma_0$ and $-\gamma_0-\pi/12$. Six values of $\gamma_0$ (corresponding to different colors) are investigated in equal increments of $\pi/12$ ranging from $\pi/12$ to $\pi/3$ for each value of the ratio $\kappa/K$.}
	\label{Fig4-2-1}
\end{figure}

\begin{figure}[h!]
	\begin{center}
	\vskip 0.3in
		\includegraphics[width=2.75 in]{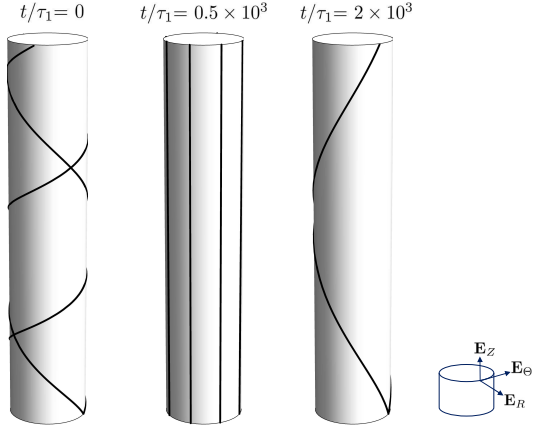}
	\end{center}
	\vspace{-0.2cm} \caption{Remodeling of fiber orientation for a monoclinic bar under finite extension  shown in the reference configuration as a function of time with $-\kappa_{M}/K= 100$ and $\lambda_0=1.1$. At $t/\tau_1=0$, when the applied stretch is $1$, both fiber families are in their initial orientation $\pi/6$ and $-\pi/4$. As the applied stretch starts to increase, the fibers align almost along the direction of loading. After $t=1$, when the applied stretch stops increasing, the fibers again remodel to $\pi/3$. }
	\label{Fig4-2-2}
\end{figure}

Fig.~\ref{Fig4-2-1} shows the results for the remodeling process as a function of normalized time for $\lambda_0=1.1$ and various values of $\gamma_0$ and $\kappa_{M}$. Final fiber orientation for both family of fibers is the same even though the initial orientations are different, and it increases in absolute value from $\pi/5$ to $\pi/2$ as the ratio $-\kappa_{M}/K$ is increased. The value of the coupling parameter $K_3$ does not affect the final orientation. The quantitative impact of $\lambda_0$ is similar to the previous example. Of particular interest is the non-monotonicity of the fiber remodeling process as a function of time as is clearly visible in Fig.~\ref{Fig4-2-1}(b). During ${t}/{\tau_1}\in(0,1]\times 10^{3}$, the applied stretch is increasing from $1$. Initially, when stretch is low, the remodeling energy is dominant and for large values of $-\kappa_{M}$, fibers can quickly remodel themselves to be almost aligned with the direction of loading. However, as the stretch stops increasing, strain energy becomes more dominant and fibers suddenly reorient themselves in a different direction. A visual representation of these two changes in the final fiber orientation with time are shown in Fig.~\ref{Fig4-2-2}.
%

\subsection{Example 3: Finite torsion of a transversely isotropic circular cylindrical bar}

In this example we consider a remodeling solid circular cylindrical bar that in its undeformed configuration has radius $R_0$ and length $L$ and is reinforced by a family of fibers with distribution given in \eqref{Fiber-Initial}. The remodeling tensor \eqref{Cylinder-Remodeling-Tensor} is assumed, and hence, at time $t$ the fiber distribution is given in \eqref{Fiber-Initial}. 
Let us assume the following deformation mappings
\begin{equation}  \label{Deformation-Torsion}
   r=r(R,t)\,,\quad \theta=\Theta+\psi(t) Z\,,\quad z=\lambda(t) Z\,,
\end{equation}
where $\psi(t)$ is twist per unit length, and $\lambda(t)$ is the axial stretch.\footnote{It should be noted that \eqref{Deformation-Torsion} is a subset of Family $3$ universal deformations \citep{Ericksen1954}, and the fiber distribution \eqref{Fiber-Initial} are universal material preferred directions \citep{YavariGoriely2021}. This means that the deformations \eqref{Deformation-Torsion} can be maintained in the absence of body forces for any incompressible isotropic solid cylinder reinforced by fibers with distribution given in \eqref{Fiber-Initial}.}  
In a twist-control loading $\psi(t)$ is given while $\lambda(t)$ is an unknown to be calculated. In a torque-control loading both $\psi(t)$ and $\lambda(t)$ are unknown functions. The deformation gradient reads
\begin{equation}
   \mathbf{F}=\mathbf{F}(R,t)=\begin{bmatrix}
  r'(R,t) & 0  & 0  \\
  0 & 1  & \psi(t)  \\
  0 & 0  & \lambda(t)
\end{bmatrix}\,,
\end{equation}
where $r'(R,t)=\frac{\partial r(R,t)}{\partial R}$. The incompressibility implies that
\begin{equation}
	J=\sqrt{\frac{\det\mathbf{g}}{\det\mathbf{G}}}\det\mathbf{F}=\frac{\lambda(t)\,r(R,t)\,r'(R,t)}{R}=1\,.
\end{equation}
Assuming that $r(0,t)=0$, we have $r(R,t)=\frac{R}{\sqrt{\lambda(t)}}$. 
The right Cauchy-Green strain reads
\begin{equation}
   \mathbf{C}=[C^A{}_B]=\begin{bmatrix}
 \frac{1}{\lambda(t)} & 0 & 0 \\
 0 & \frac{1}{\lambda(t)} & \frac{\psi(t)}{\lambda(t)} \\
 0 & \frac{R^2 \psi(t)}{\lambda(t)} & \lambda^2(t)+\frac{R^2 \psi^2(t)}{\lambda(t)}
\end{bmatrix}\,.
\end{equation}
The maximum eigenvalue of $\mathbf{C}$ is 
\begin{equation}
	\frac{1+\lambda ^3+R^2 \psi^2+\sqrt{\left(\lambda^3-1\right)^2
	+R^2 \psi^2 \left(2 \lambda^3+R^2 \psi^2+2\right)}}{2 \lambda}\,,
\end{equation}
and
\begin{equation}
	\mathbf{N}^{\mathbf{C}}_{\text{max}}=
	\begin{bmatrix}
	 0 \\
	 \frac{1-\lambda^3-R^2 \psi^2+\sqrt{\left(\lambda^3-1\right)^2
	 +2\left(\lambda^3+1\right) R^2 \psi^2+R^4 \psi^4}}{2 R^2 \psi } \\
	 1
\end{bmatrix}\,.
\end{equation}
We assume the initial fiber distribution \eqref{Fiber-Initial}.
The remodeling tensor is given in \eqref{Cylinder-Remodeling-Tensor}. The principal invariants read
\begin{equation}
\begin{aligned}
	I_1 &=\lambda^2(t)+2\lambda^{-1}(t)+\frac{R^2 \psi^2(t)}{\lambda(t)}\,,\\ 
	I_2 &=2\lambda(t)+\lambda^{-2}(t)+\frac{R^2 \psi^2(t)}{\lambda^2(t)}\,,\\
	I_4& =\left[\lambda^2(t)+R^2\psi^2\lambda^{-1}(t)\right]\sin^2\left[\alpha(R,t)-\gamma(R)\right]
	+R\,\psi(t)\lambda^{-1}(t) \,\sin\left[2(\gamma(R)-\alpha (R,t))\right]\\
	&\quad +\lambda^{-1}(t)\cos^2\left[\alpha(R,t)-\gamma(R)\right]
	 \,,\\
	I_5 &=\frac{\left(\lambda^3(t)+R^2 \psi^2(t)\right)^2+2 R^2 \psi^2(t)+1}{2\lambda^2(t)}
	-\frac{\left[\lambda^3(t)+R^2 \psi^2(t)\right]^2-1}{2 \lambda^2(t)}
	\cos\left[2(\gamma(R)-\alpha(R,t))\right]\\
	& \quad +\frac{R\psi(t) \left(\lambda^3(t)+R^2\psi^2(t)+1\right)}{\lambda^2(t)}
	\sin\left[2(\gamma(t)-\alpha(R,t))\right]
	\,.
\end{aligned}
\end{equation}
The non-zero components of the Cauchy stress are written as:\footnote{The physical components of stress are:
\begin{equation}
	\bar{\sigma}^{rr}=\sigma^{rr}\,,\quad
	\bar{\sigma}^{\theta\theta}=r^2\sigma^{\theta\theta}=\frac{R^2}{\lambda}\sigma^{\theta\theta}\,,\quad
	\bar{\sigma}^{\theta z}=r\sigma^{\theta z}=\frac{R}{\sqrt{\lambda}}\sigma^{\theta z}\,\quad
	\bar{\sigma}^{zz}=\sigma^{zz}
\end{equation}
}
\begin{equation}
	\sigma^{rr}(R,t) =-p(R,t)+2\lambda^{-1}(t)\,W_1-2\lambda(t)\,W_2\,,
\end{equation}		
\begin{align}
	\sigma^{\theta\theta}(R,t) & = -\frac{p(R,t)\lambda(t)}{R^2}
	+2 \left(\frac{1}{R^2}+\psi^2(t)\right)W_1-\frac{2 \lambda^2(t)}{R^2}W_2 \nonumber \\
	& \quad +\frac{2}{R^2}\left[R \psi(t) \sin(\gamma(R)-\alpha (R,t))
	+\cos(\gamma(R)-\alpha (R,t))\right]^2\,W_4 \nonumber \\
	& \quad +\frac{4}{\lambda(t) R^2}\left[R \psi(t) \sin(\gamma(R)-\alpha (R,t))
	+\cos(\gamma(R)-\alpha (R,t))
	\right]\\
	&\qquad \times 
	\left[R \psi(t) \left(\lambda^3(t)+R^2 \psi^2(t)+1\right) \sin(\gamma(R)-\alpha (R,t))
	+\left(R^2\psi^2(t)+1\right) \cos(\gamma(R)-\alpha(R,t))\right] W_5 \,,  \nonumber
\end{align}
\begin{equation}
\begin{aligned}
	\sigma^{zz}(R,t) & =-p(R,t)+2\lambda^2(t)\,W_1
	-\frac{2\left(R^2 \psi^2(t)+1\right)}{\lambda^2(t)}\,W_2
	+2 \lambda^2(t) \sin^2(\gamma(R)-\alpha(R,t))\,W_4 \\
	& \quad +2 \lambda(t)  \left[2\left(\lambda^3(t)+R^2 \psi^2(t)\right) 
	\sin^2(\gamma(R)-\alpha(R,t))+R \psi(t)  \sin(2(\gamma(R)-\alpha(R,t)))\right]\,W_5\,,
\end{aligned}
\end{equation}
\begin{align}
	\sigma^{\theta z}(R,t) & 
	= 2 \lambda(t)\psi(t)\,W_1+2 \psi(t) W_2 \nonumber \\
	& \quad +\frac{\lambda(t)}{R}\left[-R\psi(t) \cos(2(\gamma(R)-\alpha(R,t)))
	+\sin(2(\gamma(R)-\alpha(R,t)))+R \psi(t) \right]\,W_4 \nonumber\\
	& \quad +\Bigg[\frac{\lambda^3(t)+3 R^2 \psi^2(t)+1}{R}\,\sin(2(\gamma(R)-\alpha(R,t)))\\
	& \qquad\qquad -2 \psi(t) \left(\lambda^3(t)+R^2 \psi^2(t) \right) \cos(2(\gamma(R)-\alpha (R,t)))
	+2 \psi(t) \left(\lambda^3(t)+R^2 \psi^2(t)+1\right)\Bigg]W_5 \nonumber
	\,.
\end{align}
The only nontrivial equilibrium equation is written as 
\begin{equation}
\begin{aligned}
	\frac{\partial}{\partial R}\sigma^{rr}(R,t)=f(R,t) \,,
\end{aligned}
\end{equation}
where
\begin{align}
	f(R,t)
	&=\frac{2 R\, \psi^2(t)}{\lambda(t)}\,W_1
	+\frac{2 \left[R \,\psi(t)\sin(\gamma(R)-\alpha (R,t))
	+\cos(\gamma(R)-\alpha(R,t))\right]^2}{\lambda(t) R} \,W_4 \nonumber\\
	& \quad \frac{4}{\lambda^2(t) R} 
	\left[R \,\psi(t) \sin(\gamma(R)-\alpha (R,t))+\cos(\gamma(R)-\alpha(R,t))\right]\\
	& \quad \times \left[R \,\psi(t) \left(\lambda^3(t)+R^2 \,\psi^2(t)+1\right) \sin(\gamma(R)-\alpha(R,t))
	+\left(R^2 \,\psi^2(t)+1\right) \cos(\gamma(R)-\alpha(R,t))\right]	W_5\,. \nonumber
\end{align}
Using the boundary condition $\sigma^{rr}(R_0,t)=0$, one writes
\begin{equation}
	\sigma^{rr}(R,t) = -\int_{R}^{R_0}f(\xi,t))\,d\xi\,.
\end{equation}
This, in particular, implies that
\begin{equation}
	-p(R,t) = -\int_{R}^{R_0}f(\xi,t))\,d\xi-2\lambda^{-1}(t)\,W_1+2\lambda(t)\,W_2	\,.
\end{equation}
The other two diagonal components of stress are simplified to read	
\begin{equation}
\begin{aligned}
	\sigma^{\theta\theta}(R,t) & = -\frac{\lambda(t)}{R^2}\int_{R}^{R_0}f(\xi,t))\,d\xi
	+2\psi^2(t)\,W_1 \\
	& \quad +\frac{2}{R^2}\left[R \psi(t) \sin(\gamma(R)-\alpha (R,t))
	+\cos(\gamma(R)-\alpha (R,t))\right]^2\,W_4\\
	& \quad +\frac{4}{\lambda(t) R^2}\left[R \psi(t) \sin(\gamma(R)-\alpha (R,t))
	+\cos(\gamma(R)-\alpha (R,t))
	\right]	\,,
\end{aligned}
\end{equation}
and
\begin{equation}
\begin{aligned}
	\sigma^{zz}(R,t) & = -\int_{R}^{R_0}f(\xi,t))\,d\xi
	+2\left[\lambda^2(t)-\lambda^{-1}(t)\right]W_1
	+2\frac{\lambda^3(t)-R^2\psi^2(t)-1}{\lambda^2(t)} \,W_2\\
	&\quad  +2 \lambda^2(t) \sin^2(\gamma(R)-\alpha(R,t))\,W_4 \\
	& \quad +2 \lambda(t)  \left[2\left(\lambda^3(t)+R^2 \psi^2(t)\right) 
	\sin^2(\gamma(R)-\alpha(R,t))+R \psi(t)  \sin(2(\gamma(R)-\alpha(R,t)))\right]\,W_5 	\,.
\end{aligned}
\end{equation}
For a force-control loading at the two ends of the bar ($Z=0,L$), the axial force and torque needed to maintain the deformation are 
\begin{align}
	F(t) & =2\pi \int_{0}^{R_0}\bar{P}^{zZ}(R,t)R\,dR=0\,, \label{Axial-Force}\\
	T(t) &=2\pi \int_{0}^{R_0}\bar{P}^{\theta Z}(R,t)R^2\,dR=2\pi \int_{0}^{R_0} P^{\theta Z}(R,t)\,r(R,t)R^2\,dR\,,
	\label{Torque}
\end{align}
where $\bar{P}^{zZ}=P^{zZ}$ is the $zZ$-component of the first Piola-Kirchhoff stress and $\bar{P}^{\theta Z}=rP^{\theta Z}$ is the physical $\theta Z$ component of the first Piola-Kirchhoff stress.  
Noting that $P^{zZ}=\lambda^{-1}\sigma^{zz}$ and $P^{\theta Z}=\lambda^{-1}\sigma^{\theta z}$, we have
\begin{equation}
\begin{aligned}
	\bar{P}^{zZ}(R,t) &=  -\frac{1}{\lambda(t)} \int_{R}^{R_0} f(\xi,t)\,d\xi
	+\frac{2 \left(\lambda^3(t)-1\right)}{\lambda^2(t)} W_1
	+\frac{2}{\lambda^3(t)} \left(\lambda^3(t)-R^2 \psi^2(t)-1\right) W_2 \\
	& \quad +2 \lambda(t) \sin^2(\gamma(R)-\alpha (R,t))\,W_4 \\
	& \quad +2 \left[2\left(\lambda^3(t)+R^2 \psi^2(t)\right) \sin^2(\gamma(R)-\alpha(R,t))
	 +R\psi(t) \sin(2(\gamma(R)-\alpha(R,t)))\right] W_5\,, \\
	\bar{P}^{\theta Z}(R,t) &= \frac{2R\,\psi(t)}{\lambda^{\frac{1}{2}}(t)} \, W_1
	+\frac{2R\,\psi(t)}{\lambda^{\frac{3}{2}}(t)}\, W_2  
	+\lambda^{\frac{1}{2}}(t) \left[R \psi-R\,\psi \cos2(\gamma-\alpha)+\sin2(\gamma-\alpha)\right] 
	W_4 \\
	& \quad +\lambda^{-\frac{3}{2}}(t) 
	\Big\{\left(1+\lambda ^3+3 R^2 \psi ^2\right) \sin2(\gamma-\alpha) \\
	& \qquad\qquad\qquad+2 R \psi  \left[-\left(\lambda^3+R^2 \psi^2\right) 
	\cos2(\gamma-\alpha)+\lambda^3
	+R^2 \psi^2+1\right]\Big\} W_5 \,.
\end{aligned}
\end{equation}
Thus, Eq.~\eqref{Axial-Force} is simplified to read
\begin{align}
	& -\frac{1}{\lambda(t)} \int_{0}^{R_0} R \int_{R}^{R_0} f(\xi,t)\,d\xi\,dR 
	+\frac{2 \left(\lambda^3(t)-1\right)}{\lambda^2(t)} \int_{0}^{R_0} W_1\,R\,dR
	+\frac{2}{\lambda^3(t)} \int_{0}^{R_0}\left(\lambda^3(t)-R^2 \psi^2(t)-1\right) W_2\,R\,dR \nonumber \\
	&+2 \lambda(t) \int_{0}^{R_0} \sin^2(\gamma(R)-\alpha (R,t))\,W_4 R\,dR \\
	&+2 \int_{0}^{R_0} 
	 \left[2\left(\lambda^3(t)+R^2 \psi^2(t)\right) \sin^2(\gamma(R)-\alpha(R,t))
	 +R\psi(t) \sin(2(\gamma(R)-\alpha(R,t)))\right] R\,W_5\,dR = 0 \,. \nonumber
\end{align}
Similarly, Eq.~\eqref{Torque} is rewritten as
\begin{align}
	& \frac{2\psi(t)}{\lambda^{\frac{1}{2}}(t)} \int_{0}^{R_0} W_1 R^3\,dR
	+\frac{2\psi(t)}{\lambda^{\frac{3}{2}}(t)} \int_{0}^{R_0} W_2\,R^3\,dR 
	+\frac{1}{\lambda^{\frac{1}{2}}(t)} \int_{0}^{R_0} R^2\, W_4 
	\left[ R\psi-R\psi \cos2(\gamma-\alpha)+\sin2(\gamma-\alpha) \right]\,dR \nonumber \\
	&+\frac{1}{\lambda^{\frac{3}{2}}(t)}  \int_{0}^{R_0} 
	 \Big\{\left(\lambda^3+3 R^2 \psi ^2+1\right) \sin2(\gamma-\alpha) \\
	 & \qquad\qquad\qquad\qquad +2 R \psi  \left[-\left(\lambda^3+R^2 \psi^2\right) 
	 \cos2(\gamma-\alpha)+\lambda^3+R^2 \psi^2+1\right]\Big\} 
	 R^2\,W_5\,dR = \frac{T(t)}{2\pi} \,. \nonumber
\end{align}

The remodeling equation \eqref{Kinetic-N-M2} with $\mathbf{M}=\mathbf{N}^{\mathbf{C}}_{\text{max}}$ is simplified to read
\begin{equation} \label{KineticEq1-Example3}
\begin{aligned} 
	K\dot{\alpha} &=
	\frac{\kappa_M	}{4 R^2 \psi^2} \left[1-\lambda^3-R^2 \psi^2+\sqrt{\left(\lambda^3-1\right)^2
	+R^4 \psi^4+2  \left(\lambda^3+1\right) R^2 \psi^2}\right] \\
	&\qquad\qquad \times 
	\left[\left(\lambda^3+R^2 \psi^2-1\right) \sin 2(\gamma-\alpha)
	+2 R \psi  \cos2(\gamma-\alpha)\right]\\
	& \quad+ \frac{1}{\lambda ^2}\left[\lambda  W_4+ \left(\lambda ^3+R^2 \psi ^2+1\right)W_5 \right]
   \left[\left(\lambda^3+R^2 \psi^2-1\right) \sin2(\gamma-\alpha)+2 R \psi 
   \cos 2(\gamma-\alpha)\right]
	 \,.
\end{aligned}
\end{equation}
Similarly, the remodeling equation \eqref{Kinetic-N-M1} with $\mathbf{M}=\mathbf{N}^{\mathbf{C}}_{\text{max}}$ is simplified to read\footnote{Our numerical results show that the two remodeling equations \eqref{KineticEq1-Example3} and \eqref{KineticEq2-Example3} give very similar results. We use \eqref{KineticEq1-Example3} in our numerical examples.}
\begin{equation} \label{KineticEq2-Example3}
\begin{aligned} 
	K\dot{\alpha} &=
	\frac{\kappa_M}{2R\psi} 
	\left[\left(-1+\lambda ^3+R^2 \psi ^2-\sqrt{\left(\lambda ^3-1\right)^2+R^4 \psi ^4+2
	\left(\lambda ^3+1\right) R^2 \psi ^2} \right)\sin(\gamma-\alpha)+2 R\psi\cos(\gamma-\alpha)\right] \\
	&\quad \times \operatorname{sgn}\left[\frac{1}{2R \psi}\left(1-\lambda^3-R^2 \psi^2
	+\sqrt{R^4\psi^4+2 R^2 \left(\lambda^3+1\right) \psi^2+\left(\lambda^3-1\right)^2} \right) 
	\cos(\gamma-\alpha)+\sin(\gamma-\alpha)\right] \\
	& \quad+ \frac{1}{\lambda ^2}\left[\lambda  W_4+ \left(\lambda ^3+R^2 \psi ^2+1\right)W_5 \right]
	\left[\left(\lambda^3+R^2 \psi^2-1\right) \sin2(\gamma-\alpha)+2 R\psi \cos 2(\gamma-\alpha)\right]
	 \,.
\end{aligned}
\end{equation}


In summary we have the following two problems:
\paragraph{Twist-control loading:} For a given $\psi(t)$ in the time interval $[0,T]$ solve the following problem
\begin{align}
\begin{dcases}
	-\frac{1}{\lambda(t)}\int_{0}^{R_0} R \int_{R}^{R_0} f(\xi,t)\,d\xi\,dR 
	+\frac{2 \left(\lambda^3(t)-1\right)}{\lambda^2(t)} \int_{0}^{R_0} W_1\,R\,dR
	+\frac{2}{\lambda^3(t)} \int_{0}^{R_0}\left(\lambda^3(t)-R^2 \psi^2(t)-1\right) W_2\,R\,dR \nonumber \\
	\quad +2 \lambda(t) \int_{0}^{R_0} \sin^2(\gamma(R)-\alpha (R,t))\,W_4 R\,dR \\
	\quad +2 \int_{0}^{R_0} 
	 \left[2\left(\lambda^3(t)+R^2 \psi^2(t)\right) \sin^2(\gamma(R)-\alpha(R,t))
	 +R\psi(t) \sin(2(\gamma(R)-\alpha(R,t)))\right] R\,W_5\,dR = 0\,, \\
	K\dot{\alpha} =
	\frac{\kappa_M	}{4 R^2 \psi^2} \left[1-\lambda^3-R^2 \psi^2+\sqrt{\left(\lambda^3-1\right)^2
	+R^4 \psi^4+2  \left(\lambda^3+1\right) R^2 \psi^2}\right] \\
	\qquad\qquad \times 
	\left[\left(\lambda^3+R^2 \psi^2-1\right) \sin 2(\gamma-\alpha)
	+2 R \psi  \cos2(\gamma-\alpha)\right]\\
	 \quad+ \frac{1}{\lambda ^2}\left[\lambda  W_4+ \left(\lambda ^3+R^2 \psi ^2+1\right)W_5 \right]
   \left[\left(\lambda^3+R^2 \psi^2-1\right) \sin2(\gamma-\alpha)+2 R \psi 
   \cos 2(\gamma-\alpha)\right]\,, \\
	\lambda(0)=1\,,~ \alpha(R,0)=0	\,. 
\end{dcases}
\end{align}
For the material \eqref{MR-Material} this is simplified to
\begin{align}
\begin{dcases}
	-\frac{1}{\lambda(t)}\int_{0}^{R_0} R \int_{R}^{R_0} f(\xi,t)\,d\xi\,dR 
	+\frac{C_1 \left(\lambda^3(t)-1\right)}{\lambda^2(t)} (R_0^2-R^2)
	+\frac{C_2(R^2-R_0^2)}{2 \lambda^3}\left[2-2 \lambda^3+\psi^2 \left(R^2+R_0^2\right)\right] \nonumber \\
	\quad +2\mu_1 \lambda(t) \int_{0}^{R_0} \sin^2(\gamma(R)-\alpha (R,t))\,(I_4-1) R\,dR \\
	\quad +2\mu_2 \int_{0}^{R_0} 
	 \left[2\left(\lambda^3(t)+R^2 \psi^2(t)\right) \sin^2(\gamma(R)-\alpha(R,t))
	 +R\psi(t) \sin(2(\gamma(R)-\alpha(R,t)))\right] R\,(I_5-1)\,dR = 0\,, \\
	K\dot{\alpha} =
	\frac{\kappa_M	}{4 R^2 \psi^2} \left[1-\lambda^3-R^2 \psi^2+\sqrt{\left(\lambda^3-1\right)^2
	+R^4 \psi^4+2  \left(\lambda^3+1\right) R^2 \psi^2}\right] \\
	\qquad\qquad \times 
	\left[\left(\lambda^3+R^2 \psi^2-1\right) \sin 2(\gamma-\alpha)
	+2 R \psi  \cos2(\gamma-\alpha)\right]\\
	 \quad+ \frac{1}{\lambda ^2}\left[\lambda  W_4+ \left(\lambda ^3+R^2 \psi ^2+1\right)W_5 \right]
   \left[\left(\lambda^3+R^2 \psi^2-1\right) \sin2(\gamma-\alpha)+2 R \psi 
   \cos 2(\gamma-\alpha)\right]\,, \\
	\lambda(0)=1\,,~ \alpha(R,0)=0	\,. 
\end{dcases}
\end{align}

\paragraph{Torque-control loading:}For a given torque $T(t)$ in the time interval $[0,T]$ solve the following problem
\begin{align}
\begin{dcases}
	-\frac{1}{\lambda(t)}\int_{0}^{R_0} R \int_{R}^{R_0} f(\xi,t)\,d\xi\,dR 
	+\frac{2 \left(\lambda^3(t)-1\right)}{\lambda^2(t)} \int_{0}^{R_0} W_1\,R\,dR
	+\frac{2}{\lambda^3(t)} \int_{0}^{R_0}\left(\lambda^3(t)-R^2 \psi^2(t)-1\right) W_2\,R\,dR 
	\nonumber \\
	\quad +2 \lambda(t) \int_{0}^{R_0} \sin^2(\gamma(R)-\alpha (R,t))\,W_4 R\,dR \\
	\quad +2 \int_{0}^{R_0} 
	 \left[2\left(\lambda^3(t)+R^2 \psi^2(t)\right) \sin^2(\gamma(R)-\alpha(R,t))
	 +R\psi(t) \sin(2(\gamma(R)-\alpha(R,t)))\right] R\,W_5\,dR = 0\,, \\
	\frac{2\psi(t)}{\lambda^{\frac{1}{2}}(t)} \int_{0}^{R_0} W_1 R^3\,dR
	+\frac{2\psi(t)}{\lambda^{\frac{3}{2}}(t)} \int_{0}^{R_0} W_2\,R^3\,dR 
	+\frac{1}{\lambda^{\frac{1}{2}}(t)} \int_{0}^{R_0} R^2\, W_4 
	\left[ R\psi-R\psi \cos2(\gamma-\alpha)+\sin2(\gamma-\alpha) \right]\,dR \nonumber \\
	+\frac{1}{\lambda^{\frac{3}{2}}(t)}  \int_{0}^{R_0} 
	 \Big\{\left(\lambda^3+3 R^2 \psi ^2+1\right) \sin2(\gamma-\alpha) \\
	  \qquad\qquad\qquad\qquad +2 R \psi  \left[-\left(\lambda^3+R^2 \psi^2\right) 
	 \cos2(\gamma-\alpha)+\lambda^3+R^2 \psi^2+1\right]\Big\} 
	 R^2\,W_5\,dR = \frac{T(t)}{2\pi}\,, \\ \nonumber
	 K\dot{\alpha} =
	\frac{\kappa_M	}{4 R^2 \psi^2} \left[1-\lambda^3-R^2 \psi^2+\sqrt{\left(\lambda^3-1\right)^2
	+R^4 \psi^4+2  \left(\lambda^3+1\right) R^2 \psi^2}\right] \\
	\qquad\qquad \times 
	\left[\left(\lambda^3+R^2 \psi^2-1\right) \sin 2(\gamma-\alpha)
	+2 R \psi  \cos2(\gamma-\alpha)\right]\\
	\quad+ \frac{1}{\lambda ^2}\left[\lambda  W_4+ \left(\lambda ^3+R^2 \psi ^2+1\right)W_5 \right]
   \left[\left(\lambda^3+R^2 \psi^2-1\right) \sin2(\gamma-\alpha)+2 R \psi 
   \cos 2(\gamma-\alpha)\right]\,, \\
	\lambda(0)=1\,,~\psi(0)=0\,,~\alpha(R,0)=0	\,. 
\end{dcases}
\end{align}
For the material \eqref{MR-Material} this is simplified to
\begin{align}
\begin{dcases}
	-\frac{1}{\lambda(t)}\int_{0}^{R_0} R \int_{R}^{R_0} f(\xi,t)\,d\xi\,dR 
	+\frac{C_1 \left(\lambda^3(t)-1\right)}{\lambda^2(t)} (R_0^2-R^2)
	+\frac{C_2(R^2-R_0^2)}{2 \lambda^3}\left[2-2 \lambda^3+\psi^2 \left(R^2+R_0^2\right)\right] 
	\nonumber \\
	\quad +2\mu_1 \lambda(t) \int_{0}^{R_0} \sin^2(\gamma(R)-\alpha (R,t))\,(I_4-1) R\,dR \\
	\quad +2\mu_2 \int_{0}^{R_0} 
	 \left[2\left(\lambda^3(t)+R^2 \psi^2(t)\right) \sin^2(\gamma(R)-\alpha(R,t))
	 +R\psi(t) \sin(2(\gamma(R)-\alpha(R,t)))\right] R\,(I_5-1)\,dR = 0\,, \\
	\frac{\psi(t)}{2\lambda^{\frac{1}{2}}(t)}\left(C_1+\frac{C_1}{\lambda(t)}\right)(R_0^4-R^4) 
	+\frac{1}{\lambda^{\frac{1}{2}}(t)} \int_{R}^{R_0} R^2\, W_4 
	\left[ R\psi-R\psi \cos2(\gamma-\alpha)+\sin2(\gamma-\alpha) \right]\,dR \nonumber \\
	+\frac{1}{\lambda^{\frac{3}{2}}(t)}  \int_{0}^{R_0} 
	 \Big\{\left(\lambda^3+3 R^2 \psi ^2+1\right) \sin2(\gamma-\alpha) \\
	  \qquad\qquad\qquad\qquad +2 R \psi  \left[-\left(\lambda^3+R^2 \psi^2\right) 
	 \cos2(\gamma-\alpha)+\lambda^3+R^2 \psi^2+1\right]\Big\} 
	 R^2\,W_5\,dR = \frac{T(t)}{2\pi}\,, \\ \nonumber
	K\dot{\alpha} =
	\frac{\kappa_M	}{4 R^2 \psi^2} \left[1-\lambda^3-R^2 \psi^2+\sqrt{\left(\lambda^3-1\right)^2
	+R^4 \psi^4+2  \left(\lambda^3+1\right) R^2 \psi^2}\right] \\
	\qquad\qquad \times 
	\left[\left(\lambda^3+R^2 \psi^2-1\right) \sin 2(\gamma-\alpha)
	+2 R \psi  \cos2(\gamma-\alpha)\right]\\
	\quad+ \frac{1}{\lambda ^2}\left[\lambda  W_4+ \left(\lambda ^3+R^2 \psi ^2+1\right)W_5 \right]
   \left[\left(\lambda^3+R^2 \psi^2-1\right) \sin2(\gamma-\alpha)+2 R \psi 
   \cos 2(\gamma-\alpha)\right]\,, \\
	\lambda(0)=1\,,~ \alpha(R,0)=0	\,. 
\end{dcases}
\end{align}

\paragraph{Numerical results.}

\begin{figure}[h!]
	\begin{center}
		\includegraphics[width=0.9\textwidth]{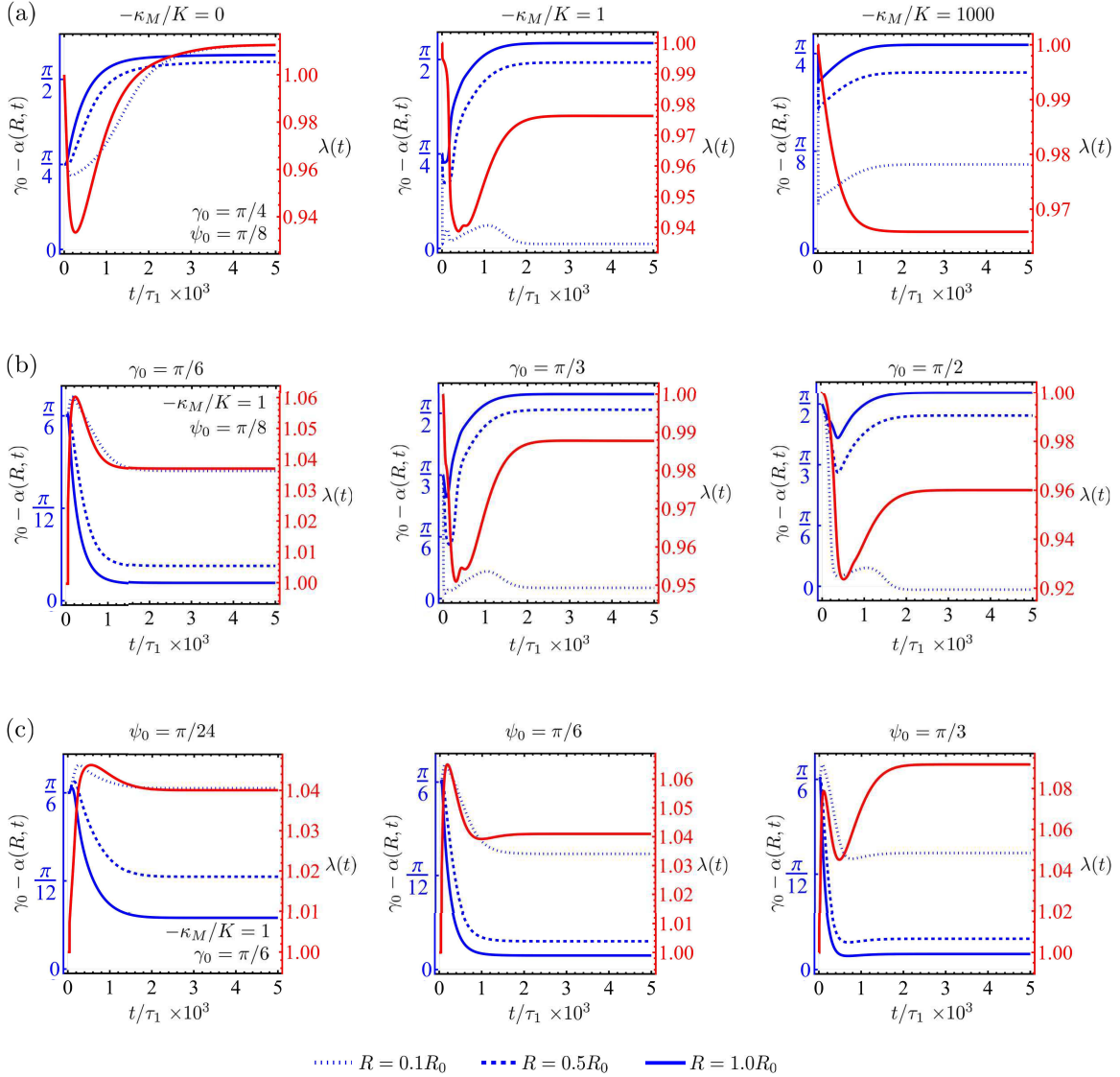}
	\end{center}
	\vspace{-0.2cm} \caption{Finite torsion of a transversely isotropic cylindrical bar under twist-control loading. The applied twist is $\psi(t)=\psi_0\,\erf\big(\frac{t}{t_0}\big)$ with $t_0=1$. Remodeled fiber orientation $\gamma_0-\alpha(R,t)$ is plotted on the left y-axis for three values of $R$ and the observed longitudinal extension $\lambda(t)$ is plotted on the right y-axis as a function of $t/\tau_1$, where $\tau_1$ is the material's relaxation time. Part (a) shows the effect of the ratio $-\kappa_M/K$, part (b) shows the effect of the initial fiber orientation, and part (c) shows the results for different values of maximum twist $\psi_0$.  }
	\label{Fig4-3-1}
\end{figure}

For the numerical parametric study, we use the same material constants as the previous two examples. Furthermore, we again assume $\gamma(R)=\gamma_0$. However, with this choice for $\gamma(R)$, unlike previous examples, the remodeling variable $\alpha$ will still depend on the spatial variable $R$ in addition to $t$. 

We first consider a twist-control loading $\psi(t)=\psi_0\,\erf\big(\frac{t}{t_0}\big)$ with $t_0=1$. Similar to the previous examples, the effect of $-\kappa_M/K$,  $\gamma_0$, and the maximum twist $\psi_0$ on the remodeled fiber orientation $\gamma_0-\alpha(R,t)$ is studied and is shown in Fig.~\ref{Fig4-3-1} for three values of $R$, namely, $R=0.1, 0.5, 1.0$ $R_0$. Furthermore, we are also interested in studying the effects of various parameters in the model on the longitudinal extension of the twisted bar. Those results are also included in Fig.~\ref{Fig4-3-1}.

\begin{figure}[h!]
	\begin{center}
		\includegraphics[width=4.5 in]{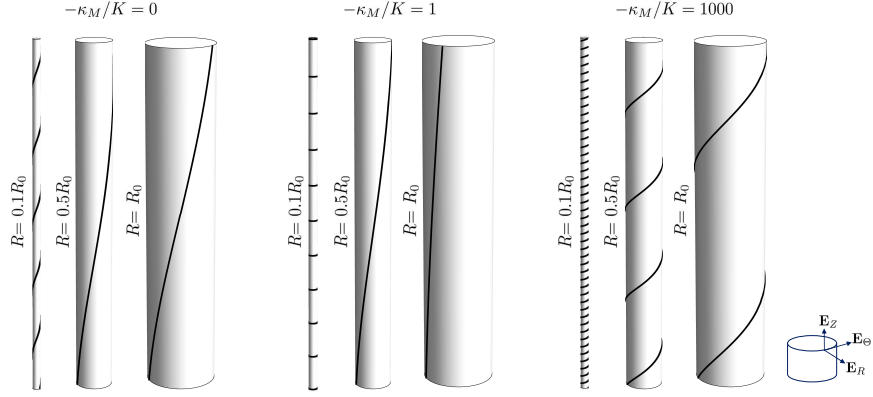}
	\end{center}
	\vspace{-0.2cm} \caption{Remodeled fiber orientation $\gamma_0-\alpha(R,t)$ for a transversely isotropic cylindrical bar under torsion at three values of $R$, namely, $R=0.1, 0.5, 1.0$ $R_0$ for three values of ratio $-\kappa_M/K$.}
	\label{Fig4-3-2}
\end{figure}

A variety of interesting behavior is observed. First, irrespective of the choice of values for the parameters, a spatially inhomogeneous fiber orientation is achieved after remodeling. Second, the results in Fig.~\ref{Fig4-3-1}  show that remodeled fiber orientation is \emph{typically} larger, that is, they align more longitudinally, for larger values of initial fiber orientation, smaller values of $-\kappa_M/K$, and smaller values of $\psi_0$. Moreover, the fibers also align more longitudinally for larger values of $R$ in the cylinder as also shown visually in Fig.~\ref{Fig4-3-2}. However, the spatial inhomogeneity means that there are exceptions where the above mentioned trends are not followed as visible in both Fig.~\ref{Fig4-3-1} and Fig. \ref{Fig4-3-2}. Third, as shown in Fig.~\ref{Fig4-3-1}(a), for large values of $-\kappa_M/K$---when the remodeling energy is dominant---the fibers remodel such that the bar unexpectedly shortens in length instead of elongating. For $-\kappa_M/K=10^3$, a maximum stretch of 0.965 is observed compared to a stretch of 1.015 for $-\kappa_M/K=0$. This effect is strongest for larger values of initial fiber orientation as further highlighted in Fig. \ref{Fig4-3-1}(b). 

\begin{figure}[h!]
	\begin{center}
	\vskip 0.3in
		\includegraphics[width=\textwidth]{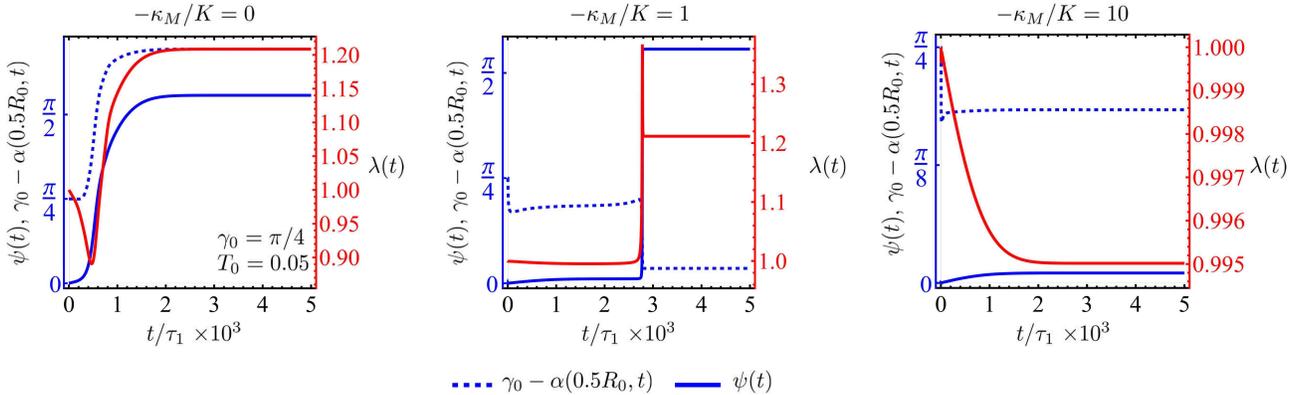}
	\end{center}
	\vspace{-0.2cm} \caption{Finite torsion of a transversely isotropic cylindrical bar under torque-control loading. The applied torque is $T(t)=T_0\,\erf\big(\frac{t}{t_0}\big)$ with $t_0=1$. Remodeled fiber orientation $\gamma_0-\alpha(R,t)$ for $R=0.5$ $R_0$ and the observed twist $\psi(t)$ are plotted on the left y-axis  and the observed longitudinal extension $\lambda(t)$ is plotted on the right y-axis as a function of $t/\tau_1$, where $\tau_1$ is the material's relaxation time, for different ratios of $-\kappa_M/K$.}
	\label{Fig4-3-3}
\end{figure}

We next consider a torque-control loading in the same form $T(t)=T_0 \,\erf\big(\frac{t}{t_0}\big)$ with $t_0=1$. The dependence of the magnitude of loading and initial fiber orientation on remodeling variable and longitudinal extension is similar for torque-control loading as for twist-control loading. The effect of the ratio $-\kappa_M/K$ on the observed twist $\psi(t)$ is more interesting. Fig.~\ref{Fig4-3-3} shows the remodeled fiber orientation at $R=0.5R_0$, $\psi(t)$ and $\lambda(t)$ for three values of $-\kappa_M/K$. We observe that, as expected, for large values of $-\kappa_M/K$, additional torsional stiffness provided by the remodeling of fibers in the direction of maximum strain results in a close to zero value of observed twist. However, for intermediate values of $-\kappa_M/K$,  a remodeling instability ensues once the applied torque stops increasing resulting in a sharp jump in the remodeled fiber orientation, longitudinal extension, and twist.

\begin{figure}[h!]
	\begin{center}
	\vskip 0.3in
		\includegraphics[width=0.8\textwidth]{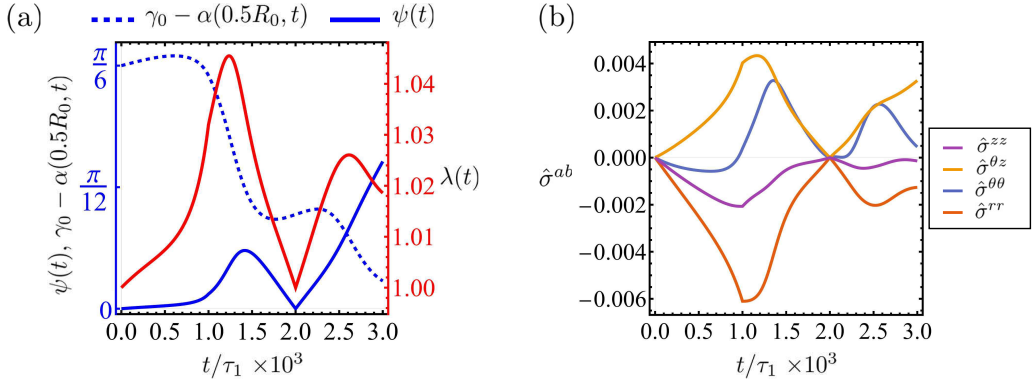}
	\end{center}
	\vspace{-0.2cm} \caption{Evolution of (a) twist $\psi(t)$ and remodeled fiber  orientation $\gamma_0-\alpha(0.5 R_0, t)$, and (b) stress components $\hat{\sigma}^{ab} (0.5 R_0, t)$, during one and a half cycles of torque-control loading-unloading.}
	\label{Fig4-3-4}
\end{figure}

Lastly, we consider a cycle of torque loading-unloading followed by a second phase of loading similar to (\ref{F-loadingunloading}) considered in Example 1. The results for observed twist, remodeled fiber orientation, stretch and stress are presented in Fig.~\ref{Fig4-3-4} and once again show a loading-history-dependence but no residual stresses (see Remark \ref{Stress-Free}). Notably, the observed twist $\psi(t)$ is much higher at the end of second phase of loading than at the end of first phase.

\section{Linearized Remodeling Mechanics}  \label{Sec:LinearizedTheory}

In this section we linearize the governing equations of the nonlinear remodeling theory. 
The motivation for linearization of the remodeling theory is applications in which strains are small, e.g., bone remodeling.
For the sake of simplicity, we restrict the analysis to isotropic solids.
Let us consider a stress-free body $\mathcal{B}$ with its flat material metric $\mathring{\mathbf{G}}$. 
We linearize with respect to the initial deformation map $\mathring{\varphi}=\iota$ and the trivial remodeling tensor $\mathring{\Fr}=\mathbf{I}$, where $\iota$ is the inclusion map, and $\mathbf{I}$ is the identity map on $T_X\mathcal{B}$.\footnote{One can also linearize with respect to a stressed and remodeled body, i.e., the small-on-large theory of remodeling.} Note that $\mathring{\mathbf{F}}=\mathbf{I}$, and $\mathring{\mathbf{C}}^{\flat}=\mathring{\mathbf{G}}$.

\subsection{Linearized kinematics}

Let us consider a one-parameter family of motions and remodeling tensors $\varphi_{\epsilon}$ and $\Fr_{\epsilon}$ such that $\varphi_{\epsilon=0}=\mathring{\varphi}$, and $\Fr_{\epsilon=0}=\mathring{\Fr}$.
The variation fields are defined as
\begin{equation}
	\delta\varphi=\frac{d}{d\epsilon}\Big|_{\epsilon=0}\varphi_{\epsilon}\,,\qquad
	\delta\Fr=\frac{d}{d\epsilon}\Big|_{\epsilon=0}\Fr_{\epsilon}
	\,.
\end{equation}
Recall that $\Fr_{\epsilon}:T_X\mathcal{B}\to T_X\mathcal{B}$ for all $\epsilon$, and hence the above derivative is well defined. 
Let $\mathbf{U}=\delta\varphi$. The vector $\mathbf{u}=\delta\varphi\circ\phio$ is the displacement field of the classical theory of linear elasticity. 
Similarly, we call $\UFr=\delta\Fr$ the \emph{remodeling displacement}, which is a material $\binom{1}{1}$-tensor. Its spatial counterpart is denoted by $\uFr={\phio}_*\UFr=\UFr\circ \phio^{-1}$.

The right Cauchy-Green deformation tensor $\delta \mathbf{C}^\flat$ is linearized as
\begin{equation}
	\delta\mathbf{C}^\flat
	= \varphi_{t,\epsilon=0}^* \mathbf{L}_{\mathbf{u}}\mathbf{g}
	= \phio^*\left(\nabla^{\mathbf{g}} \mathbf{u}^\flat + \left[ \nabla^{\mathbf{g}} 
	\mathbf{u}^\flat \right]^{\star}\right)
	= 2 \phio^*\boldsymbol \epsilon=2\boldsymbol{\varepsilon}\,,
\end{equation}
where $\boldsymbol \epsilon = \frac{1}{2}( \nabla^{\mathbf g} \mathbf u^\flat + [ \nabla^{\mathbf g} \mathbf u^\flat ]^{\star} )$ is the linearized total strain, $\mathbf{L}$ is the Lie derivative operator, and  $\boldsymbol{\varepsilon}=\phio^*\boldsymbol{\epsilon}$.	
The linearization of the elastic right Cauchy-Green strain $\delta \Ce$ is calculated as
\begin{equation}
	\delta \Ce^\flat =  \left.\left(2\Fe^\star \boldsymbol \epsilon \Fe - \Ce^\flat(\delta\Fr)\Fr^{-1} 
	- \Fr^{-\star}(\delta \Fr)^\star\Ce^\flat\right)\right|_{\epsilon=0}
	= 2\phio^*\boldsymbol \epsilon - \mathring{\mathbf{G}} \UFr - \UFr^\star \mathring{\mathbf{G}}
	= 2\phio^*\left(\boldsymbol \epsilon - \epsr\right)=2\phio^*\epse=2\vepse\,,
\end{equation}
where $\vepsr=\phio^*\epsr=\phio^*\epsr = \frac{1}{2}\left(\mathring{\mathbf{G}} \UFr+\UFr^{\star} \mathring{\mathbf{G}}\right)$ is the linearized remodeling strain, and $\vepse=\phio^*\epse$. 
The linearized elastic strain is defined as $\epse=\frac{1}{2}{\phio}_*\delta \Ce^\flat$.  
It is observed that the linearized strain is additively decomposed as $\boldsymbol\epsilon=\epse+\epsr$.
The linearized Jacobian, $\delta J$ is calculated as
\begin{equation}
	\delta J
	= \frac{1}{2}J|_{\epsilon=0}\, \mathbf C^{-1}_{\epsilon=0}\!:\!\delta\mathbf C
	= \Jo {\phio}^*\mathbf g^\sharp \!:\! {\phio}^*\boldsymbol \epsilon
	= \Jo \mathbf g^\sharp \!:\! \boldsymbol \epsilon
	= \Jo \operatorname{tr} \boldsymbol \epsilon\,.
\end{equation}
From conservation of mass $\rho J=\rho_o$, linearization of the spatial mass density is calculated as $\delta \rho = -\rhoo \operatorname{tr} \boldsymbol \epsilon$.
Knowing that $\det\Fr_{\epsilon}=1$, we can write
\begin{equation}
	\delta\Jr=\frac{d}{d\epsilon}\Big|_{\epsilon=0}\det\Fr_{\epsilon}
	=\Fr^{-1}_{\epsilon}\Big|_{\epsilon=0}\!:\!\delta\Fr
	=\mathbf{I}\!:\!\delta\Fr=\operatorname{tr}\delta\Fr
	= \operatorname{tr} \UFr = \operatorname{tr} \epsr=0\,,
\end{equation}
i.e., both the remodeling displacement and strain are traceless.
The material metric $\mathbf{G}=\Fr^*\mathring{\mathbf{G}}=\Fr^{\star}\mathring{\mathbf{G}}\Fr$ is linearized as follows.
\begin{equation}
	\delta\mathbf{G}=\frac{d}{d\epsilon}\Big|_{\epsilon=0}\mathbf{G}_{\epsilon}
	=\delta\Fr^{\star}\,\mathring{\mathbf{G}}\,\Fr_{\epsilon}\Big|_{\epsilon=0}
	+\Fr^{\star}_{\epsilon}\Big|_{\epsilon=0}\,\mathring{\mathbf{G}}\,\delta\Fr
	=\delta\Fr^{\star}\,\mathring{\mathbf{G}}\,\mathbf{I}+\mathbf{I}\,\mathring{\mathbf{G}}\,\delta\Fr
	=\UFr^{\star}\mathring{\mathbf{G}}+\mathring{\mathbf{G}}\UFr
	=2\vepsr
	\,.
\end{equation}

\subsection{Linearized stress}

For linearization purposes, the convected form of the balance of linear momentum \eqref{Linear-Momentum-Convected} is more convenient as it is entirely written with respect to the reference configuration. In other words, as the parameter $\epsilon$ varies all the terms lie in a fixed tangent space $T_X\mathcal B$. Recall that for an isotropic solid $W=\hat{W}(X,\mathbf{C}^{\flat},\mathbf{G})$.
We first compute the variation of the convected stress tensor as
\begin{equation}
\begin{split}
	\delta\boldsymbol{\Sigma}
	&= \frac{d}{d\epsilon}\left[\frac{2}{J}\frac{\partial \hat{W}}{\partial \mathbf{C}^\flat} \right]
	\Big|_{\epsilon=0}\\
	&= -\delta J\left.\left(\frac{1}{J} \boldsymbol{\Sigma}\right)\right|_{\epsilon=0}
	+\left(\frac{2}{J}\frac{\partial^2 \hat{W}}{\partial \mathbf C^\flat 
	\partial \mathbf C^\flat}\right)\Big|_{\epsilon=0} \!:\! \delta \mathbf{C}^\flat
	 +\left(\frac{2}{J}\frac{\partial^2 \hat{W}}{\partial \mathbf{C}^\flat \partial \mathbf{G}}\right)
	 \Big|_{\epsilon=0}\!:\! \delta \mathbf{G} \\
	&= -(\operatorname{tr} \boldsymbol{\epsilon}) \Sigmao
	+\left.\left(\frac{2}{J}\frac{\partial^2 \hat{W}}{\partial \mathbf{C}^\flat 
	\partial \mathbf{C}^\flat}\right)\right|_{\varphi=\iota,\Fr=\mathbf I} \!:\! \delta \mathbf{C}^\flat 
	+\left(\frac{2}{J}\frac{\partial^2 \hat{W}}{\partial \mathbf{C}^\flat \partial \mathbf{G}}\right)
	\Big|_{\varphi=\iota,\Fr=\mathbf I} \!:\! (\UFr^{\star}\mathring{\mathbf{G}}
	+\mathring{\mathbf{G}}\UFr)\\
	 &=\left.\left(\frac{2}{J}\frac{\partial^2 \hat{W}}{\partial \mathbf{C}^\flat 
	\partial \mathbf{C}^\flat}\right)\right|_{\varphi=\iota,\Fr=\mathbf I} \!:\! \delta \mathbf{C}^\flat 
	+\left(\frac{4}{J}\frac{\partial^2 \hat{W}}{\partial \mathbf{C}^\flat \partial \mathbf{G}}\right)
	\Big|_{\varphi=\iota,\Fr=\mathbf I} \!:\! \vepsr
	 \,,
\end{split}
\end{equation}
where $\Sigmao=\mathbf{0}$ was used.
Let us define the following fourth-order material elasticity tensors
\begin{equation}
	\boldsymbol{\mathbb{c}} :=4\phio_*\left[\frac{\partial^2 \hat{W}}
	{\partial \mathbf{C}^\flat \partial \mathbf{C}^\flat}\right]_{\varphi=\iota,\Fr=\mathbf I}
	\,,\quad
	\ccr :=4\phio_*\left[\frac{\partial^2 \hat{W}}{\partial \mathbf{C}^\flat \partial \mathbf{G}}
	\right]_{\varphi=\iota,\Fr=\mathbf I}\,.
\end{equation}
Thus
\begin{equation}
	\delta\boldsymbol \Sigma
	= \phio^*\left(\boldsymbol{\mathbb c} \!:\! \boldsymbol\epsilon +\ccr \!:\!  \epsilonr \right)
	=\boldsymbol{\mathbb C} \!:\! \boldsymbol\varepsilon +\CCr \!:\!  \vepsr \,,
\end{equation}
where $\CCr=\phio^*\ccr$, and $\boldsymbol{\mathbb{C}}=\phio^*\boldsymbol{\mathbb{c}}$.
Material covariance of the energy function \eqref{Material-Covariance} implies that \citep{Lu2000}
\begin{equation} \label{Infinitesimal-MC}
	\frac{\partial \hat{W}}{\partial \mathbf{C}^\flat}\cdot\mathbf{C}^{\flat}
	+\frac{\partial \hat{W}}{\partial \mathbf{G}}\cdot\mathbf{G}=\mathbf{0}\,,\quad
	\text{or~in~components}\,,\qquad
	\frac{\partial \hat{W}}{\partial C_{AM}}C_{MB}+\frac{\partial \hat{W}}{\partial G_{AM}}G_{MB}=0
	\,.
\end{equation}
Using this relation and for a stress-free reference motion one concludes that 
\begin{equation} \label{G-C-Identity}
	\frac{\partial^2 \hat{W}}{\partial\mathbf{G}\partial\mathbf{C}^\flat}
	+\frac{\partial^2 \hat{W}}{\partial\mathbf{C}^\flat\partial\mathbf{C}^\flat}=\mathbf{0}\,.
\end{equation}
Thus, $\CCr=-\boldsymbol{\mathbb C}$ \citep{Ozakin2010}, and hence 
\begin{equation}
	\delta\boldsymbol \Sigma
	= \boldsymbol{\mathbb C} \!:\! \boldsymbol\varepsilon -\boldsymbol{\mathbb C}  \!:\!  \vepsr
	= \boldsymbol{\mathbb C} \!:\! \vepse \,.
\end{equation}
As expected, the linearized stress explicitly depends on the linearized elastic strain.

\subsection{Linearized balance of linear momentum}

In remodeling problems inertial forces can be ignored.
The convected balance of linear momentum \eqref{Linear-Momentum-Convected} in the absence of inertial forces is linearized as
\begin{equation}
	\frac{d}{d\epsilon}\Big|_{\epsilon=0} \left[\operatorname{Div}_{\mathbf C_\epsilon^\flat}
	\boldsymbol \Sigma_\epsilon + \varphi_{t,\epsilon}^* 
	(\rho_\epsilon \mathbf{b}_\epsilon)\right] = \mathbf{0}\,.
\end{equation}
The body force is linearized as
\begin{equation} 
\begin{split}
\frac{d}{d\epsilon}\Big|_{\epsilon=0}\left[\varphi_{t,\epsilon}^*(\rho_\epsilon  \mathbf{b}_\epsilon)\right]
	&= \phio^*(\delta\rho \,\mathring{\mathbf{b}}) + \varphi_{t,\epsilon=0}^*
	\left[\rhoo L_{(d\varphi_{\epsilon}/d\epsilon)}  \mathbf{b}_\epsilon \right]_{\epsilon=0}
	= \phio^*(\delta\rho \,\mathring{\mathbf{b}}) 
	+ \phio^* \big(\rhoo [\mathbf{u},\mathring{\mathbf{b}}] \big)\,.
\end{split}
\end{equation}
The Christoffel symbol $\widetilde\Gamma$ of the Levi-Civita connection of $\mathbf C^\flat$ is linearized as follows
\begin{equation}
\begin{split}
\frac{d}{d\epsilon}\Big|_{\epsilon=0}\left[\widetilde\Gamma_\epsilon^A{}_{KB}\right]
	&= \frac{1}{2}\frac{d}{d\epsilon}\Big|_{\epsilon=0}\left[\mathrm C_\epsilon^{-AL}
	\left({C_\epsilon}_{LB,K}+{C_\epsilon}_{KL,B} - {C_\epsilon}_{KB,L} \right)\right] \\
	&= -\frac{1}{2}\cCo^{-AI}\delta\mathrm C_{IJ}\cCo^{-JL}\left(\cCo_{LB,K}+\cCo_{KL,B} 
	- \cCo_{KB,L} \right)\\
	&\quad+\frac{1}{2}\cCo^{-AL}\left(\delta\mathrm C_{LB,K}+\delta\mathrm C_{KL,B} 
	- \delta\mathrm C_{KB,L} \right)\\
	&=\cFo^{-A}{}_a\,\cFo^b{}_B\,\cFo^k{}_K\left[-2\mathrm g^{ai}\epsilon_{ij}\gamma^j{}_{kb}
	+\mathrm g^{al}\left(\epsilon_{lb,k}+\epsilon_{kl,b}-\epsilon_{kb,l}\right)\right]\\
	&=\cFo^{-A}{}_a\,\cFo^b{}_B\,\cFo^k{}_K\left[\mathrm g^{al}\left(\epsilon_{lb|k}
	+\epsilon_{kl|b}-\epsilon_{kb|l}\right)\right]\\
	&=\cFo^{-A}{}_a\,\cFo^b{}_B\,\cFo^k{}_K \,u^a{}_{|kb}\,.
	\end{split}
\end{equation}
Hence, the divergence term is linearized as
\begin{equation}\label{eq:del_div}
\begin{split}
	\frac{d}{d\epsilon}\left[\operatorname{Div}_{\mathbf C_\epsilon^\flat}
	\boldsymbol \Sigma_\epsilon \right]_{\epsilon=0}
	&= \operatorname{Div}_{\Co^\flat} \delta\boldsymbol \Sigma 
	+ \cSigmao^{KB}\frac{d}{d\epsilon}\left[\widetilde\Gamma_\epsilon^A{}_{KB}
	\right]_{\epsilon=0} \partial_A+ \cSigmao^{AK}\frac{d}{d\epsilon}
	\left[\widetilde\Gamma_\epsilon^B{}_{KB}\right]_{\epsilon=0} \partial_A\\
	&= \operatorname{Div}_{\Co^\flat} \left[\phio^*\left(\boldsymbol{\mathbb c} \!:\! 
	\epse \right)\right]  
	= \phio^*\operatorname{div}\left(\boldsymbol{\mathbb c} \!:\! \epse\right) \,.
\end{split}
\end{equation}
Finally, the linearized balance of linear momentum reads
\begin{equation}
	\operatorname{div}\left(\boldsymbol{\mathbb c} \!:\! \epse \right)
	+ \rhoo \,\delta \mathbf{b} =  \mathbf{0}\,,
\end{equation}
where $\delta \mathbf{b}=[\mathbf{u},\mathring{\mathbf{b}}]=\nabla^{\mathbf{g}}_{\mathbf{u}}\mathring{\mathbf{b}}-\nabla^{\mathbf{g}}_{\mathring{\mathbf{b}}}\mathbf{u}$.

\subsection{Linearized kinetic equation}

The kinetic equation \eqref{Remodeling-Equation-Metric} is linearized as follows.
\begin{equation} \label{Kinetic-Lin}
	\frac{d}{d\epsilon}\Bigg|_{\epsilon=0}\frac{\partial \phi}{\partial \dot{\Fr}_\epsilon}=
	\frac{d}{d\epsilon}\Big|_{\epsilon=0}\left[q_\epsilon\Fr_\epsilon^{-\star}\right]
	-2\frac{d}{d\epsilon}\Big|_{\epsilon=0} \left[\Fr_\epsilon^{-\star}\mathbf{G}_\epsilon
	\frac{\partial \hat{W}}{\partial \mathbf{G}_\epsilon}\right] \,.
\end{equation}
Note that
\begin{equation} 
	\frac{d}{d\epsilon}\Big|_{\epsilon=0}\left[q_\epsilon\Fr_\epsilon^{-\star}\right]
	=\delta q\,\mathbf{I}-\mathring{q}\,\UFr^{-\star}	\,,
\end{equation}
and 
\begin{equation} 
	\frac{d}{d\epsilon}\Big|_{\epsilon=0} \left[\Fr_\epsilon^{-\star}\mathbf{G}_\epsilon\right]
	=\frac{d}{d\epsilon}\Big|_{\epsilon=0} \left[\mathring{\mathbf{G}}\Fr_\epsilon\right]
	=\mathring{\mathbf{G}}\UFr\,,\qquad 
	\frac{d}{d\epsilon}\Big|_{\epsilon=0} 
	\left[\frac{\partial \hat{W}}{\partial \mathbf{G}_\epsilon}\right] 
	=\frac{\partial^2 \hat{W}}{\partial\mathbf{G}\partial\mathbf{C}^\flat}\!:\!\delta\mathbf{C}^\flat
	+\frac{\partial^2 \hat{W}}{\partial\mathbf{G}\partial\mathbf{G}}\!:\!\delta\mathbf{G}
	\,,
\end{equation}
where all the partial derivatives are evaluated at the initial configuration $(\varphi,\Fr)=(\iota,\mathbf I)$.
Thus
\begin{equation} 
\begin{aligned}
	\frac{d}{d\epsilon}\Big|_{\epsilon=0} \left[\Fr_\epsilon^{-\star}\mathbf{G}_\epsilon
	\frac{\partial \hat{W}}{\partial \mathbf{G}_\epsilon}\right]
	&=\mathring{\mathbf{G}}\UFr \,\frac{\partial \hat{W}}{\partial \mathbf{G}}
	+\mathring{\mathbf{G}}\!\cdot\!
	\left[\frac{\partial^2 \hat{W}}{\partial\mathbf{G}\partial\mathbf{C}^\flat}
	\!:\!\delta\mathbf{C}^\flat
	+\frac{\partial^2 \hat{W}}{\partial\mathbf{G}\partial\mathbf{G}}\!:\!\delta\mathbf{G} \right]\\
	& = \vepsr\cdot \frac{\partial \hat{W}}{\partial \mathbf{G}}
	+\mathring{\mathbf{G}}\!\cdot\!
	\left[-\frac{\partial^2 \hat{W}}{\partial\mathbf{C}^\flat\partial\mathbf{C}^\flat}
	\!:\!\delta\mathbf{C}^\flat
	+\frac{\partial^2 \hat{W}}{\partial\mathbf{G}\partial\mathbf{G}}\!:\!\delta\mathbf{G} \right]\\
	&=\vepsr\cdot \frac{\partial \hat{W}}{\partial \mathbf{G}}
	+\mathring{\mathbf{G}}\!\cdot\!
	\left[-2\frac{\partial^2 \hat{W}}{\partial\mathbf{C}^\flat\partial\mathbf{C}^\flat}
	\!:\! \boldsymbol{\varepsilon}
	+2\frac{\partial^2 \hat{W}}{\partial\mathbf{G}\partial\mathbf{G}}\!:\! \vepsr \right] \\
	&=\vepsr\cdot \frac{\partial \hat{W}}{\partial \mathbf{G}}
	+\frac{1}{2}\mathring{\mathbf{G}}\!\cdot\!
	\left[- \boldsymbol{\mathbb{C}}
	\!:\! \boldsymbol{\varepsilon} + \CCG \!:\! \vepsr \right]
	 \,,
\end{aligned}
\end{equation}
where
\begin{equation}
	\CCG :=4\left[\frac{\partial^2 \hat{W}}{\partial \mathbf{G} \partial \mathbf{G}}
	\right]_{\varphi=\iota,\Fr=\mathbf I}\,.
\end{equation}
Taking derivative with respect to $\mathbf{G}$ of both sides of \eqref{Infinitesimal-MC} one obtains
\begin{equation} 
	\frac{\partial^2 \hat{W}}{\partial \mathbf{G}\partial \mathbf{C}^\flat}\cdot\mathbf{C}^{\flat}
	+\frac{\partial^2 \hat{W}}{\partial \mathbf{G}\partial \mathbf{G}}\cdot\mathbf{G}
	+\frac{\partial \hat{W}}{\partial \mathbf{G}}\otimes\mathbf{I}
	=\mathbf{0}\,.
\end{equation}
Evaluating this at the initial configuration one obtains (recall that $\mathring{\mathbf{C}}^{\flat}=\mathring{\mathbf{G}}$)
\begin{equation} \label{G-derrivative}
	\frac{\partial \hat{W}}{\partial \mathbf{G}}\otimes\mathbf{I}+
	\left[-\frac{\partial^2 \hat{W}}{\partial\mathbf{C}^\flat\partial\mathbf{C}^\flat}
	+\frac{\partial^2 \hat{W}}{\partial\mathbf{G}\partial\mathbf{G}}\right]\!\cdot\!
	\mathring{\mathbf{G}}
	=\mathbf{0}\,,
\end{equation}
where \eqref{G-C-Identity} was used. Thus
\begin{equation} 
	\frac{\partial \hat{W}}{\partial \mathbf{G}}\cdot \vepsr
	=\frac{1}{4} \vepsr \!:\! (\boldsymbol{\mathbb{C}}-\CCG )\!\cdot\!\mathring{\mathbf{G}}	 \,.
\end{equation}
Using the minor and major symmetries of $\boldsymbol{\mathbb{C}}$ and $\CCG$, the above relationship is equivalent to 
\begin{equation} 
	\vepsr \cdot  \frac{\partial \hat{W}}{\partial \mathbf{G}}
	=\frac{1}{4} \mathring{\mathbf{G}} \!\cdot\! (\boldsymbol{\mathbb{C}}-\CCG )\!:\! \vepsr	 \,.
\end{equation}
Thus
\begin{equation} 
\begin{aligned}
	\frac{d}{d\epsilon}\Big|_{\epsilon=0} \left[\Fr_\epsilon^{-\star}\mathbf{G}_\epsilon
	\frac{\partial \hat{W}}{\partial \mathbf{G}_\epsilon}\right]
	&=\frac{1}{4} \mathring{\mathbf{G}} \!\cdot\! (\boldsymbol{\mathbb{C}}-\CCG )\!:\! \vepsr
	+\frac{1}{2}\mathring{\mathbf{G}}\!\cdot\!
	\left[-\boldsymbol{\mathbb{C}}
	\!:\! \boldsymbol{\varepsilon} + \CCG \!:\! \vepsr \right]
	= \frac{1}{4} \mathring{\mathbf{G}} \!\cdot\!
	\left[-2\boldsymbol{\mathbb{C}}	\!:\! \boldsymbol{\varepsilon} 
	+(\boldsymbol{\mathbb{C}}+\CCG)\!:\! \vepsr  \right]	 \,.
\end{aligned}
\end{equation}
Hence, the right-hand side of \eqref{Kinetic-Lin} is written as
\begin{equation} 
	\delta q\,\mathbf{I}-\mathring{q}\,\UFr^{-\star}
	+\mathring{\mathbf{G}}\!\cdot\! \boldsymbol{\mathbb{C}}\!:\! \boldsymbol{\varepsilon}
	-\frac{1}{2}\mathring{\mathbf{G}}\!\cdot\!(\CCG+\boldsymbol{\mathbb{C}}) \!:\!\vepsr\,.
\end{equation}

The initial Lagrange multipliers $\mathring{q}$ is calculated as follows.
The remodeling equation \eqref{Remodeling-Equation-Metric} at the initial configuration is written as
\begin{equation} 
	\frac{\partial \phi}{\partial \dot{\Fr}}\Bigg|_{\dot{\Fr}=\mathbf{0}}=
	\mathring{q}\,\mathbf{I}-2\mathring{\mathbf{G}}\frac{\partial W}{\partial \mathbf{G}} \,.
\end{equation}
It is assumed that $\mathring{\boldsymbol{B}}_r=\mathbf{0}$, and hence
\begin{equation} 
	\mathring{q}=\frac{2}{3}\frac{\partial W}{\partial \mathbf{G}}\!:\!\mathring{\mathbf{G}} \,.
\end{equation}
Using \eqref{G-derrivative}  one obtains
\begin{equation} 
	\frac{\partial \hat{W}}{\partial \mathbf{G}}\!:\!\mathring{\mathbf{G}}
	= \frac{1}{4}\mathring{\mathbf{G}}\!:\! \boldsymbol{\mathbb{C}} \!:\!\mathring{\mathbf{G}}
	-\frac{1}{4}\mathring{\mathbf{G}}\!:\!\CCG \!:\!\mathring{\mathbf{G}}  \,.
\end{equation}
Thus
\begin{equation} 
	\mathring{q}=\frac{1}{6}
	\mathring{\mathbf{G}}\!:\! (\boldsymbol{\mathbb{C}}-\CCG) \!:\!\mathring{\mathbf{G}} \,.
\end{equation}

Recall that $\phi=\hat{\phi}(X,\mathbf{C}^{\flat},\Fr,\dot\Fr,\mathbf{G})$, and hence
\begin{equation}
\begin{split}
	\frac{d}{d\epsilon}\Bigg|_{\epsilon=0}\frac{\partial \phi}{\partial \dot{\Fr}_\epsilon}
	&= \frac{\partial^2\hat\phi}{\partial \dot\Fr \partial \mathbf{C}^{\flat}}\!:\!
	\delta\mathbf{C}^{\flat}
	+ \frac{\partial^2\hat\phi}{\partial \dot\Fr \partial \Fr} \!:\! \delta\Fr
	+ \frac{\partial^2\hat\phi}{\partial \dot\Fr\partial \dot\Fr} \!:\!\delta\dot\Fr
	+\frac{\partial^2\hat\phi}{\partial \dot\Fr \partial \mathbf{G}}\!:\! \delta\mathbf{G}  \\
	&= 2\frac{\partial^2\hat\phi}{\partial \dot\Fr \partial \mathbf{C}^{\flat}}\!:\!
	\boldsymbol{\varepsilon}
	+ \frac{\partial^2\hat\phi}{\partial \dot\Fr \partial \Fr} \!:\! \UFr
	+ \frac{\partial^2\hat\phi}{\partial \dot\Fr\partial \dot\Fr} \!:\!\dot\UFr
	+2\frac{\partial^2\hat\phi}{\partial \dot\Fr \partial \mathbf{G}}\!:\! \vepsr \,,
\end{split}
\end{equation}
where all the partial derivatives are evaluated at the initial configuration corresponding to $(\varphi,\Fr)=(\iota,\mathbf I)$.
Let us define
\begin{equation}
	\hat{\boldsymbol{\mathbb{A}}}:=\frac{\partial^2\phi}{\partial \dot\Fr\partial \dot\Fr}\,,\qquad
	\hat{\boldsymbol{\mathbb{B}}}:=\frac{\partial^2\phi}{\partial \dot\Fr\partial \Fr}\,,\qquad
	\hat{\boldsymbol{\mathbb{C}}}:=\frac{\partial^2\phi}{\partial \dot\Fr \partial \mathbf{C}^{\flat}}\,,
	\qquad
	\hat{\boldsymbol{\mathbb{D}}}:=\frac{\partial^2\phi}{\partial\dot{\Fr} \partial\mathbf{G}} \,,
\end{equation}
evaluated at the initial configuration. Hence
\begin{equation}
	\delta \frac{\partial \phi}{\partial \dot{\Fr}}
	= 2\hat{\boldsymbol{\mathbb{C}}}\!:\!	\boldsymbol{\varepsilon}
	+ \hat{\boldsymbol{\mathbb{B}}} \!:\! \UFr
	+ \hat{\boldsymbol{\mathbb{A}}}\!:\!\dot\UFr
	+2\hat{\boldsymbol{\mathbb{D}}} \!:\! \vepsr \,.
\end{equation}
The dissipation potential is materially covariant (for anisotropic solids structural tensors need to be included for material covariance to hold), and hence 
\begin{equation} 
	\frac{\partial \hat{\phi}}{\partial \mathbf{C}^\flat}\cdot\mathbf{C}^{\flat}
	+\frac{\partial \hat{\phi}}{\partial \Fr}\cdot \Fr
	+\frac{\partial \hat{\phi}}{\partial \dot\Fr}\cdot \dot\Fr 
	+\frac{\partial \hat{\phi}}{\partial \mathbf{G}}\cdot\mathbf{G}=\mathbf{0}\,.
\end{equation}
Thus
\begin{equation} 
	\frac{\partial^2 \hat{\phi}}{\partial \dot\Fr \partial \mathbf{C}^\flat}\cdot\mathbf{C}^{\flat}
	+\frac{\partial^2 \hat{\phi}}{\partial \dot\Fr \partial \Fr}\cdot \Fr
	+\frac{\partial^2 \hat{\phi}}{\partial \dot\Fr \partial \dot\Fr}\cdot \dot\Fr 
	+\frac{\partial \hat{\phi}}{\partial \dot\Fr}\otimes \mathbf{I}
	+\frac{\partial^2 \hat{\phi}}{\partial \dot\Fr \partial \mathbf{G}}\cdot\mathbf{G}=\mathbf{0}\,.
\end{equation}
With respect to the initial configuration $(\mathbf{C}^\flat,\Fr,\dot\Fr,\mathbf{G})=(\mathring{\mathbf{G}},\mathbf I,\mathbf 0,\mathring{\mathbf{G}})$, this is simplified to read
\begin{equation} 
	\frac{\partial^2 \hat{\phi}}{\partial\dot\Fr \partial \mathbf{C}^\flat}\cdot\mathring{\mathbf{G}}
	+\frac{\partial^2 \hat{\phi}}{\partial\dot\Fr \partial \Fr}
	+\frac{\partial^2 \hat{\phi}}{\partial\dot\Fr \partial \mathbf{G}}\cdot\mathring{\mathbf{G}}
	=\mathbf{0}\,,
\end{equation}
i.e., $\hat{\boldsymbol{\mathbb{B}}}=-\hat{\boldsymbol{\mathbb{C}}}\cdot\mathring{\mathbf{G}}-\hat{\boldsymbol{\mathbb{D}}}\cdot\mathring{\mathbf{G}}$.
Thus
\begin{equation}
	\delta \frac{\partial \phi}{\partial \dot{\Fr}}
	= \hat{\boldsymbol{\mathbb{A}}}\!:\!\dot\UFr
	+2\hat{\boldsymbol{\mathbb{C}}}\!:\! \boldsymbol{\varepsilon}
	+(\hat{\boldsymbol{\mathbb{D}}}-\hat{\boldsymbol{\mathbb{C}}})\!:\! \vepsr
	\,.
\end{equation}
Therefore, the linearized remodeling equation is written as
\begin{equation}
	\hat{\boldsymbol{\mathbb{A}}}\!:\!\dot\UFr
	+2\hat{\boldsymbol{\mathbb{C}}}\!:\! \boldsymbol{\varepsilon}
	+(\hat{\boldsymbol{\mathbb{D}}}-\hat{\boldsymbol{\mathbb{C}}})\!:\! \vepsr
	=\delta q\,\mathbf{I}-\frac{1}{6}\left[\mathring{\mathbf{G}}\!:\! (\boldsymbol{\mathbb{C}}-\CCG) 
	\!:\!\mathring{\mathbf{G}}\right]\UFr^{-\star}
	+\mathring{\mathbf{G}}\!\cdot\! \boldsymbol{\mathbb{C}}\!:\! \boldsymbol{\varepsilon}
	-\frac{1}{2}\mathring{\mathbf{G}}\!\cdot\!(\CCG+\boldsymbol{\mathbb{C}}) \!:\!\vepsr
	\,.
\end{equation}
This can be written in terms of $\UFr$ as
\begin{equation}
	\hat{\boldsymbol{\mathbb{A}}}\!:\!\dot\UFr
	+\left[(\hat{\boldsymbol{\mathbb{D}}}-\hat{\boldsymbol{\mathbb{C}}})
	+\frac{1}{2}\mathring{\mathbf{G}}\!\cdot\!(\CCG+\boldsymbol{\mathbb{C}})  \right]
	\!:\! \mathring{\mathbf{G}}\UFr
	+\frac{1}{6}\left[\mathring{\mathbf{G}}\!:\! (\boldsymbol{\mathbb{C}}-\CCG) 
	\!:\!\mathring{\mathbf{G}}\right]\UFr^{-\star}
	+(2\hat{\boldsymbol{\mathbb{C}}}-\mathring{\mathbf{G}}\!\cdot\! \boldsymbol{\mathbb{C}})
	\!:\! \boldsymbol{\varepsilon}
	=\delta q\,\mathbf{I}
	\,.
\end{equation}
The kinetic equation is more compactly written as
\begin{equation}
	\hat{\boldsymbol{\mathbb{A}}}\!:\!\dot\UFr
	+\hat{\boldsymbol{\mathbb{H}}}\!:\!\UFr
	+\hat{\boldsymbol{\mathbb{K}}}\!:\!\UFr^{-\star}
	+\hat{\boldsymbol{\mathbb{L}}}\!:\! \boldsymbol{\varepsilon}
	=\delta q\,\mathbf{I}
	\,.
\end{equation}
It is seen even in the linear approximation the kinetic equation has a contribution from elastic deformations.
$\delta q$ can be eliminated by taking the trace of both sides, and hence
\begin{equation}
	\hat{\boldsymbol{\mathbb{A}}}\!:\!\dot\UFr
	+\hat{\boldsymbol{\mathbb{H}}}\!:\!\UFr
	+\hat{\boldsymbol{\mathbb{K}}}\!:\!\UFr^{-\star}
	+\hat{\boldsymbol{\mathbb{L}}}\!:\! \boldsymbol{\varepsilon}
	-\frac{1}{3}\operatorname{tr}\left[
	\hat{\boldsymbol{\mathbb{A}}}\!:\!\dot\UFr
	+\hat{\boldsymbol{\mathbb{H}}}\!:\!\UFr
	+\hat{\boldsymbol{\mathbb{K}}}\!:\!\UFr^{-\star}
	+\hat{\boldsymbol{\mathbb{L}}}\!:\! \boldsymbol{\varepsilon}\right]\mathbf{I}=\mathbf{0}
	\,.
\end{equation}

\section{Conclusions}  \label{Sec:Conclusions}

In this paper we formulated the nonlinear mechanics of material remodeling as a special class of anelastic processes with an internal constraint, namely volume and mass conserving material evolutions. Remodeling alters the local stress-free configuration of the body and the time dependence of the energy function is through a remodeling tensor $\Fr$. The remodeling tensor changes the material metric of the body and makes the structural tensors time dependent. However, the symmetry of the material is preserved in the sense that the material symmetry groups at different times are related to that of the initial body in the form of conjugacy through $\Fr$. We specifically studied remodeling for isotropic, transversely isotropic, orthotropic, and monoclinic solids. We derived the governing equations of a remodeling body variationally using the Lagrange-d'Alembert principle within a two-potential setting. An energy function is assumed that depends on strain, material metric, and some time-dependent structural tensors. The dissipation potential, in addition to those fields, depends on the rate of remodeling tensor as well. 
In addition to the energy function and dissipation potential, we introduced a remodeling energy that quantifies the tendency of local material remodeling, e.g., fiber reorientation, in response to the local strain  (stress). 

We derived an explicit remodeling equation for a general remodeling process for both isotropic and anisotropic solids. We also considered general $SO(3)$-remodeling and the special case of fiber reorientation when the body is reinforced with one or two families of fibers. Our kinetic equation is a generalization of \citet{Menzel2005}'s reorientation equation. In addition to the remodeling energy, the elastic strain energy naturally contributes to the remodeling equation. 
In the case of one family of fibers we showed that as long as the dissipation potential does not have a term linear in the rate of fiber tangent vector, the principal directions of the right Cauchy-Green strain are equilibrium points for the remodeling equation. 
We briefly discussed the first and second laws of thermodynamics and the restrictions they impose on the dissipation potential.
We studied three examples of remodeling in fiber-reinforced solids under some finite (universal) deformations. 
Finally the governing equations of the nonlinear theory were linearized with respect to an initial stress-free configuration in order to derive a linearized theory.

The numerical results for the three examples showed a wide variety of possible behaviors with the proposed remodeling framework. For all the three examples, assuming an initially helical family of fibers in a solid cylinder, we observe that based on the applied loading and the value of the remodeling energy parameter, the fibers can remodel to align along different directions. In the first two examples involving finite extension of cylinders, the remodeling was found to be independent of the initial fiber orientation or the radial coordinate. However, for the third example involving torsion, remodeling depends on both the initial fiber orientation and the radial coordinate. The remodeling process was often found to be non-monotonic with the loading. Under force-control loading, it showed an unstable transition between two finitely separated states. Moreover, it was observed that the stress-deformation response evolves upon cyclic loading. While $SO(3)$-remodeling does not induce residual stresses, it was shown that the resulting stress state in the remodeled material under constant loading can be uniaxial or triaxial. There is no particular preferred stress that was found in our analysis for both types of loading.

All of the above observations were explained through the competition between the action of internal strain energy function and a remodeling energy (governed by the motivation to provide the material extra stiffness or strength). The dissipation potential only affects the time scale over which remodeling occurs. For a given material, a remodeling process dominated by strain energy, such as a material under very high loading, works to align fibers in a direction that minimizes strain energy. On the other hand, a remodeling process dominated by remodeling energy, such as a material under small loading, works to align fibers in the direction of maximum principal strain according to our constitutive choice. 
Observations of collagen fibers in biological tissues remodeling themselves into a state of non-zero stress or helical orientation under uniaxial stretch have been widely reported in the literature. However, previously, only empirical models were proposed to describe these observations. The energetic competition proposed in this work provides a plausible physical explanation for the experimental observations.

The macroscopic remodeling framework presented in this paper involves three constitute inputs: (i) the material strain energy function, (ii) the dissipation potential, (iii) a remodeling energy. For a given material of interest, e.g., a soft tissue containing collagen fibers, calibration of a model based on this framework would critically require the knowledge of the first and third inputs. Many experimental and analytical methods exist to characterize strain energy function of a fiber-reinforced material. Characterization of the remodeling energy would likely have to be conducted by fitting the model to structural-level non-homogeneous experimental observations of remodeling in living tissues. Numerical results showed that remodeling energy is dominant under small loading, thus, a small loading phase is likely to provide the most suitable data.

\section*{Acknowledgement}

This work was partially supported by NSF -- Grant No. CMMI 1939901.

\bibliographystyle{abbrvnat}
\bibliography{ref}

\end{document}